\documentclass[twocolumn,pra,twocolumn,superscriptaddress]{revtex4}
\usepackage{color}
\usepackage{amsmath}
\usepackage{amssymb}
\usepackage{graphicx}
\usepackage{tikz}
\usepackage{xr}
\usepackage{float}
\usepackage{bbm}
 \usepackage{mathrsfs}
\usepackage{epsfig} 
\usepackage{verbatim}
\usepackage{array}
\usepackage{setspace}
\usepackage{multirow}

\newcommand*\xbar[1]{%
	\hbox{%
		\vbox{%
			\hrule height 0.5pt 
			\kern0.4ex
			\hbox{%
				\kern-0.1em
				\ensuremath{#1}%
				\kern-0.1em
			}%
		}%
	}%
}

\newcommand*\xbarl[1]{%
	\hbox{%
		\vbox{%
			\hrule height 0.5pt 
			\kern0.25ex
			\hbox{%
				\kern-0.1em
				\ensuremath{#1}%
				\kern-0.1em
			}%
		}%
	}%
}

\usepackage[unicode=true,
 bookmarks=true,bookmarksnumbered=false,bookmarksopen=false,
 breaklinks=false,pdfborder={0 0 1},backref=false,colorlinks=true]
 {hyperref}
\hypersetup{
 linkcolor=magenta, urlcolor=blue, citecolor=blue, pdfstartview={FitH}, hyperfootnotes=false, unicode=true}
 
\usepackage{color}

\makeatletter
\@ifundefined{textcolor}{}
{%
 \definecolor{BLACK}{gray}{0}
 \definecolor{WHITE}{gray}{1}
 \definecolor{RED}{rgb}{1,0,0}
 \definecolor{GREEN}{rgb}{0,1,0}
 \definecolor{BLUE}{rgb}{0,0,1}
 \definecolor{CYAN}{cmyk}{1,0,0,0}
 \definecolor{MAGENTA}{cmyk}{0,1,0,0}
 \definecolor{YELLOW}{cmyk}{0,0,1,0}
}

\usepackage{amsfonts}\usepackage{tabularx}\usepackage{dcolumn}\usepackage{bm}\usepackage{graphicx}\usepackage{epstopdf}

\hypersetup{urlcolor=blue}

\makeatother

\newcolumntype{C}[1]{>{\centering\arraybackslash$}p{#1}<{$}}
\newcommand\scalemath[2]{\scalebox{#1}{\mbox{\ensuremath{\displaystyle #2}}}}

\usepackage{algorithm}
\usepackage{algorithmic}

\begin{document}

\widetext

\title{Charge noise suppression in capacitively coupled singlet-triplet spin qubits under magnetic field}

\author{Guo Xuan Chan}
\affiliation{Department of Physics, City University of Hong Kong, Tat Chee Avenue, Kowloon, Hong Kong SAR, China, and City University of Hong Kong Shenzhen Research Institute, Shenzhen, Guangdong 518057, China}

\author{J.~P.~Kestner}
\affiliation{Department of Physics, University of Maryland Baltimore County, Baltimore, Maryland 21250, USA}

\author{Xin Wang}
\email{x.wang@cityu.edu.hk}
\affiliation{Department of Physics, City University of Hong Kong, Tat Chee Avenue, Kowloon, Hong Kong SAR, China, and City University of Hong Kong Shenzhen Research Institute, Shenzhen, Guangdong 518057, China}
\date{\today}

\begin{abstract}
	Charge noise is the main hurdle preventing high-fidelity operation, in particular that of two-qubit gates, of semiconductor-quantum-dot-based spin qubits. While certain sweet spots where charge noise is substantially suppressed have been demonstrated in several types of spin qubits, the existence of one for coupled singlet-triplet qubits is unclear. We theoretically demonstrate, using full configuration-interaction calculations, that a range of nearly sweet spots appear in the coupled singlet-triplet qubit system when a strong enough magnetic field is applied externally. We further demonstrate that ramping to and from the judiciously chosen nearly sweet spot using sequences based on the shortcut to adiabaticity offers maximal gate fidelities under charge noise and phonon-induced decoherence. These results should facilitate realization of high-fidelity two-qubit gates in singlet-triplet qubit systems.
\end{abstract}

\maketitle

\section{Introduction} Singlet-triplet qubits, defined by two-electron spin states confined in semiconductor double-quantum-dot (DQD) devices, are promising candidates for realization of large-scale quantum-dot quantum computation \cite{Petta.05,Shulman.12,Levy.02,Wu.14,Maune.12,Barthel.10,Shi.11,Takeda.20,Cerfontaine.20,Eng.15,Noiri.18,Harvey.17}. In these systems, the charge noise directly affects the control over the spin qubits and is thus the key obstacle preventing high fidelity quantum control \cite{Cao.13,Shinkai.09,Hayashi.03,Petersson.10,Dovzhenko.11,Gorman.05,Shi.13}. A useful strategy to mitigate charge noise is to operate the qubits near the so-called ``sweet spots'' where the control (e.g. the exchange interaction between spins) is first-order insensitive to charge noise \cite{Reed.16,Martins.16,Abadillo.19,Yang.17,Yang.18,Medford.13,Taylor.13,Kim.15,Cao.16,Shi.14,Russ.17,Malinowski.17,Shim.18}. While this strategy has been successfully demonstrated in a variety of single-qubit devices, the existence of any sweet spot, in particular for two singlet-triplet qubits, is far less obvious.

Entangling operations between singlet-triplet qubits are typically carried out by exploiting either the capacitive interaction \cite{Shulman.12,Taylor.05,Nichol.17,Nielsen.12,Hiltunen.14,Buterakos.19,Ramon.11,Calderon.15,Wolfe.17,Stepanenko.07,Yang.11,Srinivasa.15} or exchange coupling \cite{Li.12,Klinovaja.12,Mehl.14,Wardrop.14,Buterakos.18,Buterakos.18.2,Cerfontaine.20.2} between two DQD devices. Capacitive gates are achieved when the tunneling between the two DQDs is suppressed, while the Coulomb interaction mediates the inter-qubit interaction. Exchange gates, on the other hand, are mediated by the exchange coupling between two neighbouring spins between two DQDs, which can be manipulated by inter-dot tunneling and energy detuning between the two spins. In this work, we focus on capacitively coupled singlet-triplet qubits.

Gate operations on two singlet-triplet qubits coupled by capacitive interactions typically have fidelities $\sim$72\% \cite{Shulman.12} and can be improved to $\sim$90\% \cite{Nichol.17} by applying large magnetic gradient. However, to meet the stringent requirement for quantum error correction, suppression of charge noise becomes emergent. Theoretical calculations \cite{Taylor.05}, particularly using variations of the configuration interaction (CI) method \cite{Stepanenko.07,Srinivasa.15,Li.12,Yang.11}, are widely employed to search for the sweet spots. Ref.~\cite{Yang.11} proposes that there exists a sweet spot when the two singlet-triplet qubits are aligned at an appropriate angle, while Ref.~\cite{Wolfe.17} claims that a sweet spot may appear at a certain detuning value. However, these results are obtained from the Hund-Mulliken approximation keeping the lowest orbital in each quantum-dot, and it is unclear whether the results hold when higher orbitals are taken into account.  Furthermore, Ref.~\cite{Wolfe.17} assumed that the charge states of each qubit are independent of each other, but that assumption breaks down in the parameter regime where the sweet spot was claimed to occur. Refs.~\cite{Nielsen.12,Hiltunen.14,Buterakos.19}, using a more sophisticated CI method either by involving excited orbitals or populating the quantum-dot system with $s$-type Gaussian functions, have shown that,
while a sweet spot may exist for the capacitive two-qubit coupling,  it is \emph{not} at the same time a sweet spot for single-qubit exchange interactions, which limits the usefulness of those prior results in experiments.

All these previous CI calculations were performed without an external magnetic field. In this Letter, we show, using full CI calculations, that a range of \emph{nearly} sweet spots appear in the coupled singlet-triplet qubit system,
when a strong enough magnetic field is applied externally. Around these nearly sweet spots, both the capacitive coupling and the single-qubit exchange interactions are very weakly dependent on the charge noise, making possible high fidelity manipulations. We demonstrate that operating in the nearly-sweet-spot regime yields the entangling gate with fidelity much higher compared to the previous proposals \cite{Nielsen.12,Nichol.17}. Moreover, the extended range of this nearly-sweet-spot regime allows for application of shortcuts to adiabaticity for the ramping pulses to and from the operating point, which leads to about one order of magnitude improvement in the gate fidelity. In contrast to \cite{Nichol.17} by which high fidelity entangling gate results from application of large magnetic gradient on singly occupied dots, our model benefits from strong capacitive coupling with weak coupling to charge-noise. Our results should facilitate realization of high-fidelity two-qubit gates in singlet-triplet qubit systems.


\section{Model}
\begin{figure}[t]
	\centering
	\includegraphics[width=0.9\columnwidth]{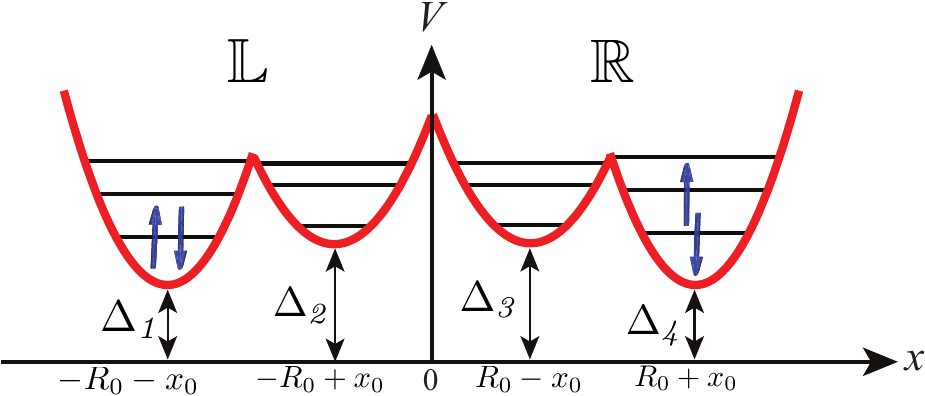}
	\caption{Schematic illustration of the model potential given in Eq.~\eqref{eq:Vpotential}.}
	\label{fig:Vpotential}
\end{figure}

We consider an $n$-electron system $H=\sum h_j + \sum{e^2}/\epsilon\left|\mathbf{r}_j-\mathbf{r}_k\right|$
with the single-particle Hamiltonian $h_j = {(-i\hbar \nabla_j+e \mathbf{A}/c)^2}/{2m^*}+V(\mathbf{r})+g^*\mu_B \mathbf{B}\cdot \mathbf{S}$.
The confinement potential of a double double-quantum-dot (DDQD) device can be modeled in the $xy$ plane as (cf.~Fig.~\ref{fig:Vpotential})
\begin{align}\label{eq:Vpotential}
	V(\mathbf{r}) &= \frac{1}{2} m^* \omega_0^2 \text{Min}\Big[\left(\mathbf{r}-\mathbf{R}_{\it 1}\right)^2+\Delta_{\it 1},\left(\mathbf{r}-\mathbf{R}_{\it 2}\right)^2+\Delta_{\it 2},\notag\\
	&\left(\mathbf{r}-\mathbf{R}_{\it 3}\right)^2+\Delta_{\it 3},\left(\mathbf{r}-\mathbf{R}_{\it 4}\right)^2+\Delta_{\it 4}\Big],
\end{align}
where $\mathbf{R}_j=(\pm R_0 \pm x_0,0)$ are the minima of the parabolic wells \cite{Nielsen.10}. The inter-dot distance is $2 x_0$ while the inter-DQD distance is $2 R_0$.

With each DQD hosting one singlet-triplet qubit, the DDQD defines a pair of capacitively coupled singlet-triplet qubits. The two-qubit logical states are $|SS\rangle$,$|ST\rangle$,$|TS\rangle$ and $|TT\rangle$, where $|S\rangle$ and $|T\rangle$ are spin-singlet and unpolarized spin-triplet ($S_z=0$) states respectively. Without a magnetic field gradient, the system Hamiltonian, $H_\text{int}$, is diagonal in the bases of logical states as \cite{Ramon.11,Calderon.15,Stepanenko.07,Wolfe.17},
\begin{equation}
	H_\text{int} =  J_\mathbb{L}^\text{eff} \sigma_z \otimes I + J_\mathbb{R}^\text{eff} I \otimes \sigma_z + \alpha \sigma_z \otimes \sigma_z,
\end{equation}
where
\begin{subequations}
	\begin{align}
		\begin{split}
			\alpha &= \frac{1}{4}(E_{|SS\rangle}-E_{|ST\rangle}-E_{|TS\rangle}+E_{|TT\rangle}),
			\label{eq:alpha}
		\end{split}\\
		\begin{split}
			J^{\text{eff}}_\mathbb{L} &= \frac{1}{4}\left[E_{|TT\rangle}-E_{|SS\rangle}-(E_{|ST\rangle}-E_{|TS\rangle})\right],
			\label{eq:Jeff1}
		\end{split}\\
		\begin{split}
			J^{\text{eff}}_\mathbb{R} &= \frac{1}{4}\left[E_{|TT\rangle}-E_{|SS\rangle}+(E_{|ST\rangle}-E_{|TS\rangle})\right].
			\label{eq:Jeff2}
		\end{split}
	\end{align}
	\label{eq:twoQubitHam}
\end{subequations}
The effective exchange energies $J^{\text{eff}}_\mathbb{L}$ and $J^{\text{eff}}_\mathbb{R}$ for the qubit defined in the left ($\mathbb{L}$) and right  ($\mathbb{R}$) DQD   respectively, contain both the individual exchange energy of the DQD in absence of the other, as well as a capacitive shift caused by the neighboring DQD. $\alpha$ is the capacitive inter-qubit coupling.

We solve the problem using the full configuration interaction (Full-CI) technique \cite{Barnes.11}, detailed in Sec.~I of the Supplemental Material \cite{sm}. We use parameters appropriate for GaAs, where the permitivity $\epsilon= 13.1\epsilon_0$, effective electron mass $m^*=0.067m_e$, confinement strength of the quantum dots $\hbar \omega_0=1$meV, effective Bohr radius $a_B=\sqrt{\hbar/m^*\omega_0}\approx 34$nm, $x_0=2.5a_B$ and $R_0=9a_B$. The inter-qubit distance $R_0$ is chosen such that the tunneling between qubits is negligible thus only the capacitive coupling remains. The parameters are summarized in Sec.~II in Supplemental Material. Practically, we truncate the Full-CI calculation using a cutoff scheme \cite{Barnes.11}, keeping orbitals up to $n=4$ Fock-Darwin states.

\begin{figure}[t]
	\centering
	\includegraphics[width=0.98\columnwidth]{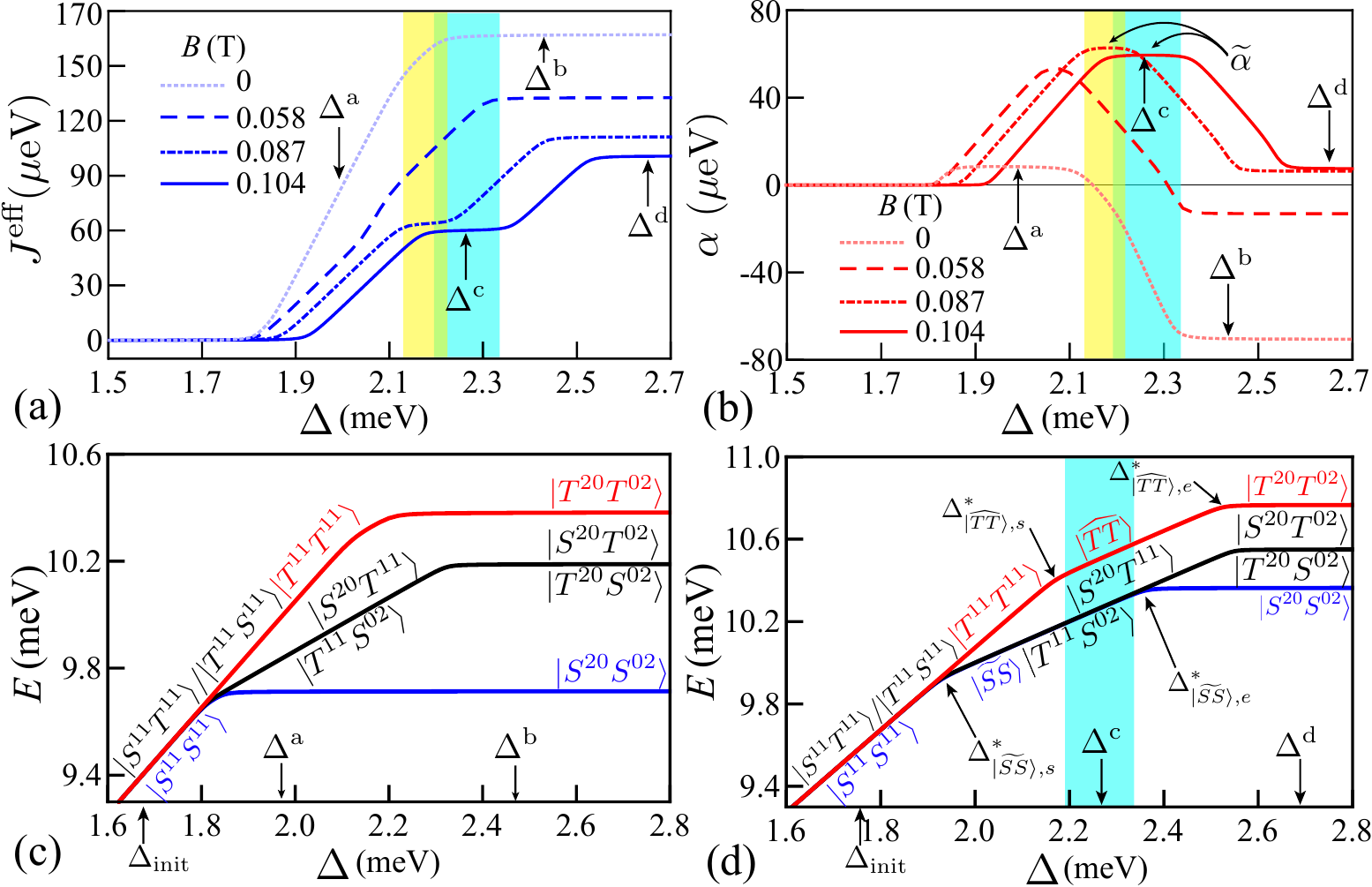}
	\caption{Nearly sweet spots in the ``Outer'' detuning scheme. (a) and (b): $J^\text{eff}$ and $\alpha$ v.s. detuning $\Delta$ for several magnetic field strengths $B$ as indicated. (c) and (d): Energy levels v.s. detuning $\Delta$ for (c) $B=0$ and (d) $B=0.104$ T. $\Delta^\mathrm{a,b,c,d}$ ($\Delta_\mathrm{init}$) are proposed operating (initialization) points which will be discussed later. The yellow and cyan area indicate the nearly-sweet-spot regime for $B=0.087$ and  0.104 T respectively, and their overlap, indicated by the green area, should be considered as belonging to both. Note that $|\widetilde{SS}\rangle$ and $|\widehat{TT}\rangle$ exist for a very small range in (c) and are therefore not indicated (see Sec. IV in \cite{sm} for more details).}
	\label{fig:JeffiBfoSSTTEvals}
\end{figure}

\section{Results}
\subsection{Nearly sweet spot}
We only consider symmetric detuning of two qubits, i.e. the detuning values on both qubits are equal. There are thus three possibilities:
\begin{equation}\label{eq:detuningScheme}
	\begin{split}
		\text{``Outer'': } &\Delta_{\it 2} = \Delta_{\it 3} = \Delta >0, \Delta_{\it 1} = \Delta_{\it 4} = 0,\\
		\text{``Center'': }&\Delta_{\it 1} = \Delta_{\it 4} = \Delta>0, \Delta_{\it 2} = \Delta_{\it 3} = 0,\\
		\text{``Right'': } &\Delta_{\it 1} = \Delta_{\it 3} = \Delta>0, \Delta_{\it 2} = \Delta_{\it 4} =0.
	\end{split}
\end{equation}
In the main text, we focus on the ``Outer'' scheme where $J_{\mathbb{L}}^\text{eff}=J_{\mathbb{R}}^\text{eff}\equiv J^\text{eff}$, and a discussion on others can be found in Sec.~V of the Supplemental Material \cite{sm}.

\begin{figure}[t]
	\centering
	\includegraphics[width=0.98\columnwidth]{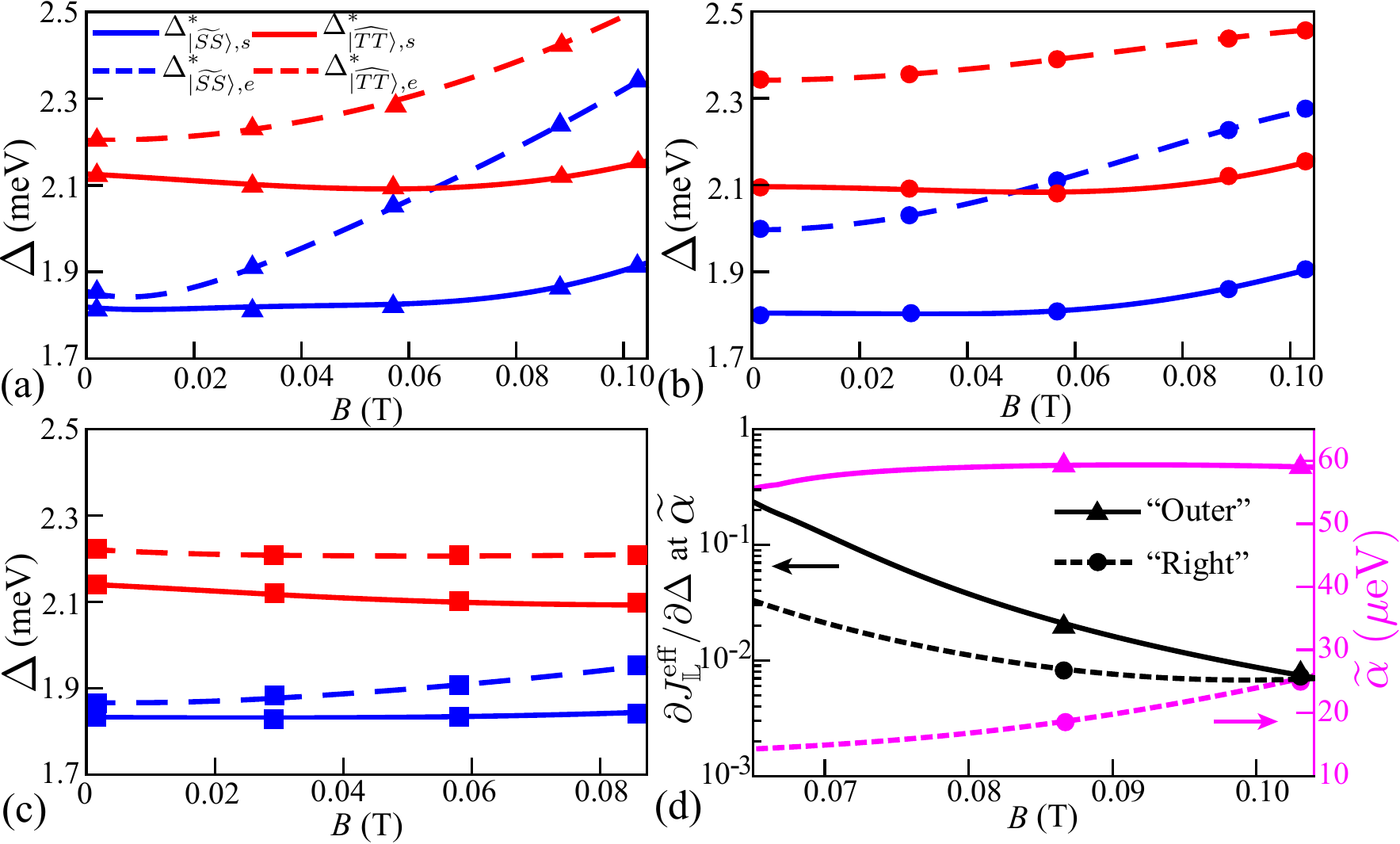}
	\caption{$\Delta_{|\widetilde{SS}\rangle,s}^*,\Delta_{|\widetilde{SS}\rangle,e}^*,\Delta_{|\widehat{TT}\rangle,s}^*$ and $\Delta_{|\widehat{TT}\rangle,e}^*$ as a function of magnetic field strength $B$ in the (a)``Outer'', (b) ``Right'' and (c) ``Center'' detuning scheme. (d) Black curves (using the left $y$-axis): the values of $\partial J_\mathbb{L}^\mathrm{eff}/\partial\Delta$ evaluated at the $\Delta$ value where $\alpha$ reaches its maximal value $\widetilde{\alpha}$. Magenta curves (using the right $y$-axis): the maximal values of $\alpha$ (i.e. $\widetilde{\alpha}$) vs the magnetic field. The symbols in magenta and black represent results from the same detuning scheme.}
	\label{fig:deltaRangeOfABOND}
\end{figure}

Fig.~\ref{fig:JeffiBfoSSTTEvals}(a) and (b) show the dependence of $J^\text{eff}$ and $\alpha$ on detuning $\Delta$ under different magnetic fields, which is the key result of this paper. When $B=0$, $\alpha$ develops two flat regimes. A sweet spot exists for $\alpha$ around $\Delta\approx2$ meV, but the same $\Delta$ range does not give any nearly sweet spots in $J^\text{eff}$. This result is consistent with Refs.~\cite{Nielsen.12,Hiltunen.14,Buterakos.19,Shulman.12}. Another regime where both $J^\text{eff}$ and $\alpha$ have nearly sweet spots is for $\Delta\gtrsim2.3$ meV. This was envisaged by \cite{Dial.13} based on single DQD results in far detuned regime, but ramping to such high detuning would expose the qubit to severe leakage or decoherence, which is therefore impractical. Increasing $B$ moves the sweet spot for $\alpha$ at $\Delta\approx2$ meV to the right, while a nearly-sweet-spot regime gradually appears for $J^\text{eff}$ at $B\gtrsim0.087$ T. At $B=0.104$ T, the nearly-sweet-spot regime where both $J^\text{eff}$ and $\alpha$ are very weakly dependent on $\Delta$ is quite extended, as indicated by the cyan area. At the same time, the $\alpha$ value is enhanced so as to reduce the gate time and minimize the accumulation of gate error. We shall see later that the detuning $\Delta^\mathrm{c}$ yields the highest gate fidelity. We also note that when $\alpha$ reaches its maximal value, $\widetilde{\alpha}$, $\partial\alpha/\partial\Delta=0$, while at the same $\Delta$ value $\partial J^\mathrm{eff}/\partial\Delta$ is small ($\sim 10^{-2}$) but not exactly zero (cf. Fig.~\ref{fig:deltaRangeOfABOND}(d)). This is the reason we call the region \emph{nearly} sweet spots. It is also found that the nearly-sweet-spot region exists for asymmetric cases, e.g. elliptical confinement potential or asymmetric confinement strengths, where the details can refer to Secs.~XIII and XV in the Supplemental Material respectively.

Fig.~\ref{fig:JeffiBfoSSTTEvals}(c) and (d) show the energy level structure of the system as the detuning is varied. The states are labeled using a Dirac ket with the first entry being the state of the left DQD and the second the right DQD. The state of one qubit (i.e., one DQD) is either a singlet (S) or a triplet (T) with the superscript showing the charge configurations. For example, the four-electron state shown in Fig.~\ref{fig:Vpotential} can be understood as $|S^{20}T^{02}\rangle$ \cite{s20t02Note}. Detailed discussions of all relevant states in terms of the extended Hubbard model can be found in Sec.~IV of the Supplemental Material \cite{sm}.

Fig.~\ref{fig:JeffiBfoSSTTEvals}(c) shows the energy levels at zero magnetic field. All levels are parallel for $\Delta\gtrsim2.3$ meV, consistent with the observation that both $J^\text{eff}$ and $\alpha$ are weakly dependent on $\Delta$ in this range. Around $\Delta\approx2$ meV, the slopes of the curves can be combined in the fashion of Eq.~\eqref{eq:alpha}, implying that $\partial\alpha/\partial\Delta\approx 0$, but not for $J^\text{eff}$ (Eqs.~\eqref{eq:Jeff1} and \eqref{eq:Jeff2}), consistent with the observations from Fig.~\ref{fig:JeffiBfoSSTTEvals}(a) and (b). When a magnetic field $B=0.104$ T is applied, however, the situation changes. Two new states becomes significant: a bonding state $|\widetilde{SS}\rangle = (|S^{11}S^{02}\rangle+|S^{20}S^{11}\rangle)/\sqrt{2}$ and an anti-bonding state $|\widehat{TT}\rangle=(|T^{11}T^{02}\rangle-|T^{20}T^{11}\rangle)/\sqrt{2}$ \cite{s20t02Note}. These two states covers an extended $\Delta$ range in the energy levels. We can find the starting (s) and ending (e) points of these ranges by setting equal the energies of the states admixed at the avoided crossing points. For example, the starting point $\Delta^*_{|\widetilde{SS}\rangle,s}$ is found by setting the energies of $|\widetilde{SS}\rangle$ and $|S^{11}S^{11}\rangle$ equal, while the ending point $\Delta^*_{|\widehat{TT}\rangle,e}$ is found by setting equal energies of $|\widehat{TT}\rangle$ and $|T^{20}T^{02}\rangle$ equal. It is interesting to note that there exists a $\Delta$ range (the cyan area) where levels $|\widehat{TT}\rangle$, $|\widetilde{SS}\rangle$, $|S^{20}T^{11}\rangle$ and $|T^{11}S^{02}\rangle$ share \emph{almost} the same slope with respect to $\Delta$ ($\approx 0.996$), making the $\Delta$ derivatives of the r.h.s. of Eqs.~\eqref{eq:alpha}-\eqref{eq:Jeff2} almost vanish altogether. This is the origin of the nearly-sweet-spot range for both $J^\text{eff}$ and $\alpha$. The existence of this range is actually not specific to the parameters chosen here. A discussion on the generality of its existence is presented in Sec.~VI of the Supplemental Material \cite{sm}.

The relevant lowest energy levels of the DDQD system can be interpreted well using the extended Hubbard model \cite{sm}, allowing us to interpolate the Full-CI results to cover a range of parameters. Figure~\ref{fig:deltaRangeOfABOND}(a), (b) and (c) show the values of $\Delta_{|\widetilde{SS}\rangle,s}^*,\Delta_{|\widetilde{SS}\rangle,e}^*,\Delta_{|\widehat{TT}\rangle,s}^*$ and $\Delta_{|\widehat{TT}\rangle,e}^*$ as functions of magnetic field in the ``Outer'', ``Right'' and  ``Center'' detuning scheme, respectively. The symbols are data points extracted from the Full-CI calculation, and the lines are interpolations using the extended Hubbard model. We see that only the ``Outer'' and ``Right'' detuning scheme gives $\Delta_{|\widetilde{SS}\rangle,e}^* > \Delta_{|\widehat{TT}\rangle,s}^*$ for sufficiently strong magnetic field, implying an overlapping region of $|\widetilde{SS}\rangle$ and $|\widehat{TT}\rangle$. No overlapping for the ``Center'' detuning scheme as $\Delta_{|\widetilde{SS}\rangle,e}^*$ is  less sensitive to magnetic field. In addition, the $\Delta$ range of overlap for the ``Outer'' scheme increases roughly linearly with the magnetic field for the values concerned, while there is a moderate increase for the ``Right'' scheme, resulting in $\approx0.2$ meV for the former while $\approx0.1$ meV for the latter at $B=0.104$ T. Fig.~\ref{fig:deltaRangeOfABOND}(d) shows the maximal value of $\alpha$, $\widetilde{\alpha}$, in the nearly-sweet-spot regime, as well as $\partial J_\mathbb{L}^\mathrm{eff}/\partial\Delta$ evaluated at the same $\Delta$ value where $\alpha$ reaches maximum ($\partial \alpha/\partial \Delta = 0$), for the ``Outer'' and ``Right''  schemes. For both schemes, $\partial J_\mathbb{L}^\mathrm{eff}/\partial\Delta$ is as small as $\sim10^{-2}$ for $B\gtrsim 0.1$ T, indicating that the susceptibility to charge noise is extremely weak. On the other hand, $\widetilde{\alpha}$ is much greater for the ``Outer'' scheme than the ``Right'' one, suggesting that the ``Outer'' scheme remains the optimal protocol to operate the coupled DDQD systems.

\subsection{CPHASE gate}
\begin{figure}[t]
	\centering{
		\includegraphics[width=0.9\columnwidth]{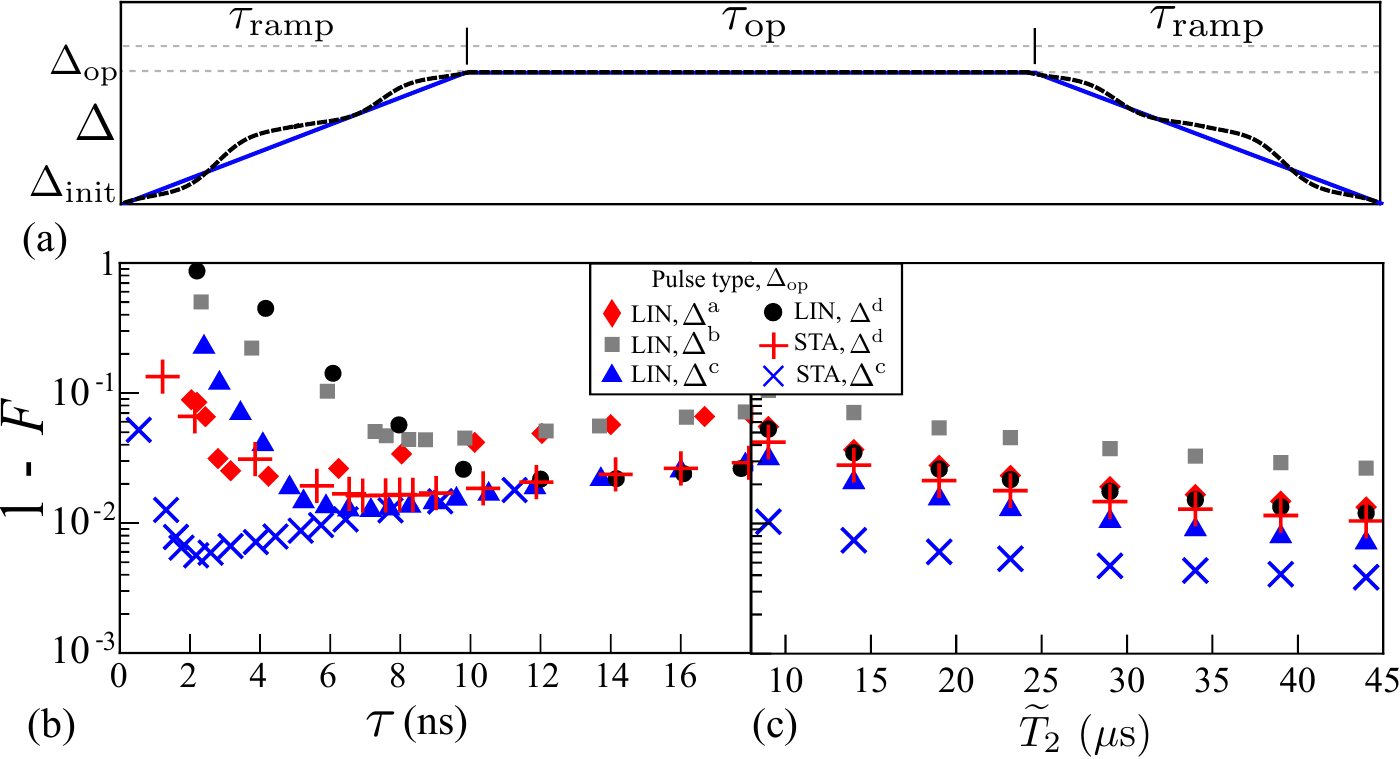}
	}
	\caption{(a) Detuning pulse sequences for the linear ramping scheme (LIN, solid line) and shortcut to adiabaticity (STA,  dashed line). (b) CPHASE gate infidelities as functions of the total gate time, $\tau$, for $\widetilde{T}_2=23$ $\mu s$. (c) CPHASE gate infidelities as functions of the reference charge-noise dephasing time $\widetilde{T}_2$ \cite{sm,Dial.13}. For each set of results, the gate time $\tau$ is chosen such that it produces the minimal gate infidelity as indicated in panel (b).
	}
	\label{fig:gateSummary}
\end{figure}

The inter-qubit coupling, $\sigma_z \otimes \sigma_z$, gives rise to a controlled-phase (CPHASE) gate \cite{Hiltunen.14,Nielsen.12}. The system is initialized at $\Delta_\mathrm{init}$ where $\alpha$ is negligible, and is then ramped to a larger detuning, $\Delta_{\text{op}}$, where the operation is performed with a reasonably strong $\alpha$. This ramping time is denoted as $\tau_\mathrm{ramp}$. After operating at $\Delta_{\text{op}}$ for a time $\tau_{\text{op}}$, the system is brought back to $\Delta_\mathrm{init}$ in $\tau_\mathrm{ramp}$ (see  Fig.~\ref{fig:gateSummary}(a)). The total gate time is therefore $\tau=2\tau_{\text{ramp}}+\tau_{\text{op}}$.

The evolution of the system in the logical subspace can be described by the master equation,
\begin{equation}\label{eq:masterEq}
	\begin{split}
		\dot{\rho} &=-i[H_{\text{int}},\rho]
		+\left(\gamma_{\varphi_\mathbb{L}}+\gamma_{\text{dep}_\mathbb{L}}\right) \mathcal{D} [\sigma_z \otimes I] \rho\\
		&+ \left(\gamma_{\varphi_\mathbb{R}}+\gamma_{\text{dep}_\mathbb{R}}\right) \mathcal{D} [I \otimes \sigma_z] \rho\\
		&+  \left(\gamma_{\varphi_{\mathbb{LR}}}+\gamma_{\text{dep}_\mathbb{LR}}\right) \mathcal{D} [\sigma_z \otimes \sigma_z] \rho+\sum_{j<k} \gamma_{\text{rel}_{jk}}\mathcal{D}[\sigma_{jk}]\rho,
	\end{split}
\end{equation}
where $\gamma_{\varphi_\mathbb{L}}$ $(\gamma_{\varphi_\mathbb{R}})$ and $\gamma_{\varphi_\mathbb{LR}}$ are the charge-noise dephasing rates  for qubit $\mathbb{L}$ $(\mathbb{R})$ and the capacitive coupling $\alpha$, respectively. All of them are proportional to a reference charge-noise dephasing time, $\widetilde{T}_2=1/\widetilde{\gamma}_\varphi$ \cite{sm}, which we shall use as our noise amplitude. $\gamma_{\text{rel}}$ ($\gamma_{\text{dep}}$) is the phonon-mediated relaxation (pure dephasing) rate. $\mathcal{D}[c]$ represents the dissipation superoperator $\mathcal{D}[c] \rho \equiv 2 c \rho c^\dagger - c^\dagger c \rho/2-\rho c^\dagger c/2$ \cite{Wiseman.09}. More details, including the derivation of the decoherence rates listed above can be found in Sec.~IX of the Supplemental Material \cite{sm}.

\begin{table}[t]
	\centering
	\begin{tabular}{|c|c|c|c|}
		\hline
		$B$ (T) & $\Delta_\text{op}$ & $\left|\alpha\right|$ ($\mu$eV)& $\partial J^{\text{eff}}/\partial \Delta$ \\
		\hline
		\hline
		\multirow{2}{*}{0} & $\Delta^\text{a}$ & $8.34$ &  $4.94\times10^{-1}$  \\
		\cline{2-4}
		& $\Delta^\text{b}$ & $70.32$ &  $1.74\times10^{-3}$  \\
		\cline{1-4}
		\multirow{2}{*}{0.104} & $\Delta^\text{c}$ & $57.2$ &  $7.50\times10^{-3}$  \\[1pt]
		\cline{2-4}
		& $\Delta^\text{d}$ & $7.52$ &  $2.83\times10^{-3}$  \\[1pt]
		\hline
	\end{tabular}
	\caption{Summary of the parameters for different $\Delta_{\text{op}}$.}
	\label{tab:tableGateSummary}
\end{table}

We have chosen  $\Delta^\mathrm{a,b,c,d}$ as candidates of $\Delta_\mathrm{op}$ (as indicated on Fig.~\ref{fig:JeffiBfoSSTTEvals}). The $\alpha$ values as well as $\partial J^{\text{eff}}/\partial \Delta$ for these points are summarized in Table~\ref{tab:tableGateSummary}. On one hand, $\partial J^{\text{eff}}/\partial \Delta$ is small for $\Delta^\mathrm{b,c,d}$, suggesting that the charge-noise-induced dephasing is suppressed. On the other hand, ramping the system to $\Delta^\text{a,c}$ requires less detuning sweeps compared to $\Delta_\text{b,d}$, suggesting that within the same $\tau_\mathrm{ramp}$, choosing $\Delta^\text{a,c}$ as the operating points limits the leakage. These considerations imply that $\Delta^\text{c}$ is the optimal choice as $\Delta_\mathrm{op}$.

We consider two ways of detuning the DDQD system from $\Delta_\mathrm{init}$ to $\Delta_\mathrm{op}$: a linear (LIN) ramping scheme where $d\Delta/dt=$ constant, as well as one based on shortcut to adiabaticity (STA) \cite{Chen.10,Chen.11} (see Fig.~\ref{fig:gateSummary}(a)). It is noticed that for $B = 0.104$ T, charge transitions of different logical states are either located at the same $\Delta$ (facilitated by the same inter-dot tunneling) or well-separated in $\Delta$ values (see Fig.~\ref{fig:JeffiBfoSSTTEvals}(d)), allowing us to apply concatenated STA pulse sequences, the details of which can be found in Sec.~X of the Supplemental Material \cite{sm}.
The application of STA pulse sequences allows the reduction of the total gate time $\tau$, without increasing the leakage, therefore suppressing decoherence. Note that STA pulse sequence is not available for $\Delta^\text{a,b}$ as the charge transitions occur very closely in $\Delta$ and cannot be individually addressed for different logical states (see Fig.~\ref{fig:JeffiBfoSSTTEvals}(c)).

We numerically simulate the master equation, Eq.~\eqref{eq:masterEq}, taking into account the leakage by expanding $H_\text{int}$ into the effective Hamiltonian block for each logical eigenstate \cite{sm}. The dephasing effect by hyperfine noise is neglected here as we found that the main limiting factors of the gate fidelity does not involve hyperfine fluctuation, for which the details are given in Sec.~XII D in the Supplemental Material.
The results of gate infidelities, $1-F$ \cite{FidelityNote}, as functions of $\tau$ and $\widetilde{T}_2$ are shown in Fig.~\ref{fig:gateSummary}(b) and (c) respectively. From Fig.~\ref{fig:gateSummary}(b), we see a reduction of infidelities at small $\tau$ for all results, but STA with operating point  $\Delta^\mathrm{c}$ gives the lowest infidelity at the shortest gate operation time, while LIN with $\Delta^\mathrm{c}$ gives the second lowest infidelity. When $\tau$ is large, the infidelities increase with $\tau$ due to accumulated exposure to various decoherence channels other than leakage, as expected. Fig.~\ref{fig:gateSummary}(c) shows the gate infidelities as functions of the reference charge-noise dephasing time $\widetilde{T}_2$, with the gate time $\tau$ for each set of results chosen such that it produces the minimal gate infidelity as indicated in panel (b). We see that in the LIN scheme, results calculated at $\Delta^{c}$ exhibits about a factor $2\sim4$ reduction in infidelity compared to other $\Delta_\mathrm{op}$ values, while using STA scheme offer another factor of $2\sim4$. Therefore the STA scheme in combination with the nearly sweet spot offers roughly an order of magnitude reduction in infidelities. We found out that similar results, including the existence of the nearly-sweet-spot region at large magnetic field and highest gate fidelity demonstrated by $\Delta^\text{c}$, are achieved for silicon DDQD device, of which the details are provided in Secs.~V and XII in the Supplemental Material.



\section{Conclusions Discussion}

We have shown, using Full CI calculations, that a range of nearly sweet spots, for both the single-qubit exchange energy as well as the capacitive coupling, appear in the coupled singlet-triplet qubit system under a strong enough external magnetic field. This range of nearly sweet spots arises due to the appearance of $|\widetilde{SS}\rangle$ and $|\widehat{TT}\rangle$ states under magnetic field, which occupy detuning ranges that increase with the magnetic field.

It is interesting to compare our capacitive gates to exchange-mediated ones studied in the literature \cite{Klinovaja.12,Li.12,Mehl.14,Wardrop.14,Buterakos.18,Buterakos.18.2,Cerfontaine.20.2}. Our proposal should be easier to implement since it only involves detuning ramping, one degree of freedom less as compared to exchange-mediated gates which involves both the inter-DQD tunneling and detuning. On the other hand, for exchange gates, leakage into states with zero $S_z$ is possible unless an additional magnetic field difference between the two DQDs is supplied. In contrast, capacitive gates are free from such leakage as the inter-dot tunneling is suppressed between two DQDs. Although leakage could occur when the detuning ramp passes through the charge transition points, it can be mitigated by pulse-shaping or adiabatic ramping.  In fact, we have demonstrated that ramping to and from the judiciously chosen nearly sweet spot using sequences based on the shortcut to adiabaticity offers maximal gate fidelities under charge noise and phonon-induced decoherence. Our results therefore should facilitate realization of high-fidelity two-qubit gates in coupled singlet-triplet qubit systems.

\section*{Acknowledgements} G.X.C.~and X.W.~are supported by the Key-Area Research and Development Program of GuangDong Province  (Grant No.~2018B030326001), the National Natural Science Foundation of China (Grant No.~11874312), the Research Grants Council of Hong Kong (Grant Nos.~11303617, 11304018, 11304920), and the Guangdong Innovative and Entrepreneurial Research Team Program (Grant No.~2016ZT06D348). J.~P.~K.~acknowledges support from the National Science Foundation under Grant No. 1915064 and the Army Research Office (ARO) under Grant Number W911NF-17-1-0287.

\bibliographystyle{apsrev4-1}

\onecolumngrid
\vspace{1cm}

\begin{center}
	{\bf\large Supplemental Material for ``Charge noise suppression in capacitively coupled singlet-triplet spin qubits under magnetic field"}
\end{center}
\vspace{0.5cm}

\setcounter{secnumdepth}{3}  
\setcounter{equation}{0}
\setcounter{figure}{0}
\setcounter{table}{0}
\setcounter{section}{0}

\renewcommand{\theequation}{S-\arabic{equation}}
\renewcommand{\thefigure}{S\arabic{figure}}
\renewcommand{\thetable}{S-\Roman{table}}
\renewcommand\figurename{Supplementary Figure}
\renewcommand\tablename{Supplementary Table}
\newcommand\citetwo[2]{[S\citealp{#1}, S\citealp{#2}]}
\newcommand\citecite[2]{[\citealp{#1}, S\citealp{#2}]}

\newcolumntype{M}[1]{>{\centering\arraybackslash}m{#1}}
\newcolumntype{N}{@{}m{0pt}@{}}

\makeatletter \renewcommand\@biblabel[1]{[S#1]} \makeatother

\makeatletter \renewcommand\@biblabel[1]{[S#1]} \makeatother


\onecolumngrid

In this Supplemental Material we provide necessary details complementary to results shown in the main text.

\section{Full Configuration Interaction (Full-CI) Calculation}\label{sec:fullCIMethod}
\subsection{Single particle basis states}
The Hamiltonian for $N$ electrons confined in a potential $V(\mathbf{r})$, in the presence of a uniform magnetic field $\mathbf{B} = B \hat{\mathbf{z}} = \nabla \times \mathbf{A}$, is
\begin{equation}
	H=\sum_{j=1}^N {\left[\frac{(-i\hbar \nabla_j+e \mathbf{A}/c)^2}{2m^*}+V(\mathbf{r}_j)\right]}+\sum_{j<k}\frac{e^2}{\epsilon |\mathbf{r}_j-\mathbf{r}_k|}+g^*\mu_B \mathbf{B}\cdot \mathbf{S},
\end{equation}
where $m^*$ is the electron effective mass, $\epsilon$ is the permitivity of the semiconductor material, $g^*$ is the effective $g$ factor, $\mu_B$ is the Bohr magneton, and $\mathbf{S}$ is the total electronic spin. Assuming the confining potential yields a quadratic in-plane potential for electrons in a lateral gate-defined confinement of a quantum dot (QD),
\begin{equation}
	V_{\mathbf{R}_0}(x,y)=\frac{1}{2}m^* \omega_0 \left[(x-x_0)^2+y^2\right],
\end{equation}
where the vector $\mathbf{R}_0=(x_0,0)$ is the position of potential minimum. The solution to the single-particle Hamiltonian, $\mathbf{p}^2/2m^*+V(\mathbf{r})$, are then the Fock-Darwin (F-D) states centered at the minimum of the potential well,
\begin{equation}\label{eq:FDSt}
	\phi_{nm} (x,y)=\frac{1}{l_0}\sqrt{\frac{\left(\frac{n-|m|}{2}\right)!}{\pi\left(\frac{n+|m|}{2}\right)!}}\left(\frac{x- x_0+i y \text{ sgn} m}{l_0}\right)^{|m|}e^{-\frac{(x- x_0)^2+y^2}{2l_0^2}+ \frac{x_0 y}{2l_B^2}}L_{\frac{n-|m|}{2}}^{|m|}\left(\frac{(x- x_0)^2+y^2}{l_0^2}\right),
\end{equation}
where $l_0=l_B/(1/4+\omega_0^2/\omega_c^2)^{1/4}$, $l_B=\sqrt{\hbar c/eB}$, $\omega_c = eB/m^* c$, $L_n^m(x)$ is the associated Laguerre polynomial and we have adopted the symmetric gauge, $\mathbf{A}=B/2\left[-y \mathbf{\hat{x}}+\left(x\mp x_0 \right)\mathbf{\hat{y}}\right]$. The corresponding single-particle energies of Eq.~\eqref{eq:FDSt} are
\begin{equation}\label{eq:FDEner}
	E_{n,m}=(n+1)\sqrt{\frac{1}{4}+\frac{\omega_0^2}{\omega_c^2}}\hbar \omega_c+\frac{m}{2}\hbar \omega_c.
\end{equation}
The derivation of the Fock-Darwin states, Eq.~\eqref{eq:FDSt}, and the corresponding energy spectrum can be found in \cite{Barnes.11}. 

Taking electron spin into account, the spin orbitals are given by
\begin{equation}
	\Phi_{nm\sigma}(\mathbf{r})=\phi_{nm}(\mathbf{r})\sigma(\omega),
\end{equation}
where $\sigma$ denotes the electron spin, $\uparrow$ or $\downarrow$, and $\omega$ is the spin variable. 

\subsection{Multi-particle Slatter determinant}
The two electron multi-particle bases for a single DQD is a Slater determinant of relevant single particle spin orbitals,
\begin{equation}
	\begin{split}
		|\Psi_\mathbb{L}\rangle &= |\Phi_{\mathbb{L}_1\uparrow}\Phi_{\mathbb{L}_2\downarrow}\rangle,\\
		|\Psi_\mathbb{R}\rangle &= |\Phi_{\mathbb{R}_1\uparrow}\Phi_{\mathbb{R}_2\downarrow}\rangle,
	\end{split}
\end{equation}
where $|\Phi_{\mathbb{L}_j\sigma}\rangle \big(|\Phi_{\mathbb{R}_j\sigma}\rangle\big)$ refers to the single-particle Fock Darwin state in the left (right) DQD of the $j$th electron with spin $\sigma$. For a DDQD device, the four-electron multi-particle bases are constructed by a direct product of the left and right DQD two-particle states,
\begin{equation}\label{eq:fourElectronState}
	|\Psi\rangle=|\Phi_{\mathbb{L}_1\uparrow}\Phi_{\mathbb{L}_2\downarrow}\rangle|\Phi_{\mathbb{R}_1\uparrow}\Phi_{\mathbb{R}_2\downarrow}\rangle.
\end{equation}

\subsection{General procedure for Full-CI calculation}

\begin{figure}[t]
	\includegraphics[width=0.6\columnwidth]{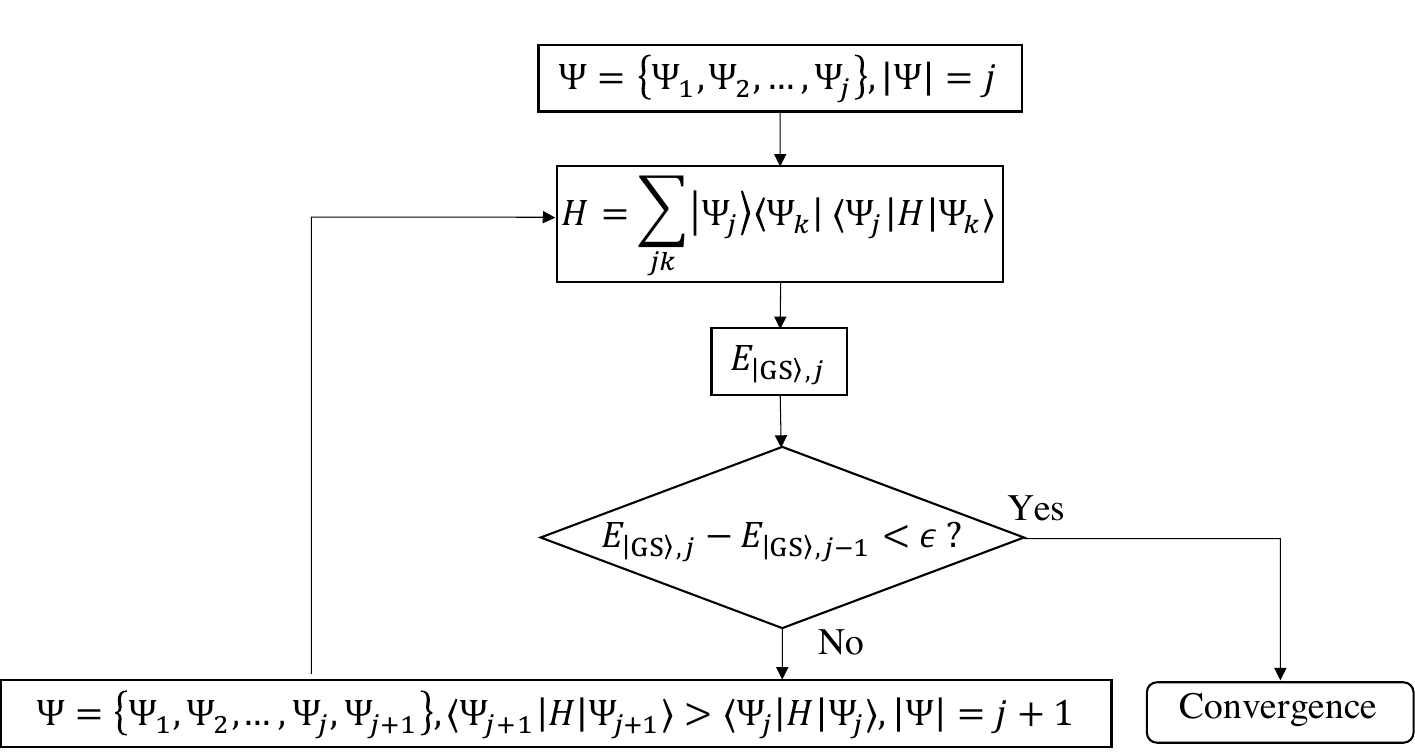}
	\caption{Schematic flow chart for the Full-CI calculation.}
	\label{fig:FullCIProcedure}
\end{figure}

The general procedure to perform Full-CI calculation is shown in Fig.~\ref{fig:FullCIProcedure}. The calculation loop is escaped when the newly obtained ground state energy is converged (see below).

\subsection{Truncation and convergence}

In principle, the Full-CI calculation should include an infinite number of electronic orbitals which is intractable. In practice, one introduces a cutoff energy, defined as the total non-interacting energy of the system above the ground state configuration \cite{Barnes.11}. The convergence is verified by raising the cutoff energy and examining the change in relevant physical quantities. 

\begin{figure}[t]
	(a)\includegraphics[width=0.43\columnwidth]{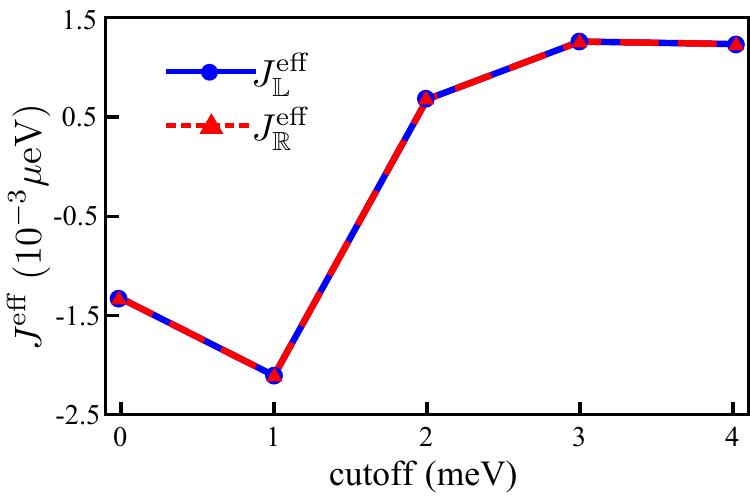}
	(b)\includegraphics[width=0.43\columnwidth]{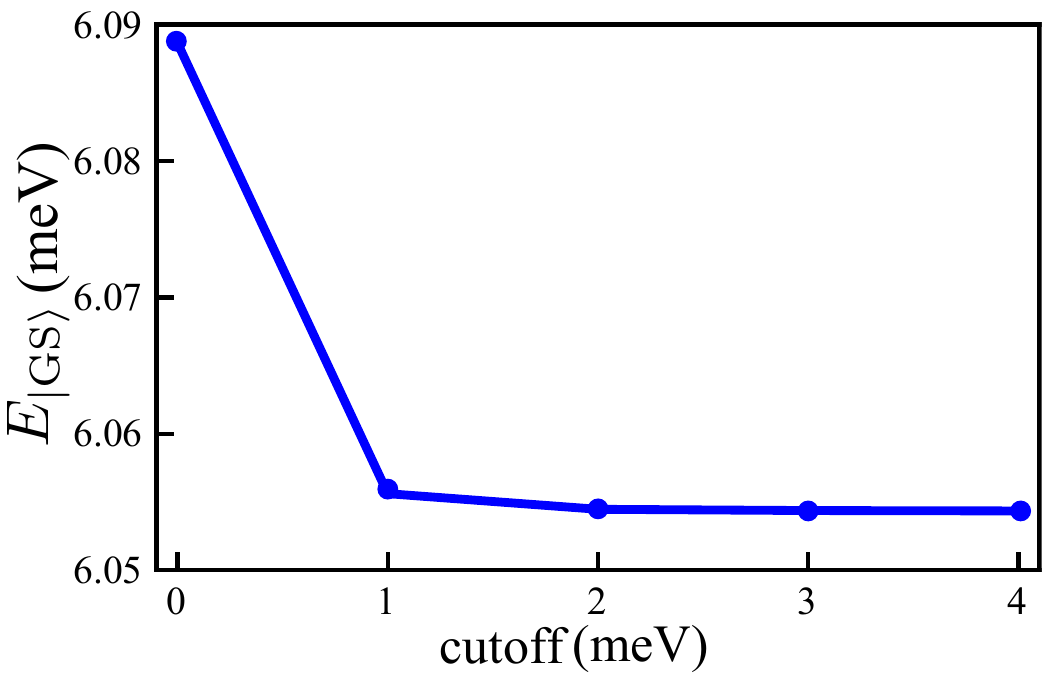}
	\caption{(a) Effective single-qubit exchange energy, $J^{\text{eff}}=J^{\text{eff}}_\mathbb{L}=J^{\text{eff}}_\mathbb{R}$ as a function of cutoff energy. (b) Ground state energy, $E_{|\text{GS}\rangle}$ as a function of the cutoff energy. $\Delta = 0$.}
	\label{fig:Jeff1-GS-n4-cutoff}
\end{figure}

Our verification of convergence is specific to the quantum-dot device parameters chosen in this paper: $\hbar \omega_0 = 1$ meV, $x_0 = 2.5a_B$, $R_0 = 9a_B$, $B=0$. In Fig.~\ref{fig:Jeff1-GS-n4-cutoff}, we demonstrate that orbitals included in the Full-CI calculation up until 4 meV is sufficient to achieve convergence. This implies that the orbitals required are up until $n=4$ Fock-Darwin states, corresponding to 60 orbitals for a DDQD system and $\approx$ 1.6 million number of Coulomb interaction terms to be evaluated. Results shown in this paper are therefore obtained by keeping relevant orbitals up to $n = 4$.

\section{Input parameters} \label{suppsec:inputParameters}
\begin{table}[!ht]
	\begin{tabular}{c|c|c|c|c|}
		\cline{2-5}
		& \multicolumn{2}{c|}{Parameter} & Value & Reference\\
		\cline{2-5}
		& \multirow{6}{*}{GaAs} &$\hbar \omega_0 $ & 1 meV & \cite{Shulman.12} \\
		\cline{3-5}
		& & $2R_0$ &  $\approx$ 608nm & \cite{Shulman.12} \\
		\cline{3-5}
		& & $2x_0$ &  $\approx$ 170nm & \cite{Shulman.12} \\
		\cline{3-5}
		& & $\epsilon$ & 13.1 &  \\
		\cline{3-5}
		& & $m^*$ & 0.067 $m_e$ &  \\
		\cline{3-5}
		& & $g^*$ & $-0.4$ &  \\
		\cline{3-5}
		(a) & & $\Delta E_z $ & $-0.14$ $\mu$eV & \cite{Bluhm.10,Dial.13}  \\
		\cline{2-5}
	\end{tabular}
	\begin{tabular}{c|c|c|c|c|}
		\multicolumn{1}{c}{} & \multicolumn{1}{c}{} & \multicolumn{1}{c}{} & \multicolumn{1}{c}{} & \multicolumn{1}{c}{} \\
		\cline{2-5}
		& \multicolumn{2}{c|}{Parameter} & Value & Reference\\
		\cline{2-5}
		& \multirow{6}{*}{Si} &$\hbar \omega_0 $ & 1 meV & \cite{Fogarty.18,Zajac.15,Yang.12} \\
		\cline{3-5}
		& & $2 x_0$ & $\approx$ 100nm & \cite{Zajac.15} \\
		\cline{3-5}
		& & $2 R_0$ & $\approx$ 360nm &  \\
		\cline{3-5}
		& & $\epsilon$ & 11.68 &  \\
		\cline{3-5}
		& & $m^*$ & 0.19 &  \\
		\cline{2-5}
		(b) & & $g^*$ & 2 &  \\
		\cline{2-5}
	\end{tabular}
	\caption{Summary of parameters used in the calculation for (a) GaAs and (b) silicon quantum-dot device.}
	\label{tab:parameters}
\end{table}
Unless stated otherwise, the parameters used are listed in Table~\ref{tab:parameters}, where $\hbar \omega_0$ is the confinement energy of the quantum dot, $2 R_0$ is the distance between two DQD centers, $2 x_0$ is the distance between two dots in a DQD, $\epsilon$ is the relative permittivity, $m^*$ is the effective mass, $g^*$ is the in-plane $g$ factor and $\Delta E_z$ is the magnetic gradient across a DQD. Note that the confinement strength for GaAs quantum-dot device, $\hbar \omega = 1 $ meV, is determined based on the inter-dot distance in correspondence to the exchange energy of an ST qubit reported in \cite{Shulman.12}. Concerning the confinement strength of silicon quantum-dot system, \cite{Fogarty.18} reported orbital excited energy within the range of 0.1$\sim$0.2 meV, \cite{Zajac.15} reported the energy yields $\sim$0.48 meV based on the dot size of $a_B$ = 29 nm and \cite{Yang.12} reported the energy within the range of 1 meV to 8 meV dependent on the number of electrons in a silicon quantum-dot. Hence, we choose $\hbar \omega_0 = 1$ as a trial value. The inter-DQD system is judiciously chosen such that inter-DQD tunneling is suppressed while inter-DQD Coulomb interaction is more pronounced.

\begin{figure}[t]
	\includegraphics[width=0.7\columnwidth]{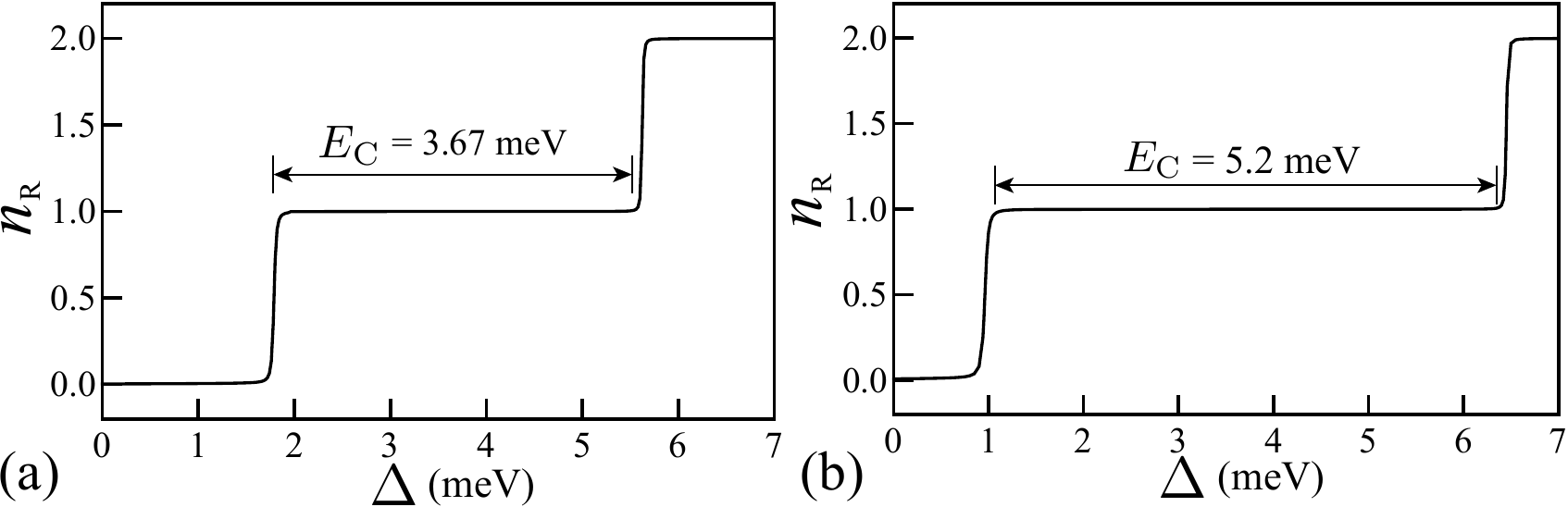}
	\caption{Single-dot electron occupation number, $n_\text{R}$, as functions of the detuning $\Delta$ for (a) GaAs and (b) silicon DQD device. The detuning is defined as the difference of the minima of potential wells. The parameters are shown in Table~\ref{tab:parameters}.}
	\label{fig:chargingEnergy}
\end{figure}

We perform a rough estimation of the charging energy, $E_\text{C}$, by using a bi-quadratic potential to model the occupation of single quantum-dot as functions of the detuning. The charging energy is extracted by evaluating the difference of detuning values at which the occupation number of single quantum-dot reaches 1 and 2 \cite{Zajac.15}, cf. Fig.~\ref{fig:chargingEnergy}. The charging energies reported for GaAs quantum-dot device varies across two orders, $10^{-1}\sim10^{1}$ \cite{Kouwenhoven.91,Gerster.18,Jung.04,Hoglund.10}. In particular, \cite{Gerster.18} and \cite{Jung.04} reported charging energy of $1.75$ meV and $2$ meV respectively, similar to Fig.~\ref{fig:chargingEnergy}(a). For silicon quantum-dot device, the observed charging energies ranges from $<10$ meV to $\sim 20$ meV \cite{Zajac.15,Fogarty.18,Yang.12}, among which \cite{Zajac.15} reported $E_\text{C} =$ 6.6 meV, similar to Fig.~\ref{fig:chargingEnergy}(b).

\section{Exchange energy of a singlet-triplet qubit in a double-quantum-dot}
\begin{figure}[t]
	\label{fig:singleQJ}
	\includegraphics[width=0.7\columnwidth]{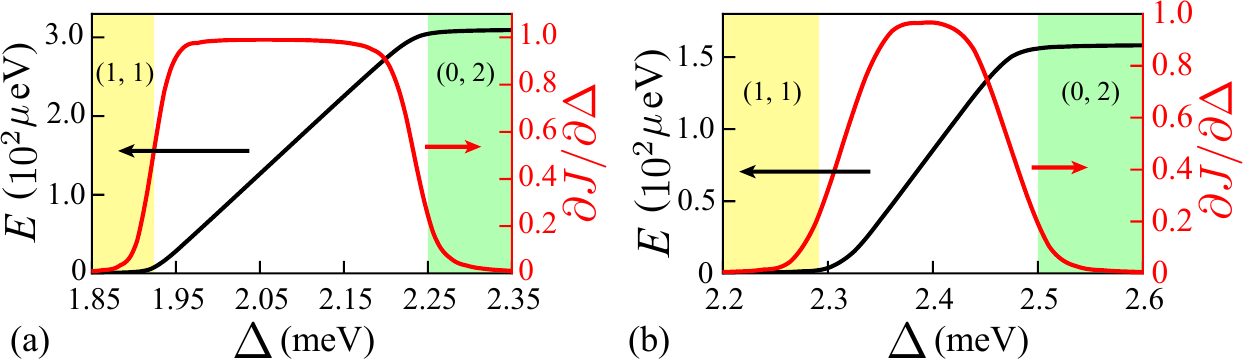}
	\caption{Exchange energy of a singlet-triplet qubit, $J$, and the corresponding $\partial J/\partial \Delta$ as functions of the detuning $\Delta$, in a GaAs DQD device for (a) $B = 0$ T and (b) $B = 0.186$ T. The electron occupation of each dot is denoted as $(n_{\it 1},n_{\it 2})$, where the first and second entry denote the electron number of left and right dot respectively.}
\end{figure}
Similar to the confinement potential adopted for DDQD device, cf. Eq.~\eqref{eq:Vpotential} in the main text, the confinement potential of a DQD device can be modeled as
\begin{align}\label{eq:VpotentialDQD}
	V(\mathbf{r}) &= \frac{1}{2} m^* \omega_0^2 \left(\text{Min}\Big[(x+x_0)^2+\frac{\Delta}{2},(x-x_0)^2-\frac{\Delta}{2}\Big]+y^2\right).
\end{align}
Supplemental Fig.~\ref{fig:singleQJ} shows the single qubit exchange energy, $J$, and the corresponding derivative, $\partial J/\partial \Delta$, as functions of the detuning $\Delta$. At small detuning, where the electron occupation is (1, 1), the susceptibility to charge-noise, $\partial J/\partial \Delta$, increases along with the increase of $J$. On the other hand, at large detuning, where the electron occupation is (0, 2), the increase of $J$ is accompanied with decrease sensitivity to charge-noise. This has been explored experimentally by \cite{Nichol.17}.

\section{Extended Hubbard Model}
\subsection{Hamiltonian}\label{subsec:HubbardModel}
\begin{figure}[t]
	\centering{
		\includegraphics[width=0.4\columnwidth]{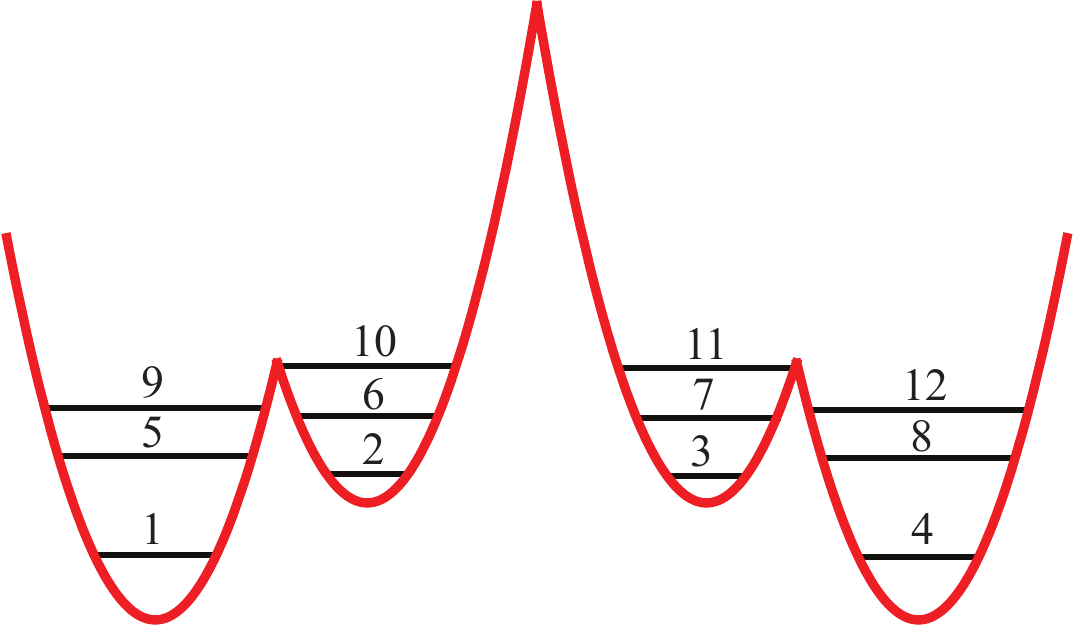}
		\caption{Schematic figure of Fock-Darwin states in a four-quantum-dot device.}
		\label{fig:FDSt1to12}
	}
\end{figure}

Although the Full-CI calculations allow us to obtain an accurate description of the energy spectrum of a DDQD device, it is computationally forbiddingly expensive to simulate a range of parameters. The extended Hubbard model on one hand gives us a computationally efficient way to calculate physical quantities of the system, and on the other hand reveals important insights on relevant eigenstates involved in our problem. The Hamiltonian is,
\begin{equation}\label{eq:HubbardModel2ndQuan}
	H = \sum_{j \sigma}\varepsilon_{j \sigma} c^\dagger_{j \sigma} c_{j \sigma}+\sum_{j<k,\sigma} \left(t_{jk\sigma} c^\dagger_{j\sigma} c_{k\sigma}+\mathrm{H. c.}\right)+\sum_{j} U_{j} n_{j\downarrow} n_{j\uparrow} +\sum_{\sigma \sigma'}\sum_{j<k} U_{jk} n_{j\sigma} n_{k\sigma'},
\end{equation}
where $j$ and $k$ are orbital indices (cf.~Fig.~\ref{fig:FDSt1to12}), while $\sigma$ and $\sigma'$ are spins. The summations over orbitals $(j,k)$ are from 1 to 12 and spins $(\sigma,\sigma')$ are for up and down. $\varepsilon_{j\sigma}$ denotes the on-site energy at dot $j$, $t_{jk\sigma}$ the tunneling between the $j$th and $k$th orbital, while $U_{j}$ denotes the on-site Coulomb interaction in the $j$th orbital and $U_{jk}$ the Coulomb interaction between the $j$th and $k$th orbital.  To avoid confusion of the Coulomb interaction between the 1st and 2nd orbital, $U_{12}$, and the on-site Coulomb interaction on the 12th orbital (also labelled as $U_{12}$), we take the symmetry into consideration. We use $U_{34}$ to label the former one (inter-site Coulomb interaction) in all situations (``Outer'',``Center'',``Right'') as they are equal to each other. For the latter one (on-site Coulomb interaction), we use $U_9$ in the  ``Outer'' and ``Center'' cases, and $U_{10}$ in the ``Right'' case. This is clear from Fig.~\ref{fig:FDSt1to12}.

Theses parameters are calculated from,
\begin{equation}
	\begin{split}
		U_{jk} &= \int{\Phi^*_j (\mathbf{r}_1)\Phi^*_k (\mathbf{r}_2) C(\mathbf{r}_1,\mathbf{r}_2) \Phi_j (\mathbf{r}_1)\Phi_k (\mathbf{r}_2) \text{d}\mathbf{r}^2},\\
		t_{jk}&=  \int{\Phi^*_j (\mathbf{r}) \left[\frac{\hbar^2}{2m^*}\nabla^2+V(\mathbf{r}) \right]\Phi_k (\mathbf{r}) \text{d}\mathbf{r}},\\
		\varepsilon_j &=  \int{\Phi^*_j (\mathbf{r}) \left[\frac{\hbar^2}{2m^*}\nabla^2+V(\mathbf{r}) \right]\Phi_j (\mathbf{r}) \text{d}\mathbf{r}},\\
		C(\mathbf{r}_1,\mathbf{r}_2)&=\frac{e^2}{\kappa|\mathbf{r}_1-\mathbf{r_2}|}.
	\end{split}
\end{equation}
For further discussion in Sec.~\ref{sec:capacitiveCoupling}, we define other Coulomb terms beyond extended Hubbard Model here.
\begin{equation}\label{eq:exchangeCoulomb}
	\begin{split}
		U^e_{jk}&=\int{\Phi^*_j (\mathbf{r}_1)\Phi^*_k (\mathbf{r}_2) C(\mathbf{r}_1,\mathbf{r}_2) \Phi_k (\mathbf{r}_1)\Phi_j (\mathbf{r}_2) \text{d}\mathbf{r}^2},\\
		V_\pm &= \frac{1}{2}\int {\left[\Phi_1^*(\mathbf{r}_1) \Phi_2^* (\mathbf{r}_2)\pm\Phi_2^* (\mathbf{r}_1)\Phi_1^* (\mathbf{r}_2) \right] C(\mathbf{r}_1,\mathbf{r}_2) \left[ \Phi_1 (\mathbf{r}_1)\Phi_2 (\mathbf{r}_2)\pm\Phi_2(\mathbf{r}_1)\Phi_1(\mathbf{r}_2)\right]} d\mathbf{r}^2.
	\end{split}
\end{equation}

Since the Hamiltonian, Eq.~\eqref{eq:HubbardModel2ndQuan}, commutes with $\mathbf{S}^2$, each logical manifold forms a diagonal block, allowing us to evaluate the eigenvalues of $|SS\rangle$, $|ST\rangle$,$|TS\rangle$ and $|TT\rangle$ by diagonalizing the corresponding individual blocks. In this section, we focus on the effective Hamiltonian blocks for the ``Outer'' detuning case. Results can also be obtained using Eq.~\eqref{eq:HubbardModel2ndQuan} for ``Center'' and ``Right''  detuning cases, but we will not present the results in this section. In the ``Outer'' detuning case, the symmetry around the center of the DDQD system allows us to simplify some parameters, e.g. $\varepsilon_1=\varepsilon_4$, inter-site Coulomb interaction $U_{12}=U_{34}$ as mentioned above, $U_{16}=U_{47}$ etc. When treating the blocks solely relevant to the singlet $|S\rangle$ (i.e the $|SS\rangle$ states ), we keep orbitals 1 through 4 only as contribution from higher orbitals can be safely neglected. However, all 12 orbitals are retained whenever the triplet state plays a role.

We assume $U_1=U_2=U_3=U_4=U_\text{os}$ (where ``os'' means on-site) and $t_{12\sigma}=t_{34\sigma}=t$ for convenience. However, $U_j \neq U_k$ for $j$ from 1 to 4 and $k$ from 5 to 12, due to different forms of wave functions with the former yielding even Gaussian functions, and the later being odd oscillating functions. Also, it should be pointed out that the single-particle tunneling between DQDs is assumed to be zero due to the large distance $R_0$ between them, i.e. $t_{jk} = 0$ when the $j$th and $k$th orbitals belong to different DQDs.

The Hamiltonian block in each logical manifold is written in the bases of four-electron multi-particle states, constructed from the direct product of two-electron Slater determinants, Eq.~\eqref{eq:fourElectronState}. We therefore write the relevant two-electron Slater determinants as:
\begin{subequations}
	\begin{align}
		\begin{split}\label{eq:SS2eSt}
			|S^{20}\rangle&=|\Phi_{1\uparrow}\Phi_{1\downarrow}\rangle,\\
			|S^{02}\rangle&=|\Phi_{4\uparrow}\Phi_{4\downarrow}\rangle,
		\end{split}\\
		\begin{split}
			|S^{11}\rangle&=\frac{1}{\sqrt{2}}|\Phi_{1\uparrow}\Phi_{2\downarrow}\rangle+|\Phi_{2\uparrow}\Phi_{1\downarrow}\rangle\\
			&=\frac{1}{\sqrt{2}}|\Phi_{3\uparrow}\Phi_{4\downarrow}\rangle+|\Phi_{4\uparrow}\Phi_{3\downarrow}\rangle,
		\end{split}\\
		\begin{split}
			|T^{11}\rangle&=\frac{1}{\sqrt{2}}|\Phi_{1\uparrow}\Phi_{2\downarrow}\rangle-|\Phi_{2\uparrow}\Phi_{1\downarrow}\rangle\\
			&=\frac{1}{\sqrt{2}}|\Phi_{3\uparrow}\Phi_{4\downarrow}\rangle-|\Phi_{4\uparrow}\Phi_{3\downarrow}\rangle,
		\end{split}\\
		\begin{split}\label{eq:TT2eSt}
			|T^{20}_\mathrm{L}\rangle &= \frac{1}{\sqrt{2}} (|\Phi_{1\uparrow}\Phi_{5\downarrow}\rangle-|\Phi_{5\uparrow}\Phi_{1\downarrow}\rangle),\\
			|T^{20}_\mathrm{H}\rangle &= \frac{1}{\sqrt{2}} (|\Phi_{1\uparrow}\Phi_{9\downarrow}\rangle-|\Phi_{9\uparrow}\Phi_{1\downarrow}\rangle),\\
			|T^{02}_\mathrm{L}\rangle &= \frac{1}{\sqrt{2}} (|\Phi_{4\uparrow}\Phi_{8\downarrow}\rangle-|\Phi_{8\uparrow}\Phi_{4\downarrow}\rangle),\\
			|T^{02}_\mathrm{H}\rangle &= \frac{1}{\sqrt{2}} \left(|\Phi_{4\uparrow}\Phi_{12\downarrow}\rangle-|\Phi_{12\uparrow}\Phi_{4\downarrow}\rangle\right).
		\end{split}
	\end{align}
\end{subequations}
The subscript L (H) in Eq.\eqref{eq:TT2eSt} and Eq.~\eqref{eq:lotsofU}  carries the meaning of ``lower'' (``higher'') as it involves the lower (higher) state for first-excited orbital, labeled as $(n,m)=(1,-1)$ $\left((n,m)=(1,1)\right)$.

As mentioned above, the effective Hamiltonian block for the $|SS\rangle$ manifold can be simplified by assuming the electrons only occupy the lowest single-particle state. This is due to the fact that for the ground $|SS\rangle$ eigenstate in large detuning, both electrons in a DQD can occupy the lowest orbitals (Eq.~\eqref{eq:SS2eSt}), in contrast to the ``repulsion'' experienced by electrons for $|TT\rangle$ eigenstate (Eq.~\eqref{eq:TT2eSt}). In the bases $|S^{20}\rangle|S^{02}\rangle, |S^{11}\rangle|S^{02}\rangle, |S^{20}\rangle|S^{11}\rangle, |S^{11}\rangle|S^{11}\rangle$, 
\begin{equation}
	H_{|SS\rangle} - E_{|S^{20}\rangle|S^{02}\rangle}=\left (
	\begin {array} {cccc}
	0 & \sqrt {2} t & \sqrt {2} t & 0 \\
	\sqrt {2} t & {U^\bigstar} - {U^\Diamond} + \varepsilon  & 0 \
	& \sqrt {2} t \\
	\sqrt {2} t & 0 & {U^\bigstar} - {U^\Diamond} + \varepsilon  \
	& \sqrt {2} t \\
	0 & \sqrt {2} t & \sqrt {2} t & {U^\bigstar} - 2 {U^\Diamond} + 
	2 \varepsilon  \\
	\end {array}
	\right),
	\label{eq:SSHamHubbard}
\end{equation}
where
\begin{equation}
	\begin{split}
		U^\bigstar &= U_{23}+U_{14}-2U_{24},\\
		U^\Diamond  &= U_\text{os}-\left(U_{34}+U_{23}-U_{14}\right),\\
		\varepsilon = \varepsilon_2 - \varepsilon_1 = \varepsilon_3 - \varepsilon_4 &= \varepsilon_6-\varepsilon_5= \varepsilon_7-\varepsilon_8= \varepsilon_{10}-\varepsilon_9= \varepsilon_{11}-\varepsilon_{12},\\
		E_{|S^{20}\rangle|S^{02}\rangle} &=2\varepsilon_1+2\varepsilon_4+2 U_{\text{os}}+4U_{14}.
	\end{split}
	\label{eq:SSparametersHubbard}
\end{equation}

Different from the $|SS\rangle$ manifold, the effective Hamiltonian block for $|TT\rangle$ includes two first-excited single-particle states in each quantum dot, similar to $n = 1$, $m = \pm 1$ Fock-Darwin states. The same holds for Hamiltonian blocks of $|ST\rangle$ and $|TS\rangle$.

In the bases of  $\left\{|T^{11}\rangle|T^{11}\rangle, |T^{11}\rangle|T^{02}_\mathrm{L}\rangle ,|T^{11}\rangle|T^{02}_\mathrm{H}\rangle , |T^{20}_\mathrm{L}\rangle|T^{11}\rangle,|T^{20}_\mathrm{H}\rangle|T^{11}\rangle, |T^{20}_\mathrm{L}\rangle|T^{02}_\mathrm{L}\rangle, |T^{20}_\mathrm{L}\rangle|T^{02}_\mathrm{H}\rangle, |T^{20}_\mathrm{H}\rangle|T^{02}_\mathrm{L}\rangle, |T^{20}_\mathrm{H}\rangle|T^{02}_\mathrm{H}\rangle\right\}$, 
\begin{equation}
	H_{|TT\rangle}=
	\left(
	\scalemath{0.85}{
		\begin{array}{ccccccccc}
			E_{|T^{11}\rangle|T^{11}\rangle} & -{t_{25}} & -{t_{29}} & {t_{25}} & {t_{29}} & 0 & 0 & 0 & 0 \\
			-{t_{25}} & E_{|T^{11}\rangle|T^{02}_\mathrm{L}\rangle}  & {0} & 0 & 0 & {t_{25}} & {t_{29}} & 0 & 0 \\
			-{t_{29}} & {0} & E_{|T^{11}\rangle|T^{02}_\mathrm{H}\rangle} & 0 & 0 & 0 & 0 & {t_{25}} & {t_{29}} \\
			{t_{25}} & 0 & 0 & E_{|T^{20}_\mathrm{L}\rangle|T^{11}\rangle} & {0} & -{t_{25}} & 0 & -{t_{29}} & 0 \\
			{t_{29}} & 0 & 0 & {0} & E_{|T^{20}_\mathrm{H}\rangle|T^{11}\rangle} & 0 & -{t_{25}} & 0 & -{t_{29}} \\
			0 & {t_{25}} & 0 & -{t_{25}} & 0 & E_{|T^{20}_\mathrm{L}\rangle|T^{02}_\mathrm{L}\rangle} & {0} & {0} & 0 \\
			0 & {t_{29}} & 0 & 0 & -{t_{25}} & {0} & E_{|T^{20}_\mathrm{L}\rangle|T^{02}_\mathrm{H}\rangle} & 0 & {0} \\
			0 & 0 & {t_{25}} & -{t_{29}} & 0 & {0} & 0 & E_{|T^{20}_\mathrm{H}\rangle|T^{02}_\mathrm{L}\rangle} & {0} \\
			0 & 0 & {t_{29}} & 0 & -{t_{29}} & 0 & {0} & {0} & E_{|T^{20}_\mathrm{H}\rangle|T^{02}_\mathrm{H}\rangle} \\
		\end{array}
	}
	\right),
	\label{eq:TTHamHubbard}
\end{equation}
where 
\begin{equation}
	\begin{split}
		E_{|T^{11}\rangle|T^{11}\rangle}-E_{|T^{20}_\mathrm{L}\rangle|T^{02}_\mathrm{H}\rangle}&= U^\bigtriangleup_\mathrm{L}+U^\bigtriangledown_\mathrm{L}-U^\square_\mathrm{M}+2 \varepsilon-\hbar \omega_{1,-1}-\hbar \omega_{1,1},\\
		E_{|T^{11}\rangle|T^{02}_\mathrm{L}\rangle}-E_{|T^{20}_\mathrm{L}\rangle|T^{02}_\mathrm{H}\rangle} &= \frac{1}{2} \left[{U^\bigtriangleup_\mathrm{L}+U^\square_\mathrm{L}}-2 (U^\square_\mathrm{M}-\varepsilon+\hbar \omega_{1,1})\right]+\varepsilon,\\
		E_{|T^{11}\rangle|T^{02}_\mathrm{H}\rangle}-E_{|T^{20}_\mathrm{L}\rangle|T^{02}_\mathrm{H}\rangle}&=\frac{1}{2} \left[{U^\bigtriangleup_\mathrm{H}}+{U^\square_\mathrm{H}}-2 (U^\square_\mathrm{M}-\varepsilon+\hbar \omega_{1,-1})\right]+\varepsilon,\\
		E_{|T^{20}_\mathrm{L}\rangle|T^{11}\rangle}-E_{|T^{20}_\mathrm{L}\rangle|T^{02}_\mathrm{H}\rangle}&=\frac{1}{2} \left[U^\bigtriangleup_\mathrm{L}+U^\square_\mathrm{L}-2 (U^\square_\mathrm{M}-\varepsilon+\hbar \omega_{1,1})\right]+\varepsilon,\\
		E_{|T^{20}_\mathrm{H}\rangle|T^{11}\rangle}-E_{|T^{20}_\mathrm{L}\rangle|T^{02}_\mathrm{H}\rangle}&=\frac{1}{2} \left[{U^\bigtriangleup_\mathrm{H}}+{U^\square_\mathrm{H}}-2 (U^\square_\mathrm{M}-\varepsilon+\hbar \omega_{1,-1})\right]+\varepsilon,\\
		E_{|T^{20}_\mathrm{L}\rangle|T^{02}_\mathrm{L}\rangle}-E_{|T^{20}_\mathrm{L}\rangle|T^{02}_\mathrm{H}\rangle}&={U^\square_\mathrm{L}}-U^\square_\mathrm{M}+\hbar \omega_{1,-1}-\hbar \omega_{1,1},\\
		E_{|T^{20}_\mathrm{H}\rangle|T^{02}_\mathrm{H}\rangle}-E_{|T^{20}_\mathrm{L}\rangle|T^{02}_\mathrm{H}\rangle}&={U^\square_\mathrm{H}}-U^\square_\mathrm{M}-\hbar \omega_{1,-1}+\hbar \omega_{1,1},\\
		E_{|T^{20}_\mathrm{L}\rangle|T^{02}_\mathrm{H}\rangle}=E_{|T^{20}_\mathrm{H}\rangle|T^{02}_\mathrm{L}\rangle}&=2 \varepsilon_1 +2 \varepsilon_4+ U_{1,12} + U_{5,12} + U_{15} + U_{19} + U_{14} + U_{45} + \hbar \omega_{1,-1} + \hbar \omega_{1,1},
	\end{split}
	\label{eq:TTparametersHubbard}
\end{equation}
and
\begin{equation}
	\begin{split}
		U^\bigtriangleup_\mathrm{L}&=2 U_{34}+2 U_{13}+2 U_{35}+ U_{14}-U_{58},\\
		U^\bigtriangleup_\mathrm{H}&=2 U_{34}+2 U_{13}+2 U_{39}+ U_{14}-U_{9,12},\\
		U^\bigtriangledown_\mathrm{L}&=U_{23}+U_{58}-2U_{35},\\
		U^\bigtriangledown_\mathrm{H}&=U_{23}+U_{9,12}-2U_{39},\\
		U^\square_\mathrm{L}&=U_{14}+(U_{15}+U_{15})+2U_{18}+U_{58},\\
		U^\square_\mathrm{M}&=U_{14}+(U_{15}+U_{19})+(U_{18}+U_{1,12})+U_{5,12},\\
		U^\square_\mathrm{H}&=U_{14}+(U_{19}+U_{19})+2U_{1,12}+U_{9,12},\\
		\hbar \omega_{p,q}&=p\hbar \omega_0 + q \hbar \omega_c.
	\end{split}
	\label{eq:lotsofU}
\end{equation}

The effective Hamiltonian block for $|ST\rangle$ and $|TS\rangle$ can be obtained using the same techniques used for $|TT\rangle$. In the bases of $\left\{|S^{ 11 }\rangle |T^{ 11 }\rangle, |S^{ 20 }\rangle |T^{ 11 }\rangle, |S^{ 11 }\rangle |T^{ 02 }_\mathrm{L}\rangle, |S^{ 11 }\rangle |T^{ 02 }_\mathrm{H}\rangle, |S^{ 20 }\rangle |T^{ 02 }_\mathrm{L}\rangle, |S^{ 20 }\rangle |T^{ 02 }_\mathrm{H}\rangle\right\}$,
\begin{equation}
	\begin{split}
		H_{|ST\rangle}=
		\left(
		\begin{array}{cccccc} E_{ |S^{ 11 }\rangle |T^{ 11 }\rangle  } & \sqrt { 2 } t & t_{ 25 } & t_{ 29 } & 0 & 0 \\
			\sqrt { 2 } t & E_{ |S^{ 20 }\rangle |T^{ 11 }\rangle  }  & 0 & 0 & t_{ 25 } & t_{ 29 } \\
			t_{ 25 } & 0 & E_{ |S^{ 11 }\rangle |T^{ 02 }_\mathrm{L}\rangle }  & 0 & \sqrt { 2 } t_{ 12 } & 0 \\
			t_{ 29 } & 0 & 0 & E_{ |S^{ 11 }\rangle |T^{ 02 }_\mathrm{H}\rangle }& 0 & \sqrt { 2 } t_{ 12 } \\
			0 & t_{ 25 } & \sqrt { 2 } t_{ 12 } & 0 & E_{ |S^{ 20 }\rangle |T^{ 02 }_\mathrm{L}\rangle}  & 0 \\
			0 & t_{ 29 } & 0 & \sqrt { 2 } t_{ 12 } & 0 & E_{ |S^{ 20 }\rangle |T^{ 02 }_\mathrm{H}\rangle }
		\end{array}
		\right),
	\end{split}
	\label{eq:STHamHubbard}
\end{equation}
where
\begin{equation}
	\begin{split}
		E_{ |S^{ 11 }\rangle |T^{ 11 }\rangle  }&=2\varepsilon_1+2\varepsilon_4+2U_{34}+2U_{13}+U_{14}+2\varepsilon,\\
		E_{ |S^{ 20 }\rangle |T^{ 11 }\rangle  }&=2\varepsilon_1+2\varepsilon_4+U_{34}+2U_{13}+2U_{14}+U_{\text{os}}+\varepsilon,\\
		E_{ |S^{ 11 }\rangle |T^{ 02 }_\mathrm{L}\rangle  }&=2\varepsilon_1+2\varepsilon_4+U_{34}+U_{13}+U_{15}+U_{35}+U_{14}+U_{45}+\varepsilon+\hbar\omega_{1,-1},\\
		E_{ |S^{ 11 }\rangle |T^{ 02 }_\mathrm{H}\rangle  }&=2\varepsilon_1+2\varepsilon_4+U_{34}+U_{13}+U_{19}+U_{39}+U_{14}+U_{49}+\varepsilon+\hbar\omega_{1,1},\\
		E_{ |S^{ 20 }\rangle |T^{ 02 }_\mathrm{L}\rangle  }&=2\varepsilon_1+2\varepsilon_4+U_{15}+2U_{14}+2U_{45}+U_{\text{os}}+\hbar\omega_{1,-1},\\
		E_{ |S^{ 20 }\rangle |T^{ 02 }_\mathrm{H}\rangle  }&=2\varepsilon_1+2\varepsilon_4+U_{19}+2U_{14}+2U_{49}+U_{\text{os}}+\hbar\omega_{1,1}.
	\end{split}
\end{equation}
$H_{|TS\rangle}$ yields a similar form as $H_{|ST\rangle}$, and we would not give its explicit form here. It is noted that when the magnetic field is absent, $B = 0$, energy splitting of excited orbitals is absent as well, giving an effective Hamiltonian block for $|TT\rangle$ $\left(|ST\rangle\right)$ spanned by $\left\{|T_{11}T_{11}\rangle,\left(|\widehat{T_{L}T_{L}}\rangle+|\widehat{T_{H}T_{H}}\rangle\right)/\sqrt{2}, \left(|T^{20}_{L}T^{20}_{L}\rangle+|T^{20}_{H}T^{20}_{H}\rangle+|T^{20}_{L}T^{20}_{H}\rangle+|T^{20}_{H}T^{20}_{L}\rangle \right)/2\right\}$ $\left(\left\{|S^{11}T^{11}\rangle,|S^{20}T^{11}\rangle,\left(|S^{11}T_\text{L}^{02}\rangle+|S^{11}T_\text{H}^{02}\rangle\right)/\sqrt{2},\left(|S^{20}T_\text{L}^{02}\rangle+|S^{20}T_\text{H}^{02}\rangle\right)/\sqrt{2} \right\}\right)$ while other energetically relevant eigenstates decouple from the logical subspace. For simplicity, we only show the dot-occupation and discard the subscripts L and H in Fig.~\ref{fig:STEvalBfo0p753DiffDetuningScheme}, \ref{fig:STEvalSi} and  Fig.~\ref{fig:JeffiBfoSSTTEvals} in the main text.

\subsection{Correspondence between Full-CI and extended Hubbard model}
\begin{figure}[t]
	\centering
	\includegraphics[width=\columnwidth]{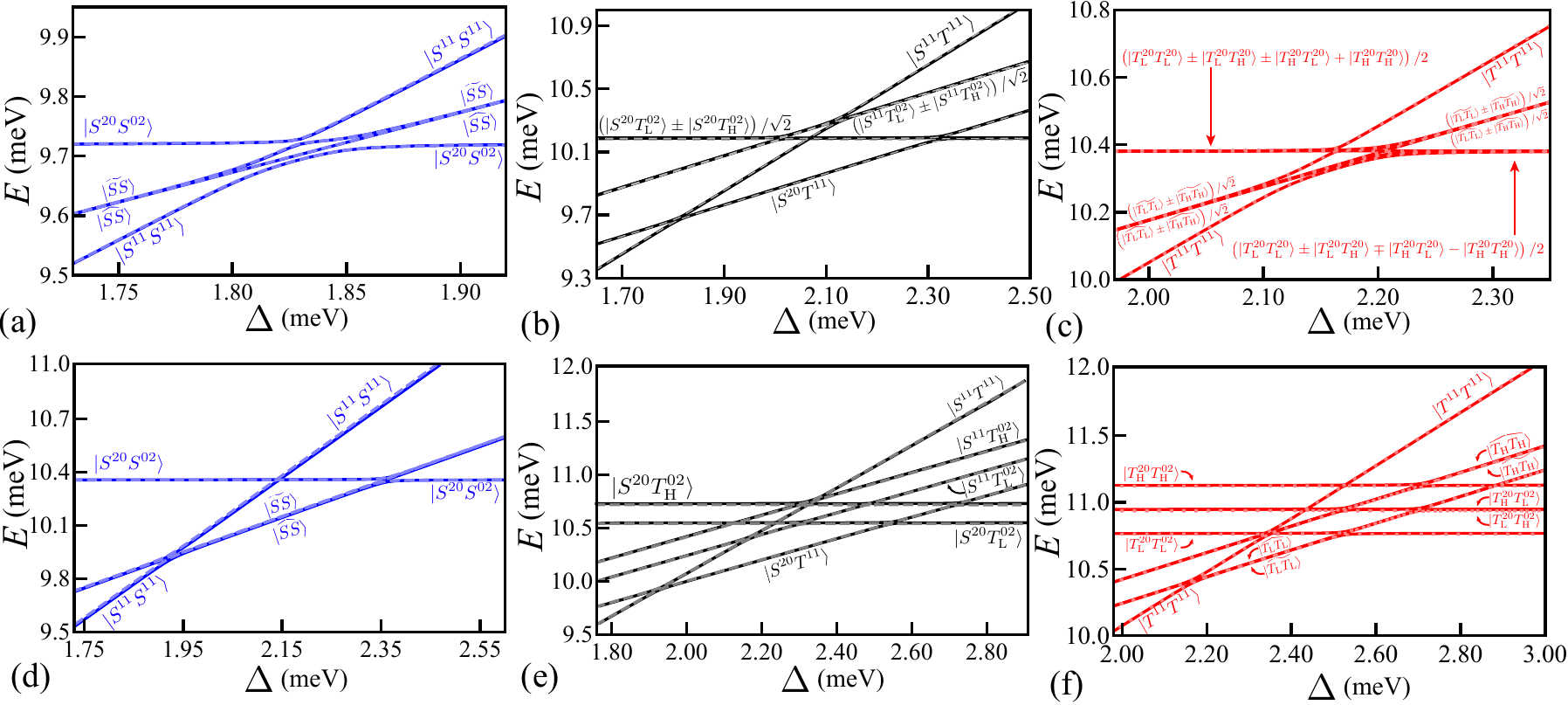}
	\caption{Comparison of energy levels calculated using the Hamiltonian given by extended Hubbard Model (dashed lighter lines) and those from the Full-CI calculation (solid colored lines) of a GaAs DDQD device for (a,d) $|SS\rangle$, (b,e) $|ST\rangle$/$|TS\rangle$ and (c,f) $|TT\rangle$. (a)-(c) are results for $B=0$ while (d)-(f) are results for $B=0.104$ T under ``Outer" detuning. Parameter used in the extended Hubbard model: $t=5.3\mu$eV, $t_{25}=t_{29}=8.0\mu$eV, $U_\text{os}=1.82$meV. }
	\label{fig:FullCIvsHubbard}
\end{figure}

Supplementary Fig.~\ref{fig:FullCIvsHubbard} compares the energy levels obtained using the Hamiltonian given by extended Hubbard Model (Sec.~\ref{subsec:HubbardModel}, dashed light blue lines) and those from the Full-CI calculation (solid colored lines) for a GaAs DDQD device. The similarity validates the accuracy of the extended Hubbard model when used in our problem.

\section{Comparison between ``Outer, ``Center'' and ``Right'' detuning schemes}

\begin{figure}[t]
	\centering{
		(a)\includegraphics[width=0.3\columnwidth]{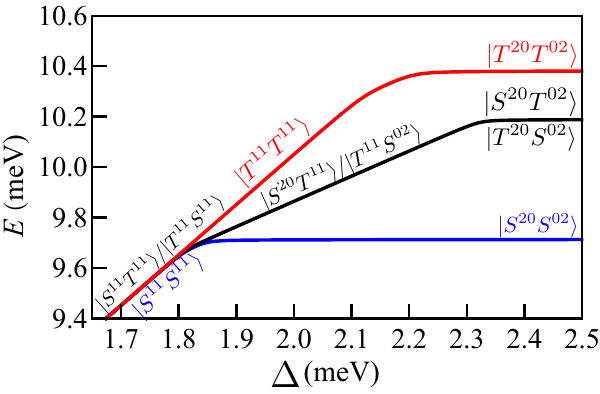}
		(b)\includegraphics[width=0.3\columnwidth]{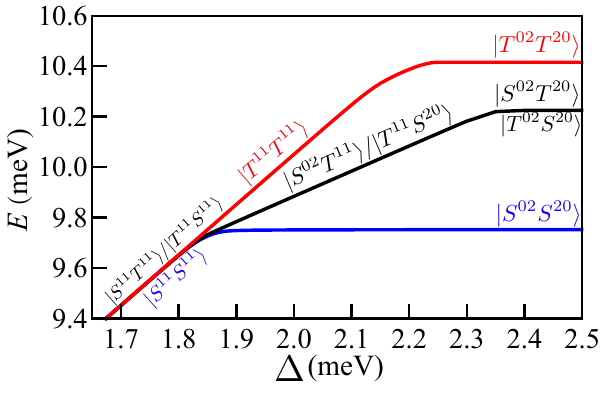}
		(c)\includegraphics[width=0.3\columnwidth]{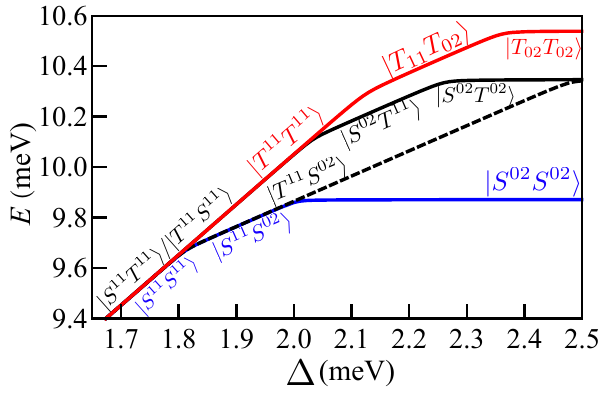}
		
		(d)\includegraphics[width=0.3\columnwidth]{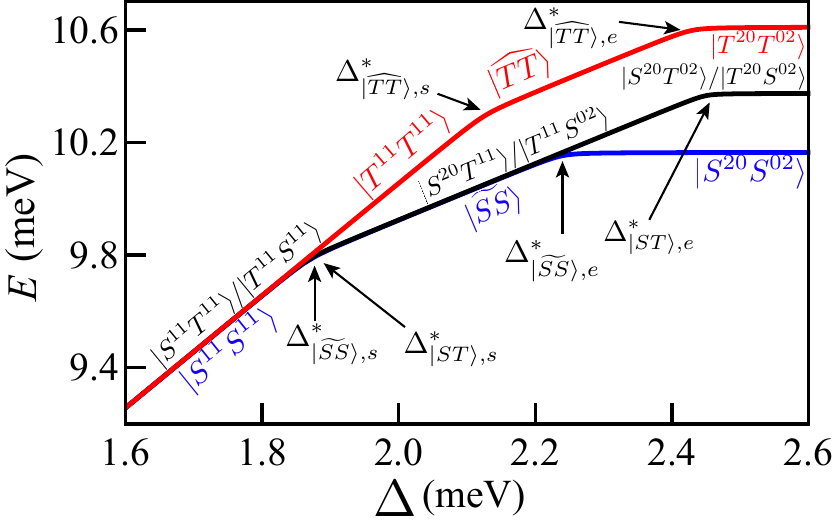}
		(e)\includegraphics[width=0.3\columnwidth]{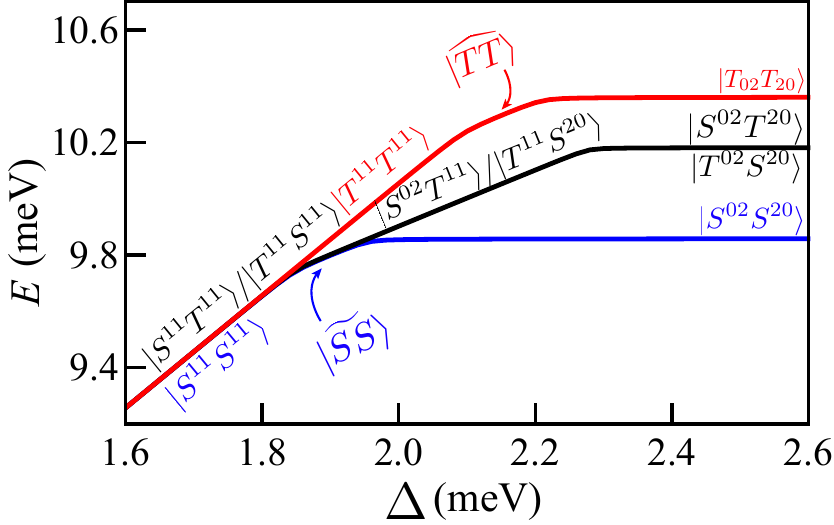}
		(f)\includegraphics[width=0.3\columnwidth]{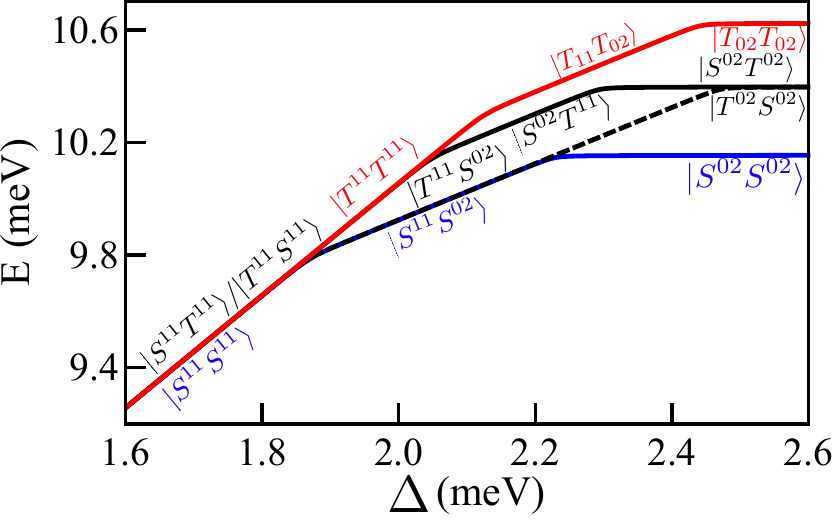}
	}
	\caption{Eigenenergies of four logical states, $|SS\rangle, |ST\rangle,|TS\rangle$ and $|TT\rangle$ for (a,d) ``Outer'', (b,e) ``Center'', (c,f) ``Right'' detuning scheme. (a)-(c) are results for $B=0$ while (d)-(f) are results for $B=0.087$ T. The results are calculated for GaAs quantum-dot device with parameters given in Table~\ref{tab:parameters}.}
	\label{fig:STEvalBfo0p753DiffDetuningScheme}
\end{figure}

\begin{figure}[t]
	\centering{
		\includegraphics[width=0.91\columnwidth]{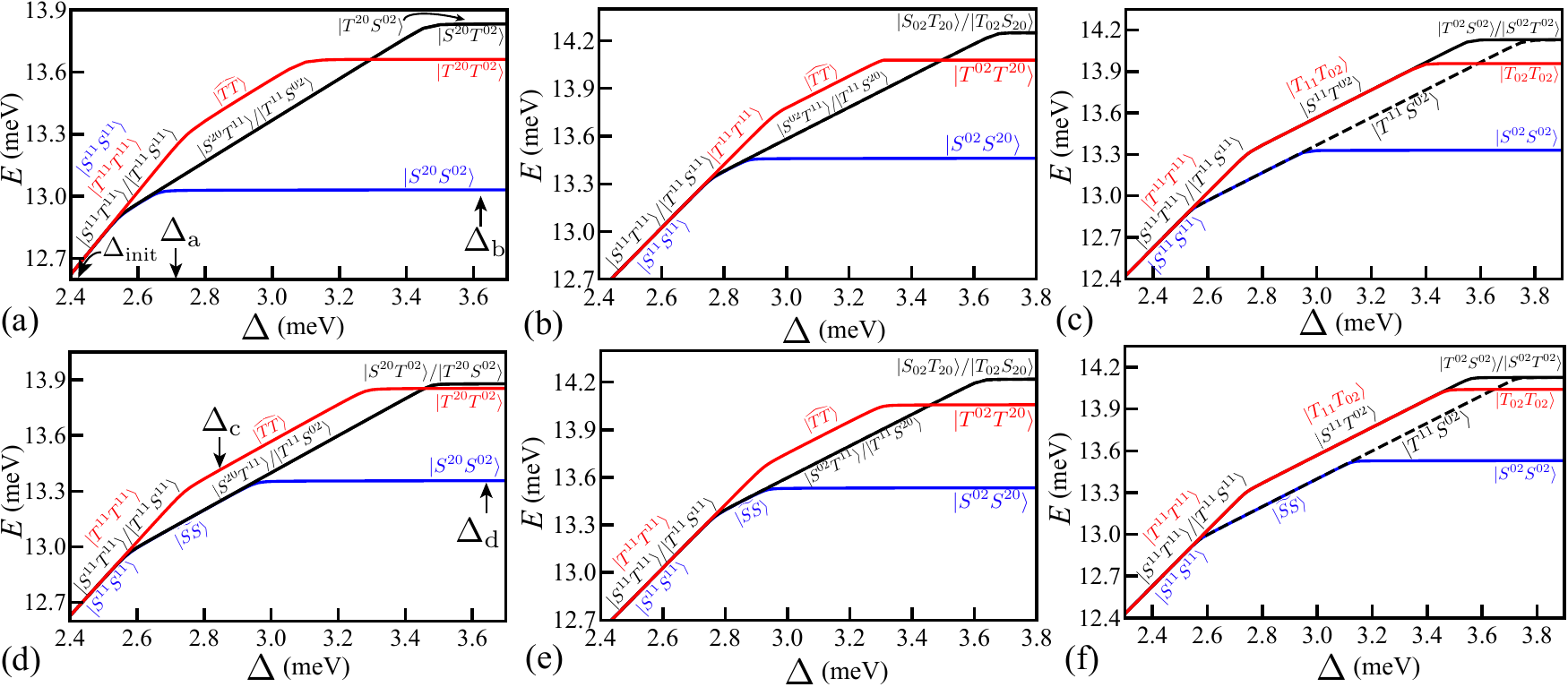}
	}
	\caption{Eigenenergies of four logical states, $|SS\rangle, |ST\rangle,|TS\rangle$ and $|TT\rangle$ for (a,d) ``Outer'', (b,e) ``Center'', (c,f) ``Right'' detuning scheme. (a)-(c) are results for $B=0$ while (d)-(f) are results for $B=0.165$ T. The results are calculated for Si quantum-dot device with parameters given in Table~\ref{tab:parameters}.}
	\label{fig:STEvalSi}
\end{figure}

\begin{figure}[t]
	\centering{
		\includegraphics[width=0.8\columnwidth]{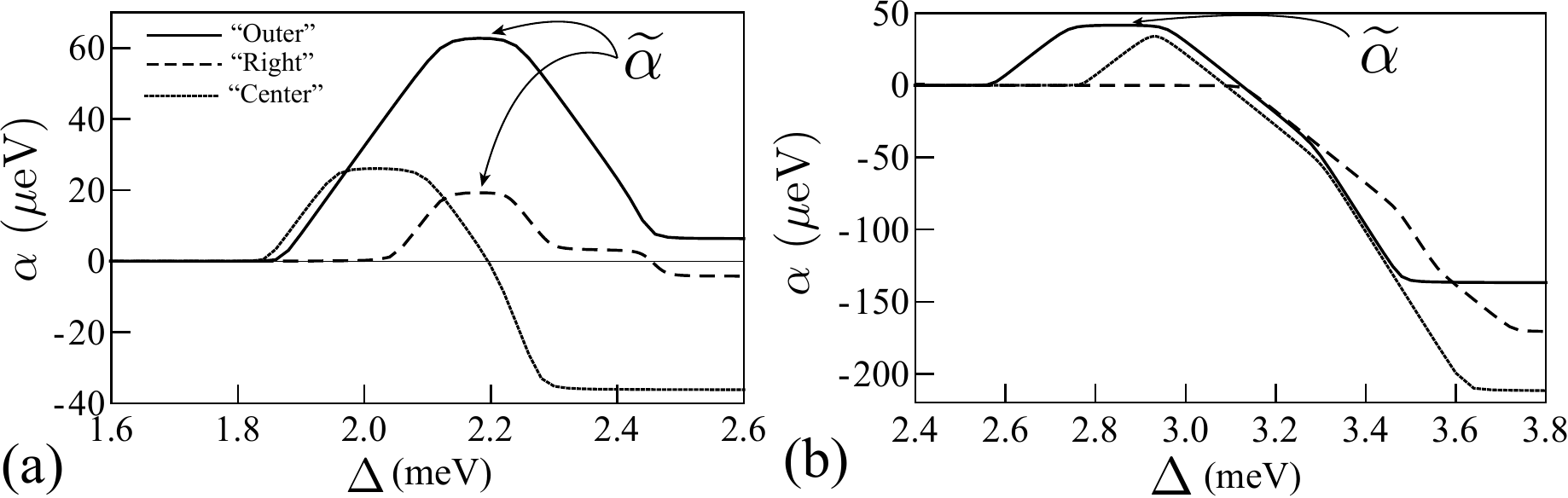}
	}
	\caption{(a) Capacitive coupling strength, $\alpha$, as a function of detuning deduced from Supplementary Fig.~\ref{fig:STEvalBfo0p753DiffDetuningScheme} (Supplementary Fig.~\ref{fig:STEvalSi}) (d)-(f) for GaAs at $B = 0.087$ T (b) Results for silicon at $B = 0.165$ T.}
	\label{fig:alphaCompre3DetuningSchemeBfo0p75}
\end{figure}

In this section we compare the ``Outer, ``Center'' and ``Right'' detuning schemes.
We first discuss the results of GaAs DDQD device (Supplementary Fig.~\ref{fig:STEvalBfo0p753DiffDetuningScheme}) and compare to results for silicon system shown in Supplementary Fig.~\ref{fig:STEvalSi}. For a GaAs DDQD device, the eigenenergies of four logical states, $|SS\rangle, |ST\rangle,|TS\rangle$ and $|TT\rangle$ are shown in Supplementary Fig.~\ref{fig:STEvalBfo0p753DiffDetuningScheme}. The left, middle and right column are for ``Outer, ``Center'' and ``Right'' schemes respectively, while the upper and lower rows are for $B=0$ and $B=0.087$ T respectively. It is found that the overlap in the $\Delta$ range between $|\widetilde{SS}\rangle$ and $|\widehat{TT}\rangle$ under external magnetic field is only found for the ``Outer'' and ``Right'' detuning schemes, but not for the "Center" scheme. Therefore the nearly-sweet-spot regime only exists for the ``Outer'' and ``Right'' , but not "Center" scheme. This is also consistent with results shown in Fig.~\ref{fig:deltaRangeOfABOND} of the main text.

Among the ``Outer'' and ``Right'' schemes, the value of capacitive coupling in the nearly-sweet-spot regime, $\widetilde{\alpha}$ is stronger for ``Outer'' as compared to the ``Right'' scheme, as shown in Supplementary Fig.~\ref{fig:alphaCompre3DetuningSchemeBfo0p75} (a), indicating that a choice of detuning value in the nearly-sweet-spot regime of the ``Outer'' scheme is superior than others.

Supplementary Fig.~\ref{fig:STEvalSi} shows the eigenenergies of logical states for a silicon DDQD device. In this calculation, we only consider the orbital degree of freedom and neglect the valley splitting. While the experimental measured value of the valley spitting ranges from $\approx0.2$ meV \cite{Wuetz.20} to $23$ meV \cite{Takashina.06,Goswami.07}, dynamics involving mixture between different branches of valleys can still be safely ignored as the inter-valley tunnel coupling and Coulomb interaction are exceedingly small \cite{Culcer.10}, which was also observed in experiment \cite{MacQuarrie.20}. The left, middle and right column of Supplementary Fig.~\ref{fig:STEvalSi} are for ``Outer, ``Center'' and ``Right'' schemes respectively, while the upper and lower rows are for $B=0$ and $B=0.165$ T respectively. The main results are similar to GaAs system, cf.~Supplementary Fig.~\ref{fig:STEvalBfo0p753DiffDetuningScheme}. It is noted that at $B = 0$ and in the ``Outer" scheme, $|\widehat{SS}\rangle$ and $|\widehat{TT}\rangle$ yield a larger $\Delta$ range as compared to the GaAs system. For the ``Right" scheme, the overlap in the $\Delta$ range between $|\widetilde{SS}\rangle$ and $|\widehat{TT}\rangle$ occur even at $B=0$. However, due to the near degeneracy between $|TT\rangle$ and $|TS\rangle$, the capacitive coupling strength is largely suppressed, cf. Supplementary Fig.~\ref{fig:alphaCompre3DetuningSchemeBfo0p75} (b). The near degeneracy between $|TT\rangle$ and $|TS\rangle$ in the nearly-sweet-spot regime is due to $|S^{11}T^{02}\rangle$ being the $|ST\rangle$ eigenstate instead of $|S^{02}T^{11}\rangle$ for GaAs system. Overall, the ``Outer" scheme at $B=0.165$ T is superior than others as it yields a nearly-sweet-spot regime (cf.~Fig.~\ref{fig:STEvalSi} (d)), while exhibiting a strong capacitive coupling, see Fig.~\ref{fig:alphaCompre3DetuningSchemeBfo0p75} (b).

\section{Existence of nearly sweet spot region}

In this section, we show that the existence of nearly-sweet-spot regime is not specific to the parameters chosen in our paper, but rather general under reasonable assumptions.
We assume that in the DDQD system, a maximum of one electron is allowed to be excited from the ground configuration, as configurations involving two or more excited electron would have much lower probability. This assumption allows us to safely discard higher energy excitations and simplifies the calculations. We will use $(n_1,n_2,n_3,n_4)$ to denote the number of electrons occupying each dot, where $n_j$ indicates number of electrons in the $j$th dot. As noted in the main text, here we focus on the ``Outer'' scheme only. Furthermore, we assume the DDQD system is symmetric under the reflection at $x=0$. This assumption leads to the same confinement strength for all the dots (cf. Eq.~\eqref{eq:Vpotential} in the main text) and symmetric detunings for both DQDs (cf. Eq.~\eqref{eq:detuningScheme} in the main text). This allows us to reduce the components to be taken into consideration for each logical state such that the relational equations can be simplified to observe their dependencies on magnetic field strength.

\subsection{$\Delta$ range of $|\widetilde{SS}\rangle$} \label{subsec:deltaRangeSS}
The relevant charge states for $|SS\rangle$ in the entire range of detuning are 
\begin{table}[t]
	\begin{tabular}{ | m{12em} | m{20em} | } 
		\hline
		Charge state& Multi-particle antisymmetrized state \\ 
		\hline
		$|(1,1,1,1)\rangle$ & $|S(\Phi_{1}\Phi_{2})\rangle|S(\Phi_{3}\Phi_{4})\rangle$\\
		\hline
		$|(1,1,1,1)\rangle^*$ & $|S(\Phi_{1}\Phi_{2})\rangle|S(\Phi_{3}\Phi_{8})\rangle$,$|S(\Phi_{2}\Phi_{5})\rangle|S(\Phi_{3}\Phi_{4})\rangle$\\
		\hline
		$|(1,1,1,1)\rangle^{**}$&$|S(\Phi_{1}\Phi_{2})\rangle|S(\Phi_{3}\Phi_{12})\rangle$,$|S(\Phi_{2}\Phi_{9})\rangle|S(\Phi_{3}\Phi_{4})\rangle$\\
		\hline
		$|(1,1,0,2)\rangle$/$|(2,0,1,1)\rangle$ & $|S(\Phi_{1}\Phi_{2})\rangle|S(\Phi_{4}\Phi_{4})\rangle$,$|S(\Phi_{1}\Phi_{1})\rangle|S(\Phi_{3}\Phi_{4})\rangle$\\ 
		\hline
		$|(1,1,0,2)\rangle^*$/$|(2,0,1,1)\rangle^*$ & $|S(\Phi_{1}\Phi_{2})\rangle|S(\Phi_{4}\Phi_{8})\rangle$,$|S(\Phi_{1}\Phi_{5})\rangle|S(\Phi_{3}\Phi_{4})\rangle$\\
		\hline
		$|(1,1,0,2)\rangle^{**}$/$|(2,0,1,1)\rangle^{**}$ &
		$|S(\Phi_{1}\Phi_{2})\rangle|S(\Phi_{4}\Phi_{12})\rangle$,$|S(\Phi_{1}\Phi_{9})\rangle|S(\Phi_{3}\Phi_{4})\rangle$\\
		\hline
		$|(2,0,0,2)\rangle$ & $|S(\Phi_{1}\Phi_{1})\rangle|S(\Phi_{4}\Phi_{4})\rangle$ \\ 
		\hline
		$|(2,0,0,2)\rangle^{*}$ & $|S(\Phi_{1}\Phi_{1})\rangle|S(\Phi_{4}\Phi_{8})\rangle$,$|S(\Phi_{1}\Phi_{5})\rangle|S(\Phi_{4}\Phi_{4})\rangle$ \\ 
		\hline
		$|(2,0,0,2)\rangle^{**}$ & $|S(\Phi_{1}\Phi_{1})\rangle|S(\Phi_{4}\Phi_{12})\rangle$,$|S(\Phi_{1}\Phi_{9})\rangle|S(\Phi_{4}\Phi_{4})\rangle$ \\ 
		\hline
	\end{tabular}
	\caption{Charge states of interest of $|SS\rangle$ in the entire range of detuning.}
	\label{tab:chargeStateSS}
\end{table}

In Supplementary Table~\ref{tab:chargeStateSS} we have defined:
\begin{align}
	\begin{split}
		|S(\Phi_j \Phi_k)\rangle&=\frac{1}{\sqrt{2}}\left(|\Phi_{j\uparrow} \Phi_{k\downarrow}\rangle + |\Phi_{k\uparrow} \Phi_{j\downarrow}\rangle\right),\\
		|S(\Phi_j \Phi_j)\rangle&=|\Phi_{j\uparrow} \Phi_{j\downarrow} \rangle,\\
		|\Phi_{j\uparrow} \Phi_{k\downarrow}\rangle&=\begin{vmatrix} \Phi_j (\mathbf{r}_1) \uparrow (\omega_1)& \Phi_j (\mathbf{r}_2) 
			\uparrow(\omega_2) \\ \Phi_k (\mathbf{r}_1) \downarrow (\omega_1)& \Phi_k (\mathbf{r}_2) \downarrow(\omega_2)\end{vmatrix},
	\end{split}
\end{align}
where $\uparrow (\downarrow)$ is the spin up (down) state and $\Phi_j$ is $j$th single-particle orbital as shown in Supplementary Fig.~\ref{fig:FDSt1to12}. The single and double asterisk(s) on the superscript of $|(n_1,n_2,n_3,n_4)\rangle$ indicate an electron occupying the orbital indexed as $5 \le j \le 8$ ($n=1,m=-1$, for a single asterisk) and $9 \le j \le 12$ ($n=1,m=1$, for double asterisks), respectively. The calculation can be further simplified by recognizing that some charge states are degenerate, e.g.~$|S(\Phi_1 \Phi_2)\rangle|S(\Phi_4 \Phi_8)\rangle$ and  $|S(\Phi_1 \Phi_5)\rangle|S(\Phi_3 \Phi_4)\rangle$. Hence, each charge state can be represented by only one of its corresponding multi-particle state, e.g.~$|(1,1,1,1)\rangle^*$ by $|S(\Phi_{1}\Phi_{2})\rangle|S(\Phi_{3}\Phi_{8})\rangle$, and $|(1,1,0,2)\rangle$ by $|S(\Phi_{1}\Phi_{2})\rangle|S(\Phi_{4}\Phi_{4})\rangle$. We denote the probability of electron excitation to higher orbitals by $\eta$, and the corresponding charge state $|\text{ex}(n_1,n_2,n_3,n_4)\rangle$. The expressions of $|S^{11}S^{11}\rangle$, $|\widetilde{SS}\rangle$ and $|S^{20}S^{02}\rangle$ are thus linear combinations of the corresponding charge states,

\begin{subequations}\label{eq:chargeStateSuperPos}
	\begin{align}
		\begin{split}
			|S^{11}S^{11}\rangle&=c_{|(1,1,1,1)\rangle} |(1,1,1,1)\rangle+c_{|(1,1,1,1)\rangle^*} |(1,1,1,1)\rangle^*+c_{|(1,1,1,1)\rangle^{**}} |(1,1,1,1)\rangle^{**}+\eta_{|(1,1,1,1)\rangle} |\text{ex}(1,1,1,1)\rangle,
		\end{split}\\
		\begin{split}
			|\widetilde{SS}\rangle&=c_{|(1,1,0,2)\rangle} |(1,1,0,2)\rangle+c_{|(1,1,0,2)\rangle^*} |(1,1,0,2)\rangle^{*}+c_{|(1,1,0,2)\rangle^{**}} |(1,1,0,2)\rangle^{**}\\
			&\qquad+\eta_{|(1,1,0,2)\rangle} |\text{ex}(1,1,0,2)\rangle,
		\end{split}\\
		\begin{split}
			|S^{20}S^{02}\rangle&=c_{|(2,0,0,2)\rangle} |(2,0,0,2)\rangle+c_{|(2,0,0,2)\rangle^*} |(2,0,0,2)\rangle^*+c_{|(2,0,0,2)\rangle^{**}} |(2,0,0,2)\rangle^{**}\\
			&\qquad+\eta_{|(2,0,0,2)\rangle} |\text{ex}(2,0,0,2)\rangle,
		\end{split}
	\end{align}
\end{subequations}
where the coefficients satisfy the normalization condition,

\begin{equation}\label{eq:normalizedSS}
	\begin{split}
		\left|c_{|(1,1,1,1)\rangle}\right|^2+\left|c_{|(1,1,1,1)\rangle^*}\right|^2+\left|c_{|(1,1,1,1)\rangle^{**}}\right|^2+\left|\eta_{|(1,1,1,1)\rangle}\right|^2 &=1,\\
		\left|c_{|(1,1,0,2)\rangle}\right|^2+\left|c_{|(1,1,0,2)\rangle^*}\right|^2+\left|c_{|(1,1,0,2)\rangle^{**}}\right|^2+\left|\eta_{|(1,1,0,2)\rangle}\right|^2 &=1,\\
		\left|c_{|(2,0,0,2)\rangle}\right|^2+\left|c_{|(2,0,0,2)\rangle^*}\right|^2+\left|c_{|(2,0,0,2)\rangle^{**}}\right|^2+\left|\eta_{|(2,0,0,2)\rangle}\right|^2 &=1.
	\end{split}
\end{equation}
The relationships between coefficients $c_{|\cdots\rangle}$ in Eq.~\eqref{eq:chargeStateSuperPos} are summarized in Eq.~\eqref{eq:coeffRelationSS} below. Eqs.~\eqref{eq:11w11}-\eqref{eq:2002w2002} are obtained by noting that electrons energetically favor the lowest orbitals. In $(1,1,0,2)/(2,0,1,1)$ charge configurations, double electron occupancy on one dot implies that  the two electrons favor to occupy different orbitals on the same dot as the Coulomb repulsion on the same orbital is stronger, e.g. $U_{15}<U_{1}$. Such repulsion effect is less pronounced in the $(1,1,1,1)$ charge configuration. Similar reasoning applies when comparing the $(1,1,0,2)$ and $(2,0,0,2)$ charge configurations, resulting in Eqs.~(\ref{eq:11wother}$-$\ref{eq:11sswother}). Also, the enhancement of the magnetic field lowers the energies of orbitals with $m<0$, increasing the probability for electrons to occupy those orbitals to mitigate strong on-site Coulomb interaction, $U_1$. Such effect is more significant for eigenstates with more doubly-occupied dots, i.e. the effect for $(1,1,0,2)$ is stronger than $(1,1,1,1)$, but the effect is strongest for $(2,0,0,2)$. These considerations lead to Eq.~\eqref{eq:etaDiff}.
\begin{subequations}\label{eq:coeffRelationSS}
	\begin{align}
		\begin{split}\label{eq:11w11}
			\left|c_{|(1,1,1,1)\rangle}\right| &\gg \left|c_{|(1,1,1,1)\rangle^*}\right|>\left|c_{|(1,1,1,1)\rangle^{**}}\right|,
		\end{split}\\
		\begin{split}\label{eq:1102w1102}
			\left|c_{|(1,1,0,2)\rangle}\right| &\gg \left|c_{|(1,1,0,2)\rangle^*}\right|> \left|c_{|(1,1,0,2)\rangle^{**}}\right|,
		\end{split}\\
		\begin{split}\label{eq:2002w2002}
			\left|c_{|(2,0,0,2)\rangle}\right| &> \left|c_{|(2,0,0,2)\rangle^*}\right|> \left|c_{|(2,0,0,2)\rangle^{**}}\right|,
		\end{split}\\
		\begin{split}\label{eq:11wother}
			\left|c_{|(1,1,1,1)\rangle}\right| &> \left|c_{|(1,1,0,2)\rangle}\right|> \left|c_{|(2,0,0,2)\rangle}\right|,
		\end{split}\\
		\begin{split}\label{eq:11swother}
			\left|c_{|(1,1,1,1)\rangle^*}\right| &> \left|c_{|(1,1,0,2)\rangle^*}\right|> \left|c_{|(2,0,0,2)\rangle^*}\right|,
		\end{split}\\
		\begin{split}\label{eq:11sswother}
			\left|c_{|(1,1,1,1)\rangle^{**}}\right|&>\left|c_{|(1,1,0,2)\rangle^{**}}\right|> \left|c_{|(2,0,0,2)\rangle^{**}}\right|,
		\end{split}\\
		\begin{split}\label{eq:etaDiff}
			\left|\eta_{|(2,0,0,2)\rangle}\right|^2-\left|\eta_{|(1,1,0,2)\rangle}\right|^2 &> \left|\eta_{|(1,1,0,2)\rangle}\right|^2-\left|\eta_{|(1,1,1,1)\rangle}\right|^2.
		\end{split}
	\end{align}
\end{subequations} 

\begin{figure}[t]
	\centering
	\includegraphics[width=0.8\columnwidth]{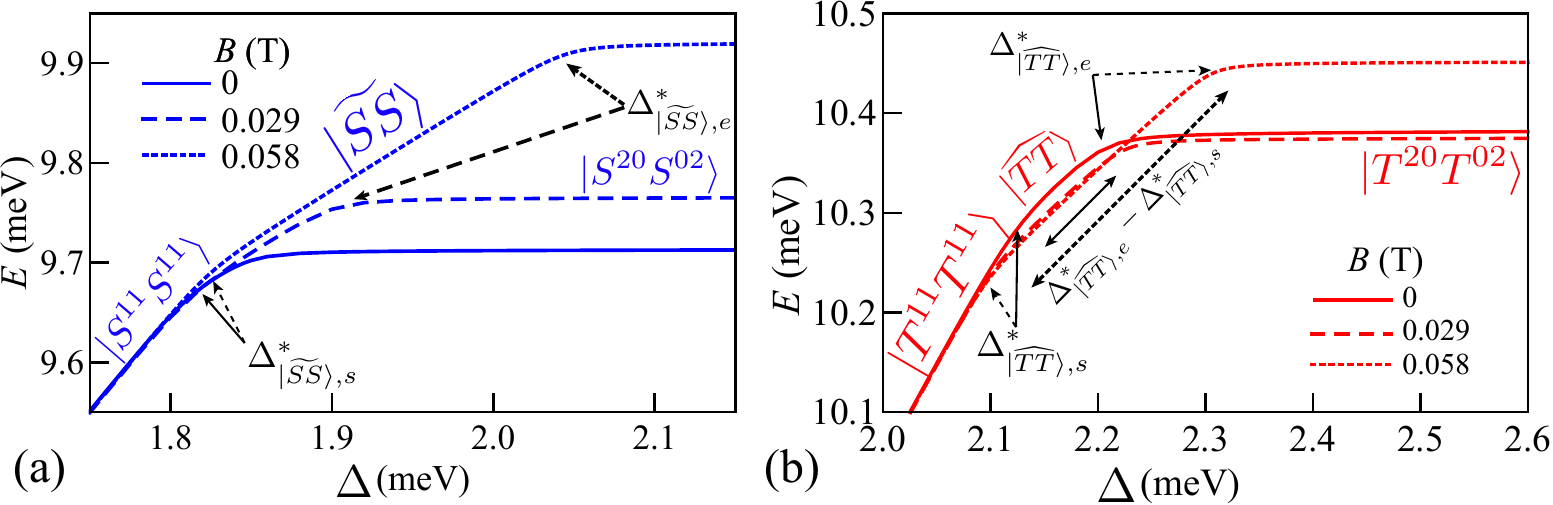}
	\caption{Energy level structures for (a) $|SS\rangle$ and (b) $|TT\rangle$ for different magnetic field strength as indicated.}
	\label{fig:SSTTEvalsDiffB}
\end{figure}

We can therefore obtain the starting, $\Delta_{{|\widetilde{SS}\rangle},s}^*$ and ending ,$\Delta_{{|\widetilde{SS}\rangle},e}^*$ points of the $\Delta$ range of $|\widetilde{SS}\rangle$, which are also the avoided-crossing points, by equating the eigenenergies of $|S^{11}S^{11}\rangle$ with $|\widetilde{SS}\rangle$ and $|\widetilde{SS}\rangle$ with  $|S^{20}S^{02}\rangle$ respectively. The results are,
\begin{subequations}\label{eq:deltaSS}
	\begin{align}
		\begin{split}\label{eq:deltaSS1}
			\Delta^*_{|\widetilde{SS}\rangle,s}&\approx\widetilde{E}_{|1111\rangle,|1102\rangle}+\hbar \omega_c \left[\left(\left|c_{|1111\rangle}\right|^2-\left|c_{|1102\rangle}\right|^2\right)+2\left(\left|c_{|1111\rangle^*}\right|^2-\left|c_{|1102\rangle^*}\right|^2\right)\right]\\
			&=\widetilde{E}_{|1111\rangle,|1102\rangle}+\hbar \omega_c \left[\left(\left|c_{|1111\rangle}\right|^2+2\left|c_{|1111\rangle^*}\right|^2\right)-\left(\left|c_{|1102\rangle}\right|^2+2\left|c_{|1102\rangle^*}\right|^2\right)\right]\\
			&\approx \widetilde{E}_{|1111\rangle,|1102\rangle}+\hbar \omega_c \left[\left(\left|c_{|1111\rangle}\right|^2+\left|c_{|1111\rangle^*}\right|^2\right)-\left(\left|c_{|1102\rangle}\right|^2+\left|c_{|1102\rangle^*}\right|^2\right)\right]\\
			&\approx \widetilde{E}_{|1111\rangle,|1102\rangle}+\hbar \omega_c \left[\left|\eta_{|(1,1,0,2)\rangle}\right|^2-\left|\eta_{|(1,1,1,1)\rangle}\right|^2\right],
		\end{split}\\
		\begin{split}\label{eq:deltaSS2}
			\Delta^*_{|\widetilde{SS}\rangle,e}&\approx\widetilde{E}_{|1102\rangle,|2002\rangle}+\hbar \omega_c \left[\left(\left|c_{|1102\rangle}\right|^2-\left|c_{|2002\rangle}\right|^2\right)+2\left(\left|c_{|1102\rangle^*}\right|^2-\left|c_{|2002\rangle^{*}}\right|^2\right)\right]\\
			&=\widetilde{E}_{|1102\rangle,|2002\rangle}+\hbar \omega_c \left[\left(\left|c_{|1102\rangle}\right|^2+2\left|c_{|1102\rangle^*}\right|^2\right)-\left(\left|c_{|2002\rangle}\right|^2+2\left|c_{|2002\rangle^{*}}\right|^2\right)\right]\\
			&\approx \widetilde{E}_{|1102\rangle,|2002\rangle}+\hbar \omega_c \left[\left(\left|c_{|1102\rangle}\right|^2+\left|c_{|1102\rangle^*}\right|^2\right)-\left(\left|c_{|2002\rangle}\right|^2+\left|c_{|2002\rangle^{*}}\right|^2\right)\right]\\
			&\approx  \widetilde{E}_{|1102\rangle,|2002\rangle}+\hbar \omega_c \left[\left|\eta_{|(2,0,0,2)\rangle}\right|^2-\left|\eta_{|(1,1,0,2)\rangle}\right|^2\right],
		\end{split}
	\end{align}
\end{subequations}
where
\begin{equation}\label{eq:energyIdentitySS}
	\begin{split}
		\widetilde{E}_{|(1,1,1,1)\rangle,|(1,1,0,2)\rangle}&=U_{15}+U_{14}-U_{34}-U_{23}+\left|c_{|(1,1,0,2)\rangle}\right|^2\left(U_{1}-U_{15}\right)+\hbar\omega_0\left(\left|c_{|(1,1,1,1)\rangle}\right|^2-\left|c_{|(1,1,0,2)\rangle}\right|^2\right),\\
		\widetilde{E}_{|(1,1,0,2)\rangle,|(2,0,0,2)\rangle}&=2U_{14}-U_{34}-2U_{13}+U_{1}\left(1-\left|c_{|(1,1,0,2)\rangle}\right|^2+\left|c_{|(2,0,0,2)\rangle}\right|^2\right)+\hbar \omega_0 \left(\left|c_{|(1,1,0,2)\rangle}\right|^2-\left|c_{|(2,0,0,2)\rangle}\right|^2\right).
	\end{split}
\end{equation}
It is reasonable to assume that although electrons may excite to higher orbitals, the probability of such excitation is low, i.e. $\left|c_{|\cdots \rangle^*}\right|\ll \left|c_{|\cdots \rangle}\right|$, as suggested by Eqs.~\eqref{eq:11w11} and \eqref{eq:1102w1102}, giving the third lines of Eq.~\eqref{eq:deltaSS1} and Eq.~\eqref{eq:deltaSS2}. The forth lines of Eq.~\eqref{eq:deltaSS1} and Eq.~\eqref{eq:deltaSS2} are obtained by applying the normalization condition, Eq.~\eqref{eq:normalizedSS}.

We are interested in the magnetic field dependence of the avoided-crossing points, i.e. $\partial \Delta^*_{|\widetilde{SS}\rangle,s}/\partial (\hbar \omega_c)$ and $\partial \Delta^*_{|\widetilde{SS}\rangle,e}/\partial (\hbar \omega_c)$. Implied by Eqs.~\eqref{eq:11wother}-\eqref{eq:11sswother}), the square brackets in the first lines of Eq.~\eqref{eq:deltaSS1} yield positive values. Also Eq.~\eqref{eq:etaDiff} suggests that the value of the square bracket in the fourth line of Eq.~\eqref{eq:deltaSS2} is greater than that in Eq.~\eqref{eq:deltaSS1}. These considerations imply that both $\Delta_{{|\widetilde{SS}\rangle},e}^*$ and $\Delta_{{|\widetilde{SS}\rangle},s}^*$ increase with increasing magnetic field, while the former increases faster. We therefore have,
\begin{equation}\label{eq:DeltaRangeSSBOND}
	\Delta_{{|\widetilde{SS}\rangle},e}^*-\Delta_{{|\widetilde{SS}\rangle},s}^*\propto \hbar\omega_c.
\end{equation}
This behavior is consistent with the result of Full-CI calculation, shown in Supplementary Fig.~\ref{fig:SSTTEvalsDiffB}(a). 

\subsection{$\Delta$ range of $|\widehat{TT}\rangle$}\label{subsec:deltaRangeTT}
The relevant charge states for $|TT\rangle$ in the entire range of detuning are 
\begin{table}[t]
	\begin{tabular}{ | m{12em} | m{20em} | } 
		\hline
		Charge state& Multi-particle antisymmetrized state \\ 
		\hline
		$|(1,1,1,1)\rangle$ & $|T(\Phi_{1}\Phi_{2})\rangle|T(\Phi_{3}\Phi_{4})\rangle$\\
		\hline
		$|(1,1,0,2)\rangle^{\mathbb{R}^*}$ & $|T(\Phi_{1}\Phi_{2})\rangle|T(\Phi_{4}\Phi_{8})\rangle$\\
		\hline
		$|(1,1,0,2)\rangle^{\mathbb{R}^{***}}$ & $|T(\Phi_{1}\Phi_{2})\rangle|T(\Phi_{8}\Phi_{12})\rangle$\\
		\hline
		$|(2,0,0,2)\rangle^{\mathbb{L}^* \mathbb{R}^*}$ & $|T(\Phi_{1}\Phi_{5})\rangle|T(\Phi_{4}\Phi_{8})\rangle$\\
		\hline
		$|(2,0,0,2)\rangle^{\mathbb{L}^*\mathbb{R}^{***}}$ & $|T(\Phi_{1}\Phi_{5})\rangle|T(\Phi_{8}\Phi_{12})\rangle$\\
		\hline
		$|(2,0,0,2)\rangle^{\mathbb{L}^{***}\mathbb{R}^{***}}$ & $|T(\Phi_{5}\Phi_{9})\rangle|T(\Phi_{8}\Phi_{12})\rangle$\\
		\hline
	\end{tabular}
	\caption{Charge states of interest of $|TT\rangle$ in the entire range of detuning.}
	\label{tab:chargeStateTT}
\end{table}

In Supplementary Table~\ref{tab:chargeStateTT} we have defined
\begin{equation}
	|T(\Phi_j \Phi_k)\rangle=\frac{1}{\sqrt{2}}\left(|\Phi_{j\uparrow} \Phi_{k\downarrow}\rangle - |\Phi_{k\uparrow} \Phi_{j\downarrow}\rangle\right).
\end{equation}

We use the superscripts to indicate which electron occupies higher energy states. In the superscripts, $\mathbb{L}$ and $\mathbb{R}$ indicate that the electron concerned is in the left and right DQD respectively. 
The number of asterisks indicates the number of excitation to the higher orbital with respect to ground orbital. For example, in the $(1,1,0,2)$ charge state, the ground state configuration of $|TT\rangle$ is $|T(\Phi_{1}\Phi_{2})\rangle|T(\Phi_{4}\Phi_{8})\rangle$, of which an electron in the right DQD occupies the first excited orbital, $\Phi_8$, is denoted as $|(1,1,0,2)\rangle^{R^*}$. If we move an electron in $\Phi_4$ in the right DQD to the second higher orbitals, $\Phi_{12}$ (state No.~12), the resultant charge state $|T(\Phi_{1}\Phi_{2})\rangle|T(\Phi_{8}\Phi_{12})\rangle$ is then denoted as $|(1,1,0,2)\rangle^{\mathbb{R}^{***}}$ as the total number of excitations of the electrons from ground orbitals is three.
For degenerate states, we keep only one of them and discard others as their coefficients can be absorbed into those shown in Table~\ref{tab:chargeStateTT}. Also, we empirically found that the Coulomb interaction between two dots in a DQD is similar regardless of the level of excitations, e.g. $U_{34}\approx U_{16}=U_{1,10}$. This allows us to neglect the excitation when each DQD is in (1,1) charge configuration, as it is energetically unfavorable to occupy higher orbitals in the inter-dot case. On the other hand, the on-site (inter-orbital) Coulomb repulsion is weaker when higher orbitals are involved, e.g. $U_{59}<U_{15}<U_{1}$. Therefore, for a doubly-occupied quantum dot, an electron will favor to occupy excited orbitals. Near the avoid-crossing between $|T^{11}T^{11}\rangle$ and $|\widehat{TT}\rangle$ at small detuning, the charge states are
\begin{subequations}
	\begin{align}
		\begin{split}
			|T^{11}T^{11}\rangle&\approx|(1,1,1,1)\rangle,
		\end{split}\\
		\begin{split}
			|\widehat{TT}\rangle&\approx|(1,1,0,2)\rangle^{\mathbb{R}^*}.
		\end{split}
	\end{align}
\end{subequations}
At larger detuning, near the avoided-crossing between $|\widehat{TT}\rangle$ and $|T^{20}T^{02}\rangle$, the charge states are
\begin{subequations}\label{eq:largeDetuningTTEffecCompt}
	\begin{align}
		\begin{split}
			|\widehat{TT}\rangle&=c_{|(1,1,0,2)\rangle^{\mathbb{R}^*}}|(1,1,0,2)\rangle^{\mathbb{R}^*}+c_{|(1,1,0,2)\rangle^{\mathbb{R}^{***}}}|(1,1,0,2)\rangle^{\mathbb{R}^{***}}\\
			&\quad+\eta_{|(1,1,0,2)\rangle}|\text{ex}(1,1,0,2)\rangle,
		\end{split}\\
		\begin{split}
			|T^{20}T^{02}\rangle&=c_{|(2,0,0,2)\rangle^{\mathbb{L}^* \mathbb{R}^*}}|(2,0,0,2)\rangle^{\mathbb{L}^* \mathbb{R}^*}+c_{|(2,0,0,2)\rangle^{\mathbb{L}^*\mathbb{R}^{**}}}|(2,0,0,2)\rangle^{\mathbb{L}^*\mathbb{R}^{**}}+c_{|(2,0,0,2)\rangle^{\mathbb{L}^{***}\mathbb{R}^{***}}}|(2,0,0,2)\rangle^{\mathbb{L}^{***}\mathbb{R}^{***}}\\
			&\quad+\eta_{|(2,0,0,2)\rangle }|\text{ex}(2,0,0,2)\rangle.
		\end{split}
	\end{align}
\end{subequations}
The relationship between the coefficients in Eq.~\eqref{eq:largeDetuningTTEffecCompt} are given as: 
\begin{subequations}
	\begin{align}
		\begin{split}\label{eq:normal1102}
			\left|c_{|(1,1,0,2)\rangle^{\mathbb{R}^*}}\right|^2&+\left|c_{|(1,1,0,2)\rangle^{\mathbb{R}^{***}}}\right|^2+\left|\eta_{|(1,1,0,2)\rangle}\right|^2=1,
		\end{split}\\
		\begin{split}\label{eq:normal2002}
			\left|c_{|(2,0,0,2)\rangle^{\mathbb{L}^* \mathbb{R}^*}}\right|^2&+\left|c_{|(2,0,0,2)\rangle^{\mathbb{L}^*\mathbb{R}^{**}}}\right|^2+\left|c_{|(2,0,0,2)\rangle^{\mathbb{L}^{***}\mathbb{R}^{***}}}\right|^2+\left|\eta_{|(1,1,0,2)\rangle}\right|^2=1,
		\end{split}\\
		\begin{split}\label{eq:1102w2002}
			\left|c_{|(1,1,0,2)\rangle^{\mathbb{R}^*}}\right|^2&>\left|c_{|(2,0,0,2)\rangle^{\mathbb{L}^* \mathbb{R}^*}}\right|^2.
		\end{split}
	\end{align}
\end{subequations}
Eqs.~\eqref{eq:normal1102}-\eqref{eq:normal2002} are obtained from the normalization condition, while Eq.~\eqref{eq:1102w2002} is found by recognizing that the Coulomb repulsion effect is more pronounced for $|(2,0,0,2)\rangle^{\mathbb{L}^* \mathbb{R}^*}$ than $|(1,1,0,2)\rangle^{\mathbb{R}^*}$, so that electrons $|(2,0,0,2)\rangle^{\mathbb{L}^* \mathbb{R}^*}$ favor occupying highly-lying orbitals, resulting in a higher energy state with lower probability than the latter. We then obtain $\Delta_{{|\widehat{TT}\rangle},s}^*$ by equating the eigenenergies of $|T^{11}T^{11}\rangle$ and  $|\widehat{TT}\rangle$ at smaller detuning and $\Delta_{{|\widehat{TT}\rangle},e}^*$ by equating the eigenenergies of $|\widehat{TT}\rangle$ and $|T^{20}T^{02}\rangle$ at larger detuning, giving
\begin{subequations}
	\begin{align}
		\begin{split}\label{eq:DeltaTTs1}
			\Delta_{{|\widehat{TT}\rangle},s}^*&\approx\widetilde{E}_{|(1,1,1,1)\rangle,|(1,1,0,2)\rangle}-\hbar\omega_c,
		\end{split}\\
		\begin{split}\label{eq:DeltaTTs2ms1}
			\Delta_{{|\widehat{TT}\rangle},e}^*-\Delta_{{|\widehat{TT}\rangle},s}^*&\approx\widetilde{E}_{|(1,1,1,1)\rangle,|(1,1,0,2)\rangle}+\widetilde{E}_{|(1,1,0,2)\rangle,|(2,0,0,2)\rangle}\\
			&\quad+\hbar\omega_c\left\{\left[1-\left(\left|c_{|(2,0,0,2)\rangle^{\mathbb{L}^* \mathbb{R}^*}}\right|^2+\left|c_{|(2,0,0,2)\rangle^{\mathbb{L}^*\mathbb{R}^{***}}}\right|^2\right)\right]+\left(\left|c_{|(1,1,0,2)\rangle^{\mathbb{R}^*}}\right|^2-\left|c_{|(2,0,0,2)\rangle}\right|^2\right)\right\},
		\end{split}
	\end{align}
\end{subequations}
where 
\begin{equation}
	\begin{split}
		\widetilde{E}_{|(1,1,1,1)\rangle,|(1,1,0,2)\rangle}&=U_{14}+U_{59}-U_{34}-U_{14}+ \left(U_{15}-U_{59}\right)\left|c_{|1,1,0,2\rangle^{\mathbb{R}^*}}\right|^2+\hbar \omega_0 \left(2-\left|c_{|1,1,0,2\rangle^{\mathbb{R}^*}}\right|^2\right),\\
		\widetilde{E}_{|(1,1,0,2)\rangle,|(2,0,0,2)\rangle} &=U_{59}+2U_{14}-U_{34}-2U_{13}\\
		&\quad+
		\left(U_{59}-U_{15}+\hbar \omega_0\right)\left(\left|c_{|(1,1,0,2)\rangle^{\mathbb{R}^*}}\right|^2-2\left|c_{|(2,0,0,2)\rangle^{\mathbb{L}^* \mathbb{R}^*}}\right|^2-\left|c_{|(2,0,0,2)\rangle^{\mathbb{L}^*\mathbb{R}^{***}}}\right|^2\right)+2\hbar \omega_0.
	\end{split}
\end{equation}
Again, we are interested in the the magnetic-field dependence of the avoided-crossing points, i.e. $\partial \Delta^*_{|\widehat{TT}\rangle,s}/\partial (\hbar \omega_c)$ and $\partial \Delta^*_{|\widehat{TT}\rangle,e}/\partial (\hbar \omega_c)$. Eq.~\eqref{eq:DeltaTTs1} shows that a stronger magnetic field leads to a smaller value of $\Delta_{{|\widehat{TT}\rangle},s}^*$. Also, Eq.~\eqref{eq:normal2002} and Eq.~\eqref{eq:1102w2002} suggest that the curly bracket in Eq.~\eqref{eq:DeltaTTs2ms1} yields a positive value, so that the $\Delta$ range of $|\widehat{TT}\rangle$ increases with the magnetic field as
\begin{equation}\label{eq:DeltaRangeTTABOND}
	\left(\Delta^*_{{|\widehat{TT}\rangle},e}-\Delta^*_{{|\widehat{TT}\rangle},s}\right) \propto \hbar\omega_c.
\end{equation}
These results are again consistent with those from the Full-CI calculation, as can be seen in Supplementary Fig.~\ref{fig:SSTTEvalsDiffB}(b).

\subsection{$\Delta$ range of $|S^{20}T^{11}\rangle$ and $|T^{11}S^{02}\rangle$}
We denote the starting and ending $\Delta$ values of $|S^{20}T^{11}\rangle$ as $\Delta^*_{|ST\rangle,s}$ and  $\Delta^*_{|ST\rangle,e}$ respectively (cf. Supplementary Fig.~\ref{fig:STEvalBfo0p753DiffDetuningScheme}(d)). Near the avoided crossing between $|S^{11}T^{11}\rangle$ and $|S^{20}T^{11}\rangle$, the relevant charge states for $|ST\rangle$ are
\begin{table}[t]
	\begin{tabular}{ | m{12em} | m{20em} | } 
		\hline
		Charge state& Multi-particle antisymmetrized state \\ 
		\hline
		$|(1,1,1,1)\rangle$ & $|S(\Phi_{1}\Phi_{2})\rangle|T(\Phi_{3}\Phi_{4})\rangle$\\
		\hline
		$|(1,1,1,1)\rangle^*$ & $|S(\Phi_{1}\Phi_{2})\rangle|T(\Phi_{3}\Phi_{8})\rangle$,$|S(\Phi_{2}\Phi_{5})\rangle|T(\Phi_{3}\Phi_{4})\rangle$\\
		\hline
		$|(1,1,1,1)\rangle^{**}$&$|S(\Phi_{1}\Phi_{2})\rangle|T(\Phi_{3}\Phi_{12})\rangle$,$|S(\Phi_{2}\Phi_{9})\rangle|T(\Phi_{3}\Phi_{4})\rangle$\\
		\hline
		$|(1,1,0,2)\rangle$/$|(2,0,1,1)\rangle$ & $|S(\Phi_{1}\Phi_{2})\rangle|T(\Phi_{4}\Phi_{4})\rangle$,$|S(\Phi_{1}\Phi_{1})\rangle|T(\Phi_{3}\Phi_{4})\rangle$\\ 
		\hline
		$|(1,1,0,2)\rangle^*$/$|(2,0,1,1)\rangle^*$ & $|S(\Phi_{1}\Phi_{2})\rangle|T(\Phi_{4}\Phi_{8})\rangle$,$|S(\Phi_{1}\Phi_{5})\rangle|T(\Phi_{3}\Phi_{4})\rangle$\\
		\hline
		$|(1,1,0,2)\rangle^{**}$/$|(2,0,1,1)\rangle^{**}$ &
		$|S(\Phi_{1}\Phi_{2})\rangle|T(\Phi_{4}\Phi_{12})\rangle$,$|S(\Phi_{1}\Phi_{9})\rangle|T(\Phi_{3}\Phi_{4})\rangle$\\
		\hline
	\end{tabular}
	\caption{Charge states of interest of $|ST\rangle$ near the avoided crossing between $|S^{11}T^{11}\rangle$ and $|S^{20}T^{11}\rangle$.}
	\label{tab:chargeStateST}
\end{table}

To avoid confusion, we emphasize that even though the notations for charge states here are same as those shown in Supplementary Sec.~\ref{subsec:deltaRangeSS}, they refer to $|ST\rangle$ instead of $|SS\rangle$. We have
\begin{subequations}\label{eq:chargeStateST}
	\begin{align}
		\begin{split}
			|S^{11}T^{11}\rangle&=c_{|(1,1,1,1)\rangle} |(1,1,1,1)\rangle+c_{|(1,1,1,1)\rangle^*} |(1,1,1,1)\rangle^*+c_{|(1,1,1,1)\rangle^{**}} |(1,1,1,1)\rangle^{**}+\eta_{|(1,1,1,1)\rangle} |\text{ex}(1,1,1,1)\rangle,
		\end{split}\\
		\begin{split}
			|S^{20}T^{11}\rangle&=c_{|(1,1,0,2)\rangle} |(1,1,0,2)\rangle+c_{|(1,1,0,2)\rangle^*} |(1,1,0,2)\rangle^{*}+c_{|(1,1,0,2)\rangle^{**}} |(1,1,0,2)\rangle^{**}\\
			&\qquad+\eta_{|(1,1,0,2)\rangle} |\text{ex}(1,1,0,2)\rangle.
		\end{split}
	\end{align}
\end{subequations}
Equating the eigenenergies of $|S^{11}T^{11}\rangle$ with $|S^{20}T^{11}\rangle$ gives that
\begin{equation}\label{eq:deltaST}
	\begin{split}
		\Delta^*_{|ST\rangle,s}&\approx\widetilde{E}_{|1111\rangle,|1102\rangle}+\hbar \omega_c \left[\left(\left|c_{|1111\rangle}\right|^2-\left|c_{|1102\rangle}\right|^2\right)+2\left(\left|c_{|1111\rangle^*}\right|^2-\left|c_{|1102\rangle^*}\right|^2\right)\right],
	\end{split}
\end{equation}
where
\begin{equation}\label{eq:energyIdentityST}
	\begin{split}
		\widetilde{E}_{|(1,1,1,1)\rangle,|(1,1,0,2)\rangle}&=U_{15}+U_{14}-U_{34}-U_{23}+\left|c_{|(1,1,0,2)\rangle}\right|^2\left(U_{1}-U_{15}\right)+\hbar\omega_0\left(\left|c_{|(1,1,1,1)\rangle}\right|^2-\left|c_{|(1,1,0,2)\rangle}\right|^2\right).
	\end{split}
\end{equation}
Eq.~\eqref{eq:deltaST} yields the same expression as the first line of Eq.~\eqref{eq:deltaSS1}, implying that 
\begin{equation}\label{eq:SSsSTsCompareLocation}
	\Delta^*_{|\widetilde{SS}\rangle,s} \approx \Delta^*_{|ST\rangle,s}.
\end{equation}

Near the avoided crossing between $|S^{20}T^{11}\rangle$ and $|S^{20}T^{02}\rangle$, we focus on the main composition of states $|ST\rangle$ and $|TT\rangle$. The relevant charge states are
\begin{subequations}\label{eq:ST11022002TT11022002}
	\begin{align}
		\begin{split}
			|S^{20}T^{11}\rangle&\approx |S(\Phi_1\Phi_1)\rangle|T(\Phi_3\Phi_4)\rangle,
		\end{split}\\
		\begin{split}
			|S^{20}T^{02}\rangle&\approx |S(\Phi_1\Phi_1)\rangle|T(\Phi_4\Phi_8)\rangle,
		\end{split}\\
		\begin{split}
			|\widehat{TT}\rangle&\approx |T(\Phi_1\Phi_5)\rangle|T(\Phi_3\Phi_4)\rangle,
		\end{split}\\
		\begin{split}
			|T^{20}T^{02}\rangle&\approx |T(\Phi_1\Phi_5)\rangle|T(\Phi_4\Phi_8)\rangle.
		\end{split}
	\end{align}
\end{subequations}
The charges states with excited orbitals as shown in Eq.~\eqref{eq:largeDetuningTTEffecCompt} are not considered here for the purpose of comparison between the $|ST\rangle$ and $|TT\rangle$ charge state. For degenerate states, we keep only one of them and discard others as their coefficients can be absorbed into those shown in Eq.~\eqref{eq:ST11022002TT11022002}. Also, we empirically found that the Coulomb interaction between two dots is similar regardless of the level of excitations, e.g. $U_{14}\approx U_{18}\approx U_{45}$. We then obtain $\Delta^*_{|\widehat{TT}\rangle,e}$ by equating the eigenenergies of $|\widehat{TT}\rangle$ and $|T^{20}T^{02}\rangle$, and $\Delta^*_{
	|ST\rangle,e}$ by equating the eigenenergies of $|S^{20}T^{11}\rangle$ and $|S^{20}T^{02}\rangle$, giving
\begin{subequations}\label{eq:STeTTeCompare}
	\begin{align}
		\begin{split}
			\Delta^*_{|ST\rangle,e}&\approx U_{15}-(U_{13}+U_{13})-U_{34}+(U_{45}+U_{45})+\hbar \omega_0 -\hbar \omega_c,
		\end{split}\\
		\begin{split}
			\Delta^*_{|\widehat{TT}\rangle,e}&\approx U_{15}-(U_{13}+U_{35})-U_{34}+(U_{45}+U_{85})+\hbar \omega_0 -\hbar \omega_c\\
			&\approx U_{15}-(U_{13}+U_{13})-U_{34}+(U_{45}+U_{45})+\hbar \omega_0 -\hbar \omega_c.
		\end{split}
	\end{align}
\end{subequations}
Eq.~\eqref{eq:STeTTeCompare} shows that
\begin{equation}\label{eq:STeTTeCompareLocation}
	\Delta^*_{|ST\rangle,e}\approx\Delta^*_{|\widehat{TT}\rangle,e}.
\end{equation}
Eq.~\eqref{eq:SSsSTsCompareLocation} and Eq.~\eqref{eq:STeTTeCompareLocation} are consistent with results from the Full-CI calculation, as can be seen in Supplementary Fig.~\ref{fig:STEvalBfo0p753DiffDetuningScheme}(d). Since the eigenenergy of $|TS\rangle$ is the same as that of $|ST\rangle$ for the ``Outer'' scheme, the starting and ending points of the $\Delta$ range of $|T^{11}S^{02}\rangle$ are identical to those of $|ST\rangle$.

\subsection{Nearly sweet spot regime and strong inter-qubit coupling}

To perform a high fidelity two-qubit gate, we need on one hand substantially suppressed charge-noises, i.e. $\partial J^{\text{eff}}_\mathbb{L}/\partial\Delta \rightarrow 0$, $\partial J^{\text{eff}}_\mathbb{R}/\partial\Delta \rightarrow 0$ and $\partial\alpha/\partial\Delta \rightarrow 0$, and on the other hand a strong inter-qubit coupling $\alpha$ to limit the exposure to other decoherence channels. To satisfy the first condition, we need a relatively wide $\Delta$ range where  $\partial E_{|SS\rangle}/\partial\Delta\approx \partial E_{|ST\rangle}/\partial\Delta\approx \partial E_{|TS\rangle}/\partial\Delta\approx \partial E_{|TT\rangle}/\partial\Delta$. 
Concerning the second condition, we note that 
an electron of $|T^{02}\rangle$ ($|T^{20}\rangle$) has to occupy an excited orbital, while both electrons of $|S^{02}\rangle$ ($|S^{20}\rangle$) can occupy the ground orbital. Therefore the difference of single-particle energy due to orbital effect, $\hbar \omega_0$ contributes to strong inter-qubit capacitive coupling, $\widetilde{\alpha}$, as
\begin{equation}
	\begin{split}
		\widetilde{\alpha} &= \frac{1}{4}\left[(U_{28}-U_{24})+(U_{18}-U_{14})+(U_{15}-U_\text{os})+(\hbar\omega_0-\hbar\omega_c)\right].
		\label{eq:alphaABOND}
	\end{split}
\end{equation}
We emphasize that $\widetilde{\alpha}$ is the $\alpha$ value where $\partial\alpha/\partial\Delta=0$, (cf. Supplementary Fig.~\ref{fig:alphaCompre3DetuningSchemeBfo0p75}).
As stated in the previous section, we found that the inter-dot Coulomb interaction is not considerably dependent on the orbital excitation, while the on-site Coulomb interaction is weaker when an electron occupies an excited orbital, e.g. $U_{15} < U_1$. Hence, Eq.~\eqref{eq:alphaABOND} predicts that the capacitive coupling in this case should primarily depend on the on-site Coulomb interaction and orbital excitation energy.
To facilitate later discussions, we note that 
\begin{subequations}\label{eq:SSs2TTs1wcDependence}
	\begin{align}
		\begin{split}\label{eq:SSs2wcDependence}
			\frac{\partial\Delta_{{|\widetilde{SS}\rangle},e}^*}{\partial(\hbar \omega_c)}>0,
		\end{split}\\
		\begin{split}\label{eq:TTs1wcDependence}
			\frac{\partial\Delta_{{|\widehat{TT}\rangle},s}^*}{\partial(\hbar \omega_c)} \approx-1,
		\end{split}
	\end{align}
	where Eq.~\eqref{eq:SSs2wcDependence} is given by the last line of Eq.~\eqref{eq:deltaSS2} while Eq.~\eqref{eq:TTs1wcDependence} is derived from Eq.~\eqref{eq:DeltaTTs1}.
\end{subequations} 

When the magnetic field increases, $\Delta_{{|\widetilde{SS}\rangle},e}^*$ increases (Eq.~\eqref{eq:SSs2wcDependence}), while $\Delta^*_{{|\widehat{TT}\rangle},s}$ decreases (Eq.~\eqref{eq:TTs1wcDependence}), making it possible an overlap between the $\Delta$ ranges of  $|\widetilde{SS}\rangle$ and $|\widehat{TT}\rangle$ above certain magnetic field strength, satisfying
\begin{equation}\label{eq:overlapCondition}
	\Delta_{{|\widetilde{SS}\rangle},e}^*-\Delta^*_{{|\widehat{TT}\rangle},s}>0.
\end{equation}
Also, since $\Delta^*_{|ST\rangle,s} \approx \Delta^*_{|\widetilde{SS}\rangle,s} < \Delta^*_{|\widetilde{SS}\rangle,e}$ (Eq.~\eqref{eq:SSsSTsCompareLocation}) and $\Delta^*_{|ST\rangle,e} \approx \Delta^*_{|\widehat{TT}\rangle,e}> \Delta^*_{|\widehat{TT}\rangle,s}$ (Eq.~\eqref{eq:STeTTeCompareLocation}), the overlap will occur simultaneously for all logical eigenstates $|\widetilde{SS}\rangle$, $|S^{20}T^{11}\rangle$, $|T^{11}S^{02}\rangle$ and $|\widehat{TT}\rangle$. In addition, the $\Delta$ ranges of $|\widetilde{SS}\rangle$ and  $|\widehat{TT}\rangle$ increase with the magnetic field, given by Eq.~\eqref{eq:DeltaRangeSSBOND} and \eqref{eq:DeltaRangeTTABOND}, making the overlap range wide enough. Therefore, we have demonstrated that at sufficiently large magnetic field, the DDQD system will exhibit a nearly-sweet-spot regime for both single-qubit exchange interactions and the inter-qubit capacitive coupling, and at the same time the capacitive coupling can be strong.

\section{Effect of strong magnetic field}
\begin{figure}[t]
	\includegraphics[width=0.75\columnwidth]{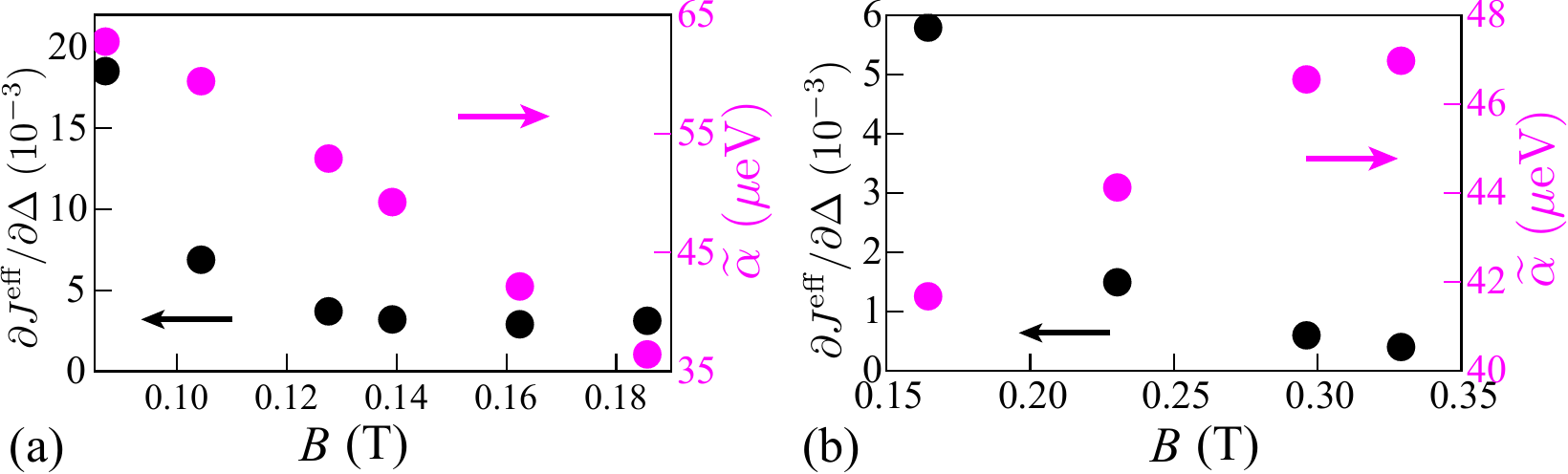}
	\caption{The values of $\partial J^\text{eff}/\partial\Delta$ evaluated at the $\Delta$ value when $\alpha$ reaches its maximal value $\tilde{\alpha}$ in the nearly-sweet-spot regime for (a) GaAs and (b) silicon. The results are obtained for ``Outer" scheme.}
	\label{fig:strongMag}
\end{figure}
Supplementary Fig.~\ref{fig:strongMag} (a) show the value of $\partial J^\text{eff}/\partial\Delta$ 
for larger magnetic field strength for of GaAs system. It is observed that the $\partial J^\text{eff}/\partial\Delta$ is further suppressed along with the increase of magnetic field strength. Although the capacitive coupling strength decreases with magnetic field strength, the value is large enough to perform a fast entangling gate. Supplementary Fig.~\ref{fig:strongMag} (b) shows the results for silicon system. In contrast to GaAs system, the suppression of $\partial J^\text{eff}/\partial\Delta$ along with magnetic field strength is accompanied with the increase of capacitive coupling strength.
\section{Capacitive coupling strength}\label{sec:capacitiveCoupling}
\begin{figure}[t]
	\includegraphics[width=0.95\columnwidth]{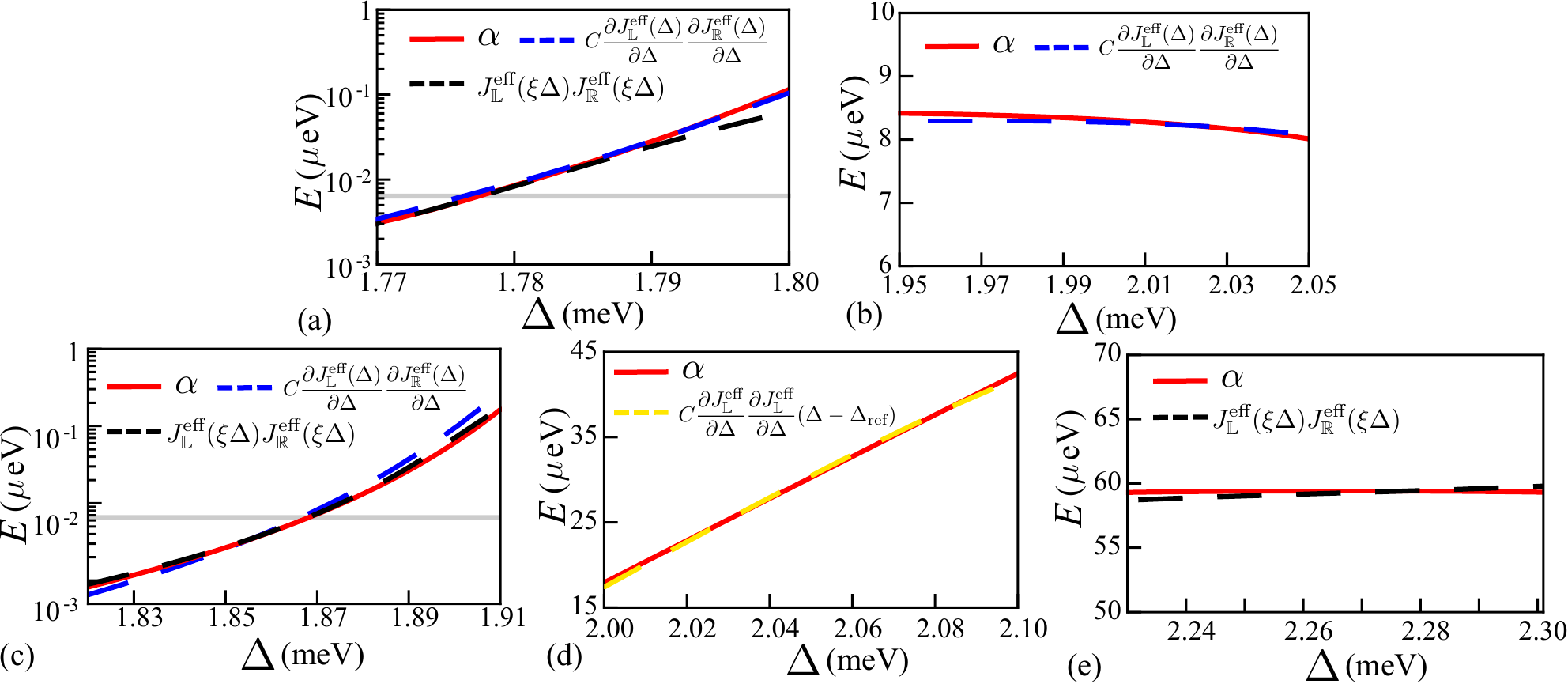}
	\caption{Relation between effective exchange energies, $J^\text{eff}_\mathbb{L}$ and $J^\text{eff}_\mathbb{L}$, and capacitive coupling strength, $\alpha$. The results are obtained for GaAs system at (a)-(b) $B=0$ T and (c)-(e) $B=0.104$T. The gray horizontal line indicates the capacitive coupling strength reported by \cite{Shulman.12}, $\tau_\text{ent}=140$ ns ($\tau_\text{ent}=\pi/(2\alpha)$), at which the calculated effective exchange energies are $911$ MHz and $608$ MHz for $B=0$ and $B=0.104$ T respectively.The parameters are (a) $C=2.1\times10^{-2}$ $\mu$eV, $\xi=1.015$, (b) $C=10^{-1}$ $\mu$eV, $\xi=1.012$, (c) $C=3.4*10^{-2}$ $\mu$eV, (d) $C=4.25$ $\mu$eV, $\Delta_\text{ref}=1.93$ meV and (e) $\xi=1$.}
	\label{fig:alphaVSJTrend}
\end{figure}
\subsection{Moderately large detuning range at both weak and strong magnetic field strength}
The capacitive coupling at moderately large detuning range can be understood qualitatively using Hund-Mulliken approximation \cite{Calderon.15}. In that detuning regime, when $|S^{02}\rangle$ is energetically inaccessible, the lowest singlet state of single DQD is a hybridized state of $|S^{11}\rangle $ and $|S^{20}\rangle$,
\begin{equation}\label{eq:Shybrid}
	|\widetilde{S}\rangle = \sin \theta |S^{20}\rangle+\cos\theta|S^{11}\rangle,
\end{equation}
while $|T\rangle = |T^{11}\rangle$. The mixing angle $\theta$ is introduced to parameterize the hybridization and is defined as 
\begin{equation}
	\tan \theta = \frac{2\sqrt{2} t}{U_\text{os}-\Delta-V_++\sqrt{\left(U_\text{os}-\Delta-V_+\right)^2+8 t^2}}.
\end{equation}
The single qubit exchange energy, $J(\Delta)$, is
\begin{equation}
	J(\Delta) = V_- - \frac{1}{2}\left[U-\Delta+V_+-\sqrt{\left(U_\text{os}-\Delta-V_+\right)^2+8t^2}\right].
\end{equation}
Projecting $H$, Eq.~\eqref{eq:HubbardModel2ndQuan}, onto the subspace spanned by $\{|\widetilde{S}\widetilde{S}\rangle, |\widetilde{S}T\rangle,|T\widetilde{S}\rangle,|TT\rangle\}$, up to $\mathcal{O}[t^2/(U_\text{os}-V_+-\Delta)^2]$, we obtain
\begin{subequations}\label{eq:JeffAlphaAna}
	\begin{align}
		\begin{split}
			J^\text{eff}_\mathbb{L} (\Delta) = J^\text{eff}_\mathbb{R} (\Delta) &=\Delta-U_\text{os}+V_-+ \frac{1}{4} \left(3U_{14}-2U_{24} - U_{23}\right) - \frac{2t^2}{\left(U_\text{os}-V_+-\Delta\right)^2} \left[\left(U_\text{os}-V_+-\Delta\right)+U_{14} - U_{24}\right]\\
			&\quad+ \mathcal{O} \left[\left(\frac{t}{U_\text{os}-V_+-\Delta}\right)^3\right],
		\end{split}\\
		\begin{split}\label{eq:alphaAna}
			\alpha (\Delta) &= \frac{1}{4}\left(U_{14}-2U_{24}+U_{23}\right)\left[1-\frac{4t^2}{\left(U_\text{os}-\Delta-V_+\right)^2}\right]+ \mathcal{O} \left[\left(\frac{t}{U_\text{os}-V_+-\Delta}\right)^3\right],
		\end{split}
	\end{align}
\end{subequations}
where $\Delta>U_\text{os}-V_+$ and $t/(U_\text{os}-V_+-\Delta)$ is small. Note that $|\widetilde{S}\rangle$ here denote hybridized single state for single qubit, instead of the $|\widehat{SS}\rangle$ at the nearly-sweet-spot region. Eq.~\ref{eq:JeffAlphaAna} is obtained for ``Outer" scheme.
From Eq.~\eqref{eq:JeffAlphaAna}, we obtain
\begin{equation}\label{eq:JeffDerivative}
	\frac{\partial J^{\text{eff}}_\mathbb{L}}{\partial\Delta} = \frac{\partial J^{\text{eff}}_\mathbb{R}}{\partial\Delta} = 1 - \frac{2t^2}{\left(U_\text{os}-V_+-\Delta\right)^2} +\mathcal{O}\left[\frac{t^2 U_{jk}}{\left(U_\text{os}-V_+-\Delta\right)^3}\right]
\end{equation}
Hence,
\begin{equation}\label{eq:alphaWithJDerivative}
	\alpha(\Delta)=\frac{1}{4}\left(U_{14}-2U_{24}+U_{23}\right)\frac{\partial J^{\text{eff}}_\mathbb{L}}{\partial\Delta}\frac{\partial J^{\text{eff}}_\mathbb{R}}{\partial\Delta},
\end{equation}
conforming with the experimental result that $\alpha(\Delta)\propto (\partial J^{\text{eff}}_\mathbb{L}/\partial \Delta)(\partial J^{\text{eff}}_\mathbb{R}/\partial \Delta ) $ \cite{Nichol.17,Shulman.12}. Also, Eq.~\eqref{eq:alphaWithJDerivative} shows that it is the dipole-dipole interaction mediating the capacitive coupling. Supplemental Fig.~\ref{fig:alphaVSJTrend} (a) and (c) shows that this relation holds for either weak and strong applied magnetic field strengths. We believe that the relation, $\alpha \propto J^{\text{eff}}_\mathbb{L} J^{\text{eff}}_\mathbb{R}$, which is widely adopted in experiments \cite{Nichol.17,Shulman.12}, is on one hand empirical as it is not evident from Eq.~\eqref{eq:JeffAlphaAna}, but on the other hand is a very good approximation since numerical results show good agreements in certain detuning range, cf. Fig.~\ref{fig:alphaVSJTrend} (a) and (c).
\subsection{Large detuning at weaker magnetic field strength}
At much larger detuning, $|SS\rangle$ enters $|S^{20}S^{02}\rangle$ region while $|TT\rangle$ persists in $|T^{11}T^{11}\rangle$ state due to Pauli exclusion principle. This corresponds to our calculation result for small magnetic field strength, at which the energy levels are similar to the one presented for $B = 0$ T, cf. Supplementary Fig.~\ref{fig:STEvalBfo0p753DiffDetuningScheme} (a) and \ref{fig:STEvalSi} (a). From Eq.~\eqref{eq:JeffAlphaAna} and Eq.~\eqref{eq:JeffDerivative},
\begin{subequations}\label{eq:muchLargerDetuningJDerivativeAlpha}
	\begin{align}
		\begin{split}
			\frac{\partial J^{\text{eff}}_\mathbb{L}}{\partial\Delta}&=\frac{\partial J^{\text{eff}}_\mathbb{R}}{\partial\Delta} \approx 1
		\end{split}\\
		\begin{split}
			\alpha(\Delta)&\approx\frac{1}{4}\left(U_{14}-2U_{24}+U_{23}\right)
		\end{split}
	\end{align}
\end{subequations}
as $\Delta \gg (U_\text{os}-V_+)$ at much larger detuning. Eq.~\eqref{eq:muchLargerDetuningJDerivativeAlpha} shows $\alpha(\Delta)\propto (\partial J^{\text{eff}}_\mathbb{L}/\partial \Delta)(\partial J^{\text{eff}}_\mathbb{R}/\partial \Delta ) $ is valid in this detuning range, which is confirmed with numerical result, cf. Supplementary Fig.~\ref{fig:alphaVSJTrend} (b). However, $\alpha \propto J^{\text{eff}}_\mathbb{L} J^{\text{eff}}_\mathbb{R}$ is not valid in this regime, as $J^{\text{eff}}_\mathbb{L}$ and $J^{\text{eff}}_\mathbb{R}$ increases along with $\Delta$ while $\alpha$ is remains constant, cf. Fig.~\ref{fig:JeffiBfoSSTTEvals} (a) and (b) in the main text.
\subsection{Large detuning at strong magnetic field strength}
When the applied magnetic field strength is strong, the behavior of capacitive coupling shows two different cases at large detuning. Using the result of GaAs DDQD device at $B = 0.087$ T as an example (Supplementary Fig.~\ref{fig:STEvalBfo0p753DiffDetuningScheme} (a)), we focus on two large detuning ranges: (1) between $\Delta^*_{|\widetilde{SS}\rangle,s}$ and $\Delta^*_{|\widehat{TT}\rangle,s}$ and (2) $\Delta^*_{|\widehat{TT}\rangle,s}$ and $\Delta^*_{|\widetilde{SS}\rangle,e}$. The values of $J^\text{eff}_\mathbb{L}$, $J^\text{eff}_\mathbb{R}$ and $\alpha$ exhibit the same behavior at $B=0.104$T (Fig.~\ref{fig:JeffiBfoSSTTEvals} (b) and (d)).

The logical subspace in case (1) is spanned by $\{|\widetilde{SS}\rangle,|S^{20}T^{11}\rangle,|T^{11}S^{02}\rangle, |T^{11}T^{11}\rangle \}$. Using the extended Hubbard Model introduced in Supplementary Sec.~\ref{subsec:HubbardModel} and include additional exchange Coulomb term, we have 
\begin{subequations}\label{eq:JeffAlphaLargeDetunStrMag}
	\begin{align}
		\begin{split}
			J^\text{eff}_\mathbb{L}&=J^\text{eff}_\mathbb{R}=\frac{1}{4}\left(\Delta-U_\text{os}+U_{12}-U_{14}+U_{23}-3U^e_{12}\right),
		\end{split}\\
		\begin{split}
			\alpha&=\frac{1}{4}\left(\Delta-U_\text{os}+U_{12}-U_{14}+U_{23}-3U^e_{12}\right),\label{sep:alphadep}
		\end{split}
	\end{align}
\end{subequations}
where we obtain
\begin{equation}\label{eq:alphaLarMagLarDelta}
	\alpha\propto\Delta\frac{\partial J^\text{eff}_\mathbb{L}}{\partial \Delta}\frac{\partial J^\text{eff}_\mathbb{R}}{\partial \Delta},
\end{equation}
conforming with the numerical result, cf. Supplementary Fig.~\ref{fig:alphaVSJTrend} (d). Eq.~\eqref{sep:alphadep} suggests the capacitive coupling strength is mediated by dipole-dipole interaction and electrostatic detuning energy.

In case (2), the logical subspace is spanned by $\{|\widetilde{SS}\rangle,|S^{20}T^{11}\rangle, |T^{11}S^{02}\rangle, |\widehat{TT}\rangle\}$. This results in
\begin{subequations}
	\begin{align}
		\begin{split}
			J^\text{eff}_\mathbb{L}&=J^\text{eff}_\mathbb{R}=\frac{1}{4}\left[\left(\hbar\omega_0-\hbar\omega_c\right)+\left(U_{15}-U_\text{os}\right)+\left(U_{28}-U_{24}\right)+\left(U_{18}-U_{14}\right)+\left(-2U^e_{12}-U^e_{15}\right)\right],
		\end{split}\\
		\begin{split}\label{eq:alphaLarMagLarDeltaNSSAlpha}
			\alpha&=\frac{1}{4}\left[\left(\hbar\omega_0-\hbar\omega_c\right)+\left(U_{15}-U_\text{os}\right)+\left(U_{28}-U_{24}\right)+\left(U_{18}-U_{14}\right)+\left(2U^e_{12}-U^e_{15}\right)\right].
		\end{split}
	\end{align}
\end{subequations}
In the limit of weak exchange Coulomb terms, $U^e_{jk}$, which are typically much smaller compared to on-site and inter-dot Coulomb interactions, $U_\text{os}$ and $U_{jk}$, we recover the result obtained from extended Hubbard Model, Eq.~\eqref{eq:alphaABOND}. Eq.~\eqref{eq:alphaLarMagLarDeltaNSSAlpha} suggests that the capacitive coupling in the nearly-sweet-spot region is mediated by the on-site Coulomb interactions and orbital excitation energy.

\section{System bath Hamiltonian}\label{sec:systemBathHam}
The Hamiltonian of a DDQD in a noisy environment can be expressed as \cite{Kornich.19}:
\begin{equation}\label{eq:wholeHam}
	\begin{split}
		H &= H_{\text{int}}+H_\text{z}+H_{\text{SOI}}+H_{\text{hyp}}+H_{\text{el-ph}}+H_{\text{ph}} +H_\varphi\\
		&= \widetilde{H} +H_{\text{el-ph}}+H_{\text{ph}}+H_\varphi,
	\end{split}
\end{equation}
where $H_\text{int}$ is the system Hamiltonian of interest, $H_\text{z}$ the Zeeman term, $H_{\text{SOI}}$ the spin-orbit coupling, $H_{\text{hyp}}$ the hyperfine coupling, $H_{\text{el-ph}}$ the electron-phonon interaction, $H_{\text{ph}}$ the phonon bath, and $H_{\varphi}$ the dephasing Hamiltonian. We have
\begin{subequations}
	\begin{align}
		\begin{split}
			\widetilde{H}&=H_\text{int}+H_\text{z}+H_\text{SOI}+H_{\text{hyp}},
		\end{split}\\
		\begin{split}
			H_\text{z} &=\frac{E_\text{z}}{2} \sigma_B,
		\end{split}\\
		\begin{split}
			H_\text{SOI} &\simeq g^* \mu_B (\boldsymbol{r} \times \boldsymbol{B}) \cdot  \boldsymbol{\sigma},
		\end{split}\\
		\begin{split}
			H_\text{hyp}&\simeq \frac{\Delta E_z}{2} |S^{11}\rangle\langle T^{11}|,
		\end{split}\\
		\begin{split}
			H_{\text{el-ph}}&=\sum_{\mathbf{q},s} W_s(\mathbf{q})a_{\mathbf{q}s}e^{i \mathbf{q}\cdot \boldsymbol{r}}+\text{H.c.},
		\end{split}\\
		\begin{split}
			H_\text{ph}&=\sum_{\mathbf{q},s}\hbar \omega_{\mathbf{q}s}\left(a^\dagger_{\mathbf{q}s}a_{\mathbf{q}s}+\frac{1}{2}\right),
		\end{split}\\
		\begin{split}
			H_\varphi&= H_{\varphi_{\mathbb{L}}}+ H_{\varphi_{\mathbb{R}}}+ H_{\varphi_{\mathbb{LR}}},
		\end{split}
	\end{align}
\end{subequations}
where $E_{z} = g^* \mu_B B$ is the Zeeman energy, $g^*$ the effective Land\'e-$g$ factor, $\mu_B$ the Bohr magneton, $B = |\boldsymbol{B}|$ the magnetic field strength, and $\sigma_B = \boldsymbol{\sigma}\cdot \boldsymbol{e_B}$ the Pauli spin operator along the direction of the magnetic field.  The hyperfine coupling strength $\Delta E_z$ couples states $|S^{11}\rangle$ and $|T^{11}\rangle$ of a DQD. $W_s (\mathbf{q})$ is the strength of electron-phonon interaction with phonon of type $s$ and wave vector $\mathbf{q}$, while $a^\dagger_{\mathbf{q}s}$ $\left(a_{\mathbf{q}s}\right)$ creates (annihilates) a phonon. The dispersion relation for the phonon is assumed to be $\omega_{\mathbf{q}s} = q \nu_s$ \cite{Van.89}. $H_{\varphi_\mathbb{L}}$, $H_{\varphi_\mathbb{R}}$ and $H_{\varphi_{\mathbb{LR}}}$ are charge-noise-induced dephasing terms on the single-qubit exchange energies, $J^{\text{eff}}_\mathbb{L}$ (for the left DQD),  $J^{\text{eff}}_\mathbb{R}$ (for the right DQD) and the capacitive coupling between the left and right DQD, $\alpha$, respectively.

To obtain the master equation, we first apply a unitary transformation, $\widetilde{U}$, that diagonalizes $\widetilde{H}$, resulting in an effective Hamiltonian, $H_\text{q} + H_\text{q-ph}+ H_\text{ph}$ as
\begin{subequations}
	\begin{align}
		\begin{split}
			\widetilde{U} \widetilde{H} \widetilde{U}^\dagger&= H_\text{q}= \frac{1}{2}g^*\mu_B \left[B^{\text{eff}}_\mathbb{L}\left(\sigma_z \otimes I \right)
			+B^{\text{eff}}_\mathbb{R} \left(I \otimes \sigma_z\right) +B^{\text{eff}}_{\mathbb{LR}} \sigma_z \otimes \sigma_z\right],
		\end{split}\\
		\begin{split}\label{eq:noisyEffHam}
			\widetilde{U} H_\text{el-ph} \widetilde{U}^\dagger&= H_\text{q-ph}= H_{\text{q-ph}_\text{dep}}+H_{\text{q-ph}_\text{rel}},
		\end{split}\\
	\end{align}
\end{subequations}
where 
\begin{subequations}\label{eq:HamPhEl}
	\begin{align}
		\begin{split}\label{eq:PhDephas}
			H_{\text{q-ph}_\text{dep}} &= \frac{1}{2}g^*\mu_B \left[\delta B_\mathbb{L} \left(\sigma_z \otimes I\right) + \delta B_\mathbb{R} \left(I \otimes \sigma_z \right) + \delta B_\mathbb{LR} \left(\sigma_z \otimes \sigma_z \right)\right],
		\end{split}
		\\
		\begin{split}\label{eq:PhRelax}
			H_{\text{q-ph}_\text{rel}} &=\frac{1}{2}g^*\mu_B \sum_{j<k} \delta B_{{jk}}(\sigma_{jk}+\text{H.c.}).
		\end{split}
	\end{align}
\end{subequations}

$B^\text{eff}_\mathbb{L}$, $B^\text{eff}_\mathbb{R}$ and $B^\text{eff}_\mathbb{LR}$ are the effective magnetic field while $\delta B_\mathbb{L}$, $\delta B_\mathbb{R}$, $\delta B_\mathbb{LR}$ and $\delta B_{jk}$ are phonon-induced noisy terms. $\sigma_{jk} = jk^\dagger$ are the coupling operators between logical states, with $\{j,k\} \in \{|TT\rangle,|TS\rangle,|ST\rangle,|SS\rangle\}$. The subscripts ``dep'' and ``rel'' refers to phonon induced pure dephasing and relaxation effect respectively.

\subsection{Charge-noise dephasing $\gamma_{\varphi}$}\label{sec:daphasingRate}
The charge-noise dephasing effect can be modeled as time-varying fluctuations on exchange energies and the capacitive coupling, i.e.
\begin{equation}\label{eq:HdephasChargeNoise}
	\begin{split}
		H_{\varphi_\mathbb{L}} &= \delta J^{\text{eff}}_\mathbb{L}(t) \left(\sigma_z \otimes I \right)=\hbar \nu_\mathbb{L} \left(\sigma_z \otimes I \right) f_\varphi (t) =\hbar \nu_\mathbb{L} \left(\sigma_z \otimes I \right) \int_{-\infty}^{\infty} f_{\varphi}(\omega) e^{i\omega t}d\omega,\\
		H_{\varphi_\mathbb{R}} &= \delta J^{\text{eff}}_\mathbb{R}(t) \left(I \otimes \sigma_z \right)=\hbar \nu_\mathbb{R} \left(I \otimes \sigma_z \right) f_\varphi (t) =\hbar \nu_\mathbb{R} \left(I \otimes \sigma_z \right) \int_{-\infty}^{\infty} f_{\varphi}(\omega) e^{i\omega t}d\omega,\\
		H_{\varphi_{\mathbb{LR}}}&=\delta \alpha (t)  \left(\sigma_z\otimes \sigma_z\right)=\hbar \nu_{\mathbb{LR}} \left(\sigma_z\otimes \sigma_z\right) f_{\nu_{\mathbb{LR}}} (t) = \hbar \nu_{\mathbb{LR}} \left(\sigma_z\otimes \sigma_z\right) \int_{-\infty}^{\infty} f_{\varphi}(\omega) e^{i\omega t}d\omega,
	\end{split}
\end{equation}
where $\delta J^{\text{eff}}_\mathbb{L}(t),\delta J^{\text{eff}}_\mathbb{R}(t)$, and $\delta J^{\text{eff}}_\mathbb{LR}(t)$ are the fluctuations of exchange energies and the capacitive coupling as functions of time, respectively, $f_\varphi (t)$ is a random function of time with zero mean, while $\nu_\mathbb{L}$,$\nu_\mathbb{R}$ and $\nu_\mathbb{LR}$ are the coupling magnitudes with exchange energies, $J^\text{eff}_\mathbb{L}$, $J^\text{eff}_\mathbb{R}$ and the capacitive coupling, $\alpha$ respectively.  Also, we use the identity $f_\varphi(t)=\int_{-\infty}^{\infty} f_\varphi (\omega) e^{i\omega t} d\omega$ to write $H_\varphi$ in the frequency space.

We apply the unitary transformation $\widetilde{U}$, on the Hamiltonian Eq.~\eqref{eq:wholeHam}, and move to the rotating frame defined by the transformation $\xbarl{U}=\exp[-i(H_\text{q}+H_\text{ph})t/\hbar]$. We have
\begin{subequations}\label{eq:noisyHam}
	\begin{align}
		\begin{split}
			\xbar{U} \widetilde{U} H_{\varphi_\mathbb{L}} \widetilde{U}^\dagger \xbar{U}\rule{0.8pt}{-18pt}^\dagger
			&=  \hbar \nu_\mathbb{L} \left(\sigma_z\otimes I\right) \int_{-\infty}^{\infty} f_\varphi (\omega) e^{i\omega t} d\omega
		\end{split},\\
		\begin{split}
			\xbar{U} \widetilde{U} H_{\varphi_\mathbb{R}} \widetilde{U}^\dagger \xbar{U}\rule{0.8pt}{-18pt}^\dagger
			&=  \hbar \nu_\mathbb{R} \left(I \otimes \sigma_z\right) \int_{-\infty}^{\infty} f_\varphi (\omega) e^{i\omega t} d\omega
		\end{split},\\
		\begin{split}
			\xbar{U} \widetilde{U} H_{\varphi_\mathbb{LR}} \widetilde{U}^\dagger \xbar{U}\rule{0.8pt}{-18pt}^\dagger 
			&=  \hbar \nu_\mathbb{LR} \left(\sigma_z \otimes \sigma_z \right)  \int_{-\infty}^{\infty} f_\varphi (\omega) e^{i\omega t} d\omega
		\end{split}.
	\end{align}
\end{subequations}
Eq.~\eqref{eq:noisyHam} implies the dephasing rates \cite{Boissonneault.09}:
\begin{equation}
	\begin{split}
		\gamma_{\varphi_\mathbb{L}} &= 2\nu_\mathbb{L}^2 S(\omega\rightarrow 0),\\
		\gamma_{\varphi_\mathbb{R}} &= 2\nu_\mathbb{R}^2 S(\omega\rightarrow 0),\\
		\gamma_{\varphi_\mathbb{LR}} &= 2\nu_\mathbb{LR}^2 S(\omega\rightarrow 0).
	\end{split}
\end{equation}
Since $\nu_\mathbb{L}$, $\nu_\mathbb{R}$ and $\nu_\mathbb{LR}$ indicate the sensitivity of exchange energies and the capacitive coupling to the charge noise, they can be parameterized by $\partial J^{\text{eff}}_\mathbb{L}/\partial \Delta$, $\partial J^{\text{eff}}_\mathbb{R}/\partial \Delta$ and $\partial\alpha/\partial\Delta$, respectively. We therefore have 
\begin{equation}
	\begin{split}
		\nu_\mathbb{L} &\propto \partial J^{\text{eff}}_\mathbb{L}/\partial\Delta,\\
		\nu_\mathbb{R} &\propto \partial J^{\text{eff}}_\mathbb{R}/\partial\Delta,\\
		\nu_{\mathbb{LR}} &\propto \partial\alpha/\partial\Delta,
	\end{split}
\end{equation}
resulting in
\begin{equation}
	\begin{split}
		\gamma_{\varphi_\mathbb{L}} &= \widetilde{\gamma}_\varphi \left( \frac{\partial J^{\text{eff}}_\mathbb{L}/\partial \Delta}{\left[\partial J/\partial \Delta\right]_\text{ref}}\right)^2,\\
		\gamma_{\varphi_\mathbb{R}} &= \widetilde{\gamma}_\varphi \left( \frac{\partial J^{\text{eff}}_\mathbb{R}/\partial \Delta}{\left[\partial J/\partial \Delta\right]_\text{ref}}\right)^2,\\
		\gamma_{\varphi_\mathbb{LR}} &= \widetilde{\gamma}_\varphi \left( \frac{\partial\alpha/\partial\Delta}{\left[\partial J/\partial\Delta\right]_\text{ref}}\right)^2,
	\end{split}
\end{equation}
where $\widetilde{T}_2=1/\widetilde{\gamma}_{\varphi}$, is the reference charge-noise dephasing time while $[\partial J/\partial \Delta]_\text{ref}$ is the reference derivative of the exchange energy, $J$, with respect to detuning.

For GaAs system, we refer to spin-echo measurement results by \cite{Dial.13}. At temperature $\mathcal{T}=50$ mK, the shortest dephasing time yields $T_2\approx 1$ $\mu s$ when $J=350$ MHz and $\left[\partial J/\partial V\right]_\text{ref} \approx 1.13 \times 10^3$ MHz/mV. Note that the notation $V$ is the gate voltage instead of the chemical potential detuning, $\Delta$, where $\Delta =V/9.4$ \cite{Dial.13}. As it was found that $T_2\propto \mathcal{T}^{-2}$ \cite{Dial.13}, we deduce the reference charge-noise dephasing time $\widetilde{T}_2$ to range from $6.25$ $\mu$s to $44.44$ $\mu$s for $ \mathcal{T}=7.5-20$ mK \cite{Scheller.14,Maradan.14,Nichol.17}, where $\mathcal{T} = 7.5$ mK has been achieved for GaAs QD device \cite{Scheller.14}. For silicon system, we refer to the Hahn-echo measurements by \cite{Jock.18}, which reported a charge-noise dephasing time of $8.4$ $\mu s$, at which $\left[\partial J/\partial \Delta\right]_\text{ref} \approx 0.24$ (inferred from the quasistatic charge-noise dephasing time of $1$ $\mu$s at the same detuning and $J\approx t^2/\Delta$ with $t = 0.7$ $\mu$eV). Therefore we set the reference charge-noise dephasing time $\widetilde{T}_2= 8.4$ $\mu s$ for the calculation of gate infidelities, unless otherwise indicated.
It was shown in \cite{Dial.13} that the charge-noise induced dephasing time extracted from spin-echo measurement is more relevant to power-law voltage noise, i.e. $1/f$ noise, rather than white noise. Therefore, a more comprehensive study would need to employ $1/f$ noise, but it is out of scope of this work and is left for future investigation.

\subsection{Phonon mediated relaxation $\gamma_{\text{rel}}$ and the pure dephasing $\gamma_{\text{dep}}$}\label{sec:relaxationRate}
Phonon mediated relaxation and pure dephasing rate can be obtained by adopting the Golden-Redfield theory \cite{Kornich.19}. They are
\begin{subequations}\label{eq:rate}
	\begin{align}
		\begin{split}
			\gamma_{\text{rel}_{jk}} =J_{jk}^+ (\omega_{jk}),
		\end{split}\\
		\begin{split}
			\gamma_{\text{dep}_{j}} =J^+_{\text{dep}_j}(0),
		\end{split}
	\end{align}
\end{subequations}
where $\gamma_{\text{rel}_{jk}}$ is the relaxation rate between the logical states $j$ and $k$, $\gamma_{\text{dep}_{j}}$ the pure dephasing rate of logical state $j$ and $\omega_{jk}=(E_k-E_j)/\hbar$. The pure dephasing rates are 
\begin{equation}
	\begin{split}
		\gamma_{\text{dep}_\mathbb{L}} &= J^+_{\text{dep}_{|TT\rangle}}(0) - J^+_{\text{dep}_{|SS\rangle}}(0) - \left(J^+_{\text{dep}_{|ST\rangle}}(0) - J^+_{\text{dep}_{|TS\rangle}}(0)\right),\\
		\gamma_{\text{dep}_\mathbb{R}} &= J^+_{\text{dep}_{|TT\rangle}}(0) - J^+_{\text{dep}_{|SS\rangle}}(0) + \left(J^+_{\text{dep}_{|ST\rangle}}(0) - J^+_{\text{dep}_{|TS\rangle}}(0)\right),\\
		\gamma_{\text{dep}_\mathbb{LR}} &= J^+_{\text{dep}_{|SS\rangle}}(0) - J^+_{\text{dep}_{|ST\rangle}}(0) - J^+_{\text{dep}_{|TS\rangle}}(0) + J^+_{\text{dep}_{|TT\rangle}}(0).
	\end{split}
\end{equation}
The r.h.s. of Eq.~\eqref{eq:rate} can be expressed as:
\begin{equation}
	\begin{split}
		J^+_{jk}(\omega)&=\text{Re}\left[J_{jk}(\omega)+J_{jk}(-\omega)\right]\\
		&=\frac{g^2\mu_B^2}{2\hbar^2}\int_{-\infty}^{\infty} \cos (\omega \tau) \langle \delta B_{jk}(0) \delta B_{jk}(\tau)\rangle d\tau,\\
		J^+_{\text{dep}_j}(0)&=2\text{Re}\left[J_{\text{dep}_j}(0)\right]\\
		&=\frac{g^2\mu_B^2}{\hbar^2}\int_{-\infty}^{\infty} \cos (\omega \tau) \langle \delta B_{j}(0) \delta B_{j}(\tau)\rangle d\tau\Bigg|_{\omega\rightarrow 0},
	\end{split}
\end{equation}
where $\delta B(\tau)= e^{i H_\text{ph}\tau/\hbar} \delta B e^{-i H_\text{ph}\tau/\hbar}$
and $\delta B_{jk}=\text{Re}\left[\delta B_{jk}\right]+\text{Im}\left[\delta B_{jk}\right]$. The temperature-dependent correlator $\langle \delta B_{j}(0) \delta B_{j}(\tau)\rangle$ can be calculated by exploiting the relation: $\langle a^\dagger_{\mathbf{q}s} a_{\mathbf{q'}s'} \rangle=\delta_{\mathbf{q},\mathbf{q'}}\delta_{s,s'}n_B(\omega_{\mathbf{q}s})$, where $n_B(\omega)$ is the Bose-Einstein distribution: 
\begin{equation}
	n_B(\omega)=\frac{1}{e^{\hbar \omega/(k_B T)}-1},
\end{equation}
$k_B$ the Boltzmann constant and $T$ the temperature.
In order to calculate $\widetilde{U} H_{\text{q-ph}_{\text{rel}}}  \widetilde{U}^\dagger$, we perform the Schrieffer-Wolff transformation up to the second order of electron-phonon coupling, corresponding to the contribution from two-phonon processes to the $\delta B$ term. The relaxation $\gamma_{\text{rel}}$, and the pure-dephasing rates $\gamma_{\text{dep}}$ are obtained for temperature $T=20$ mK \cite{Nichol.17}. The phonon parameters are extracted from \cite{Kornich.19} for the GaAs quantum-dot device.

\section{Pulse sequences for an entangling gate}
\begin{figure}[t]
	\centering
	\includegraphics[width=0.5\columnwidth]{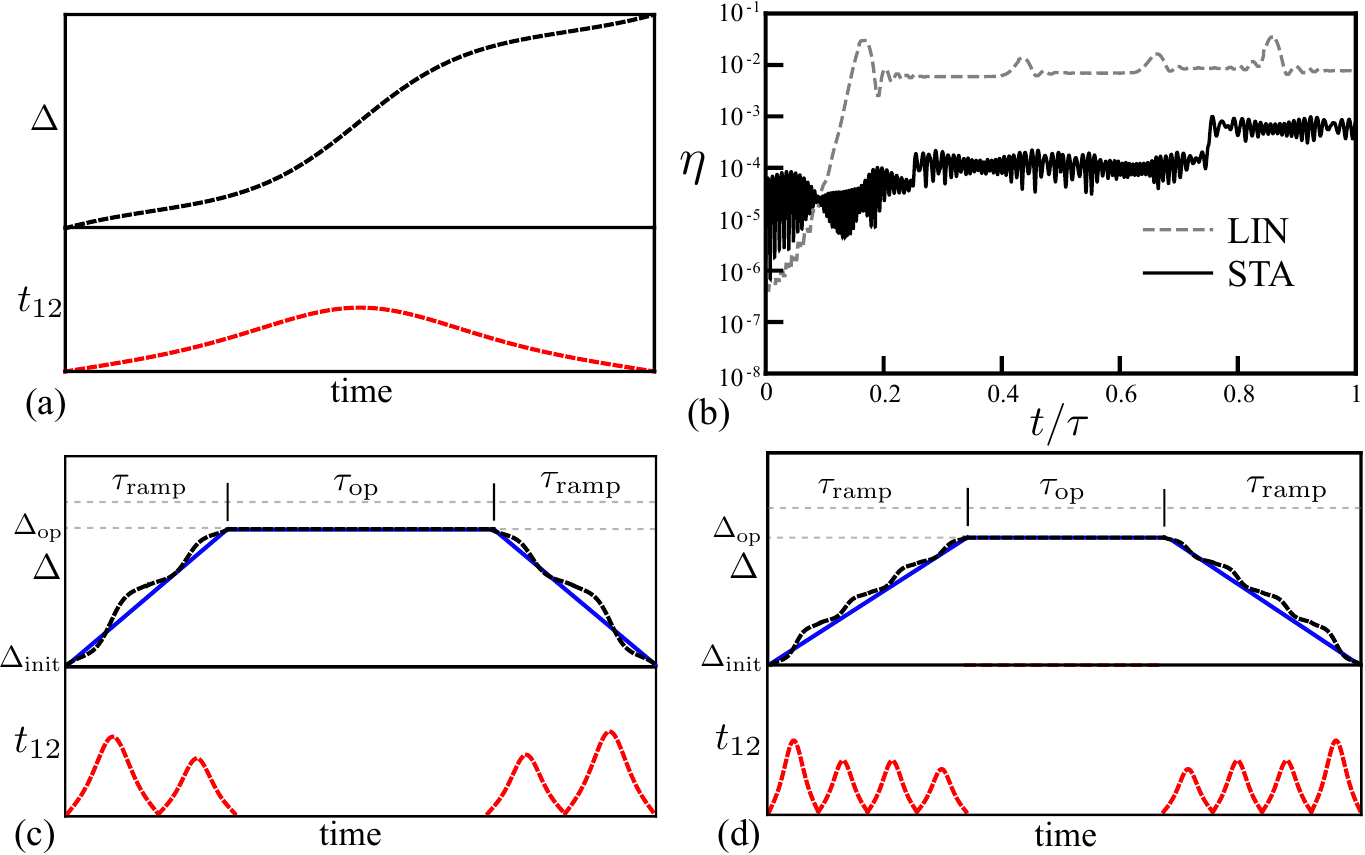}
	\caption{(a) Schematic figure of an STA pulse to perform high-fidelity state transfer in arbitrary time for a two-level system, based on Lewis-Riesenfeld invariants \cite{Chen.11}. The black (red) dashed line shows the detuning (tunnel coupling) as a function of time. 
		(b) The leakage, $\eta$, as a function of the normalized time $t/\tau$, where $t/\tau = 0$ and $1$ indicate the DDQD system residing at $(1,1,1,1)$ and $(2,0,0,2)$ charge configuration, respectively. At $t/\tau = 0$, the system is initialized as $|\Psi_0\rangle = \sum 1/2 |jk\rangle$, where the sum is over all logical eigenstates, $(j,k) = (S,T)$. The dashed line shows the leakage when linear ramping (LIN) is adopted, while the solid line shows results from concatenated STA pulses as shown in (d). To facilitate comparisons, the ramping time for both LIN and STA schemes is $\tau_\text{ramp}=7$ ns .
		(c) and (d): Concatenated STA pulses applicable for energy spectra of the logical states given by Fig.~\ref{fig:JeffiBfoSSTTEvals}(d) in the main text and Supplementary Fig.~\ref{fig:STEvalBfo0p753DiffDetuningScheme}(d). (c): Pulses performing a piecewise state transfer from (1,1,1,1) to (1,1,0,2)/(2,0,1,1)) charge regime. (d): Pulses performing piecewise state transfer from (1,1,1,1) to (2,0,0,2) charge regime.}
	\label{fig:STAPulse}
\end{figure}

In this section, we provide more details on pulse sequences performing an entangling gate. We focus on two ramping schemes: linear ramp (LIN), and shortcut to adiabaticity (STA).

We assume that the system is initialized in the (1,1,1,1) charge configuration, the detuning value for which is labeled by $\Delta_\mathrm{init}$. This is to ensure that the capacitive coupling $\alpha$ is ``off'' (i.e. negligible) \cite{Nichol.17,Shulman.12}. This setup also retains the coupling between logical states by the magnetic gradient $\Delta B$, i.e. $\langle S | H_\text{int} | T\rangle=\Delta B$. In order to perform an entangling gate operation, we need to ramp up the $\alpha$ value to another detuning point, which we denote as $\Delta_\mathrm{op}$. $\Delta_\mathrm{op}$ should be in the  $(1,1,0,2)/(2,0,1,1)$ or $(2,0,0,2)$ charge regimes, where the nearly-sweet-spot regime exists.

The simplest technique to detune the system from $\Delta_\mathrm{init}$ to $\Delta_\mathrm{op}$ is a linear ramp on the detuning, i.e. $d\Delta/dt=$constant (LIN scheme). The ramping time, denoted as $\tau_{\text{ramp}}$, must be on one hand long enough such that leakage due to the non-adiabatic effect is suppressed, and on the other hand short enough such that the exposure of the system to various decoherence channels is limited. These conflicting requirements severely limit the practicality of the LIN scheme. In this work, we use experimentally relevant ramping speed to perform numerical simulation on the master equation \cite{Dovzhenko.11,Koh.13}, Eq.~\ref{eq:masterEq} in the main text.

For a two-level system, the state transfer can be performed using shortcut to adiabaticity (STA) pulses generated by Lewis-Riesenfeld invariants \cite{Chen.11}. Although most experimental works focused on rectangular pulse with finite rise time and oscillating pulse \cite{Dovzhenko.11,Koh.13,Takeda.20,Nichol.17,Takeda.16}, it has been suggested to improve the gate performance by adopting pulse shaping operation \cite{Petit.20}. An example of the STA pulse is shown in Supplementary Fig.~\ref{fig:STAPulse}(a). To tailor the STA pulses for our purpose, it is useful to note that when the magnetic field is sufficiently strong, the avoided-crossing points for relevant charge states are well separated in detuning, making it possible to concatenate several elementary STA pulses. Taking the energy level structure shown in Supplementary Fig.~\ref{fig:STEvalBfo0p753DiffDetuningScheme}(d) as an example: The $(1,1,1,1)$ to $(1,1,0,2)/(2,0,1,1)$ transitions for $|SS\rangle$, $|ST\rangle$ and $|TS\rangle$ occur at similar detuning values, $\Delta \approx 1.85$meV; The next charge transition, $(1,1,1,1)$ to $(1,1,0,2)/(2,0,1,1)$ transitions for $|TT\rangle$ occurs at $\Delta \approx 2.15$meV; Transition into $(2,0,0,2)$ from $(1,1,0,2)/(2,0,1,1)$ for $|SS\rangle$ and $|ST\rangle/|TS\rangle/|TT\rangle$ occur at $\Delta \approx 2.24$meV and $2.43$meV respectively. In addition, the tunnel coupling between different charge configurations are all controlled by inter-dot tunneling, i.e. $t_{12}=\langle \Phi_1 |H|\Phi_2\rangle$ for $|SS\rangle$,$|ST\rangle$ and $|TS\rangle$ near $\Delta^*_{|\widetilde{SS}\rangle,s}$, and $t_{16}=\langle \Phi_1 |H|\Phi_6\rangle$ for $|ST\rangle$, $|TS\rangle$ and $|TT\rangle$ near $\Delta^*_{|\widehat{TT}\rangle,e}$,  both controlled by the barrier height between two dots within a DQD. Therefore, we place elementary STA pulses given by Supplementary Fig.~\ref{fig:STAPulse}(a) centering at the corresponding charge transition points, forming the concatenated piecewise STA pulses as shown in Supplementary Fig.~\ref{fig:STAPulse}(c) and (d). These concatenated STA pulse sequences allow state transfer of any arbitrary superposition of logical states as the input state.

We define the leakage $\eta$ as 
\begin{equation}
	\eta = \langle\Omega(t)|\Psi(t)\rangle,
\end{equation}
where $\Omega(t)$ and $\Psi(t)$ are the wavefunctions under adiabatic and non-adiabatic ramping at time $t$.
Supplementary Fig.~\ref{fig:STAPulse}(b) shows the leakage, $\eta$, as a function of the normalized time, $t/\tau$, for a DDQD system with $B = 0.104$T initialized in $(1,1,1,1)$ and evolved into the $(2,0,0,2)$ charge configuration. To illustrate the concatenated STA pulse sequences to perform state transfer on an arbitrary input state, the initial state is chosen to be the two-qubit counterpart, $|\Psi_0\rangle = \sum \frac{1}{2} |jk\rangle$, of a four-qubit maximally entangled state, $|\Psi_0\rangle = \sum \frac{1}{2} |jk,jk\rangle$ (cf. Sec.~\ref{sec:twoQGateFidelity}), where the sum is over all logical eigenstates, $(j,k) = (S,T)$. It can be observed that the leakage is substantially suppressed for STA pulses as compared to the LIN scheme. The residual leakage observed for STA pulses arises from the weak, yet non-negligible, coupling to other higher-lying states apart from the effective two-level system centered at each charge-transition point.

\section{Two-qubit Gate Fidelity}\label{sec:twoQGateFidelity}
In the main text, we evaluate the average gate fidelity  $F$ as \cite{Nielsen.02,Horodecki.99}:
\begin{equation}
	F = \frac{d F_e+1}{d+1},
\end{equation}
where $d$ is the dimension of the system ($d=4$ for a two-qubit system). The entanglement fidelity, $F_e$, for a noisy two-qubit gate is defined by setting the initial state as a maximally entangled state $|\Psi_0\rangle$ of four qubits, two of which is applied upon by the gate. To calculate $F$ for two-qubit gates on singlet-triplet qubits, the initial state is $|\Psi_0\rangle = \frac{1}{2} \sum_{j,k=S,T}{|jk,jk\rangle}$, with the initial density matrix $\rho_{\Psi_0} =|\Psi_0\rangle\langle\Psi_0|=\frac{1}{4}\sum_{j,k,m,n=S,T}|jk,jk\rangle\langle mn,mn|$. The resulting state after evolution in the noisy environment is then $|\widetilde{\Psi}\rangle = (\mathcal{N}_\chi \otimes I)[\rho_{\Psi_0}]$, where $\mathcal{N}_\chi[\rho] = \widehat{U}_\chi \rho \widehat{U}_\chi^\dagger$ and $\widehat{U}_\chi$ encapsulates the noisy effects. The entanglement fidelity is then  $F_e = \langle \Psi | (\mathcal{N}_\chi\otimes I)[\rho_{\Psi_0}] |\Psi\rangle$, where $|\Psi\rangle$ is the resulting state after an ideal evolution (without noise and leakage).

For evolution of time $\tau_\text{ent}=\pi/(2 \alpha)$, where $\alpha$ is the capacitive coupling strength, the entanglement fidelity is
\begin{equation}
	F_e(\tau_\text{ent})=\frac{1}{4}\left[1+e^{\frac{\pi}{\alpha}\left(\gamma_{\varphi_\mathbb{L}}+\gamma_{\varphi_\mathbb{LR}}\right)}+e^{\frac{\pi}{\alpha}\left(\gamma_{\varphi_\mathbb{R}}+\gamma_{\varphi_\mathbb{LR}}\right)}+e^{\frac{\pi}{\alpha}\left(\gamma_{\varphi_\mathbb{L}}+\gamma_{\varphi_\mathbb{R}}\right)}\right],
\end{equation}
where we neglect the phonon-induced relaxation and dephasing term for simplicity. It is reported by \cite{Shulman.12} that the white noise dephasing rate of left and right DQD yield $460$ ns and $510$ ns respectively when $\alpha/2\pi=0.87$ MHz. As $J^\text{eff}_\mathbb{L}/\alpha, J^\text{eff}_\mathbb{R}/\alpha \approx 300$, the two-qubit dephasing can be neglected \cite{Shulman.12}, i.e. $\gamma_\mathbb{LR}\rightarrow 0$. The above equation hence gives $F_e=0.64$, similar to the entanglement fidelity reported by \cite{Shulman.12}, i.e. 0.72.

\section{Gate Infidelity}\label{sec:GateInFResult}
\subsection{Gate Infidelity for silicon system}
\begin{figure}[t]
	\includegraphics[width=0.65\columnwidth]{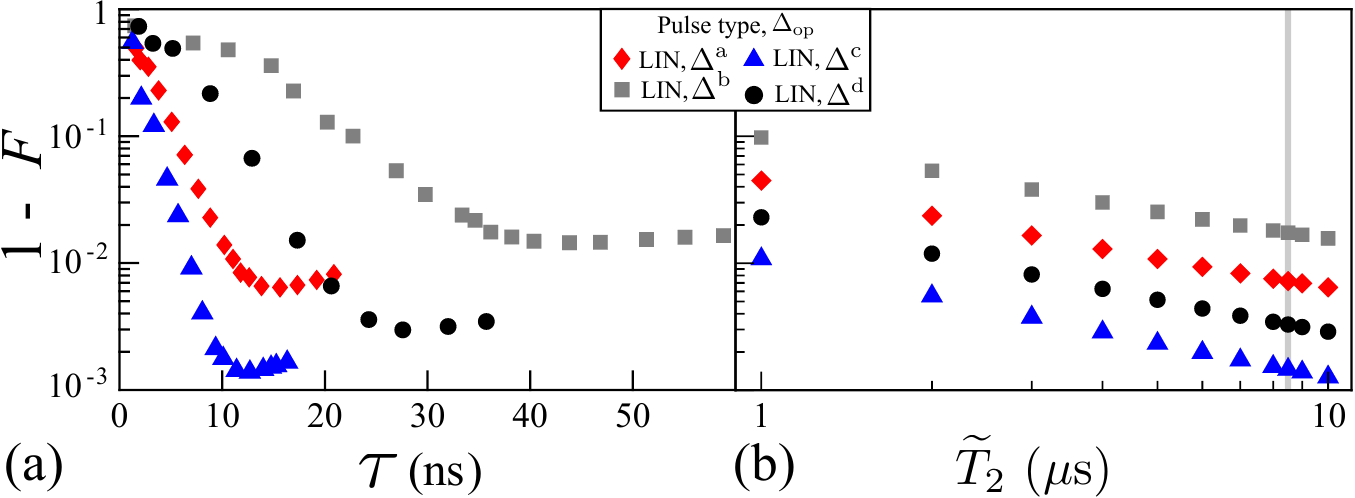}
	\caption{(a) CPHASE gate infidelities as functions of the total gate time, $\tau$, for $\widetilde{T}_2=8.4$ $\mu$s. (b) CPHASE gate infidelities as functions of the reference charge-noise dephasing time, $\widetilde{T}_2$. For each set of results, the gate time $\tau$ is chosen such that it produces the minimal gate infidelity as indicated in panel (a). The gray vertical line indicates the dephasing time of 8.4 $\mu$s reported by \cite{Jock.18}. The results are obtained for silicon system at $B = 0.296$ T.}
	\label{fig:GateInFSi}
\end{figure}
We have chosen $\Delta^\mathrm{a,b,c,d}$ as candidates of $\Delta_\mathrm{op}$ (as indicated on Supplementary Fig.~\ref{fig:STEvalSi} (a) and (d)), whose properties are the same as those chosen for GaAs DDQD device. Supplementary Fig.~\ref{fig:GateInFSi} shows the gate infidelities as functions of the (a) total gate time, $\tau$, and (b) reference charge-noise dephasing time, $\widetilde{T}_2$. Similar to case for GaAs system, $\Delta^{\text{c}}$ offers the maximum gate fidelities.

\subsection{Contribution of decoherence effect and leakage to the gate infidelity for GaAs system}
\begin{figure}[t]
	\centering{
		\includegraphics[width=0.45\columnwidth]{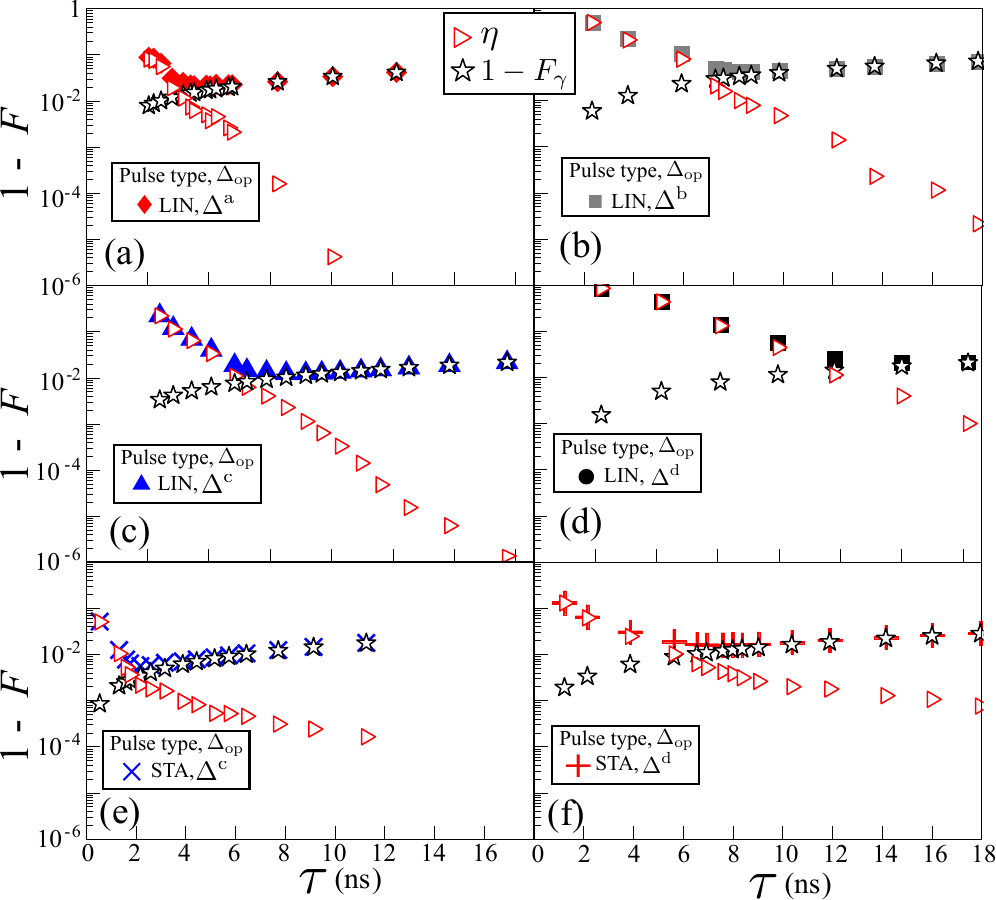}
		\includegraphics[width=0.45\columnwidth]{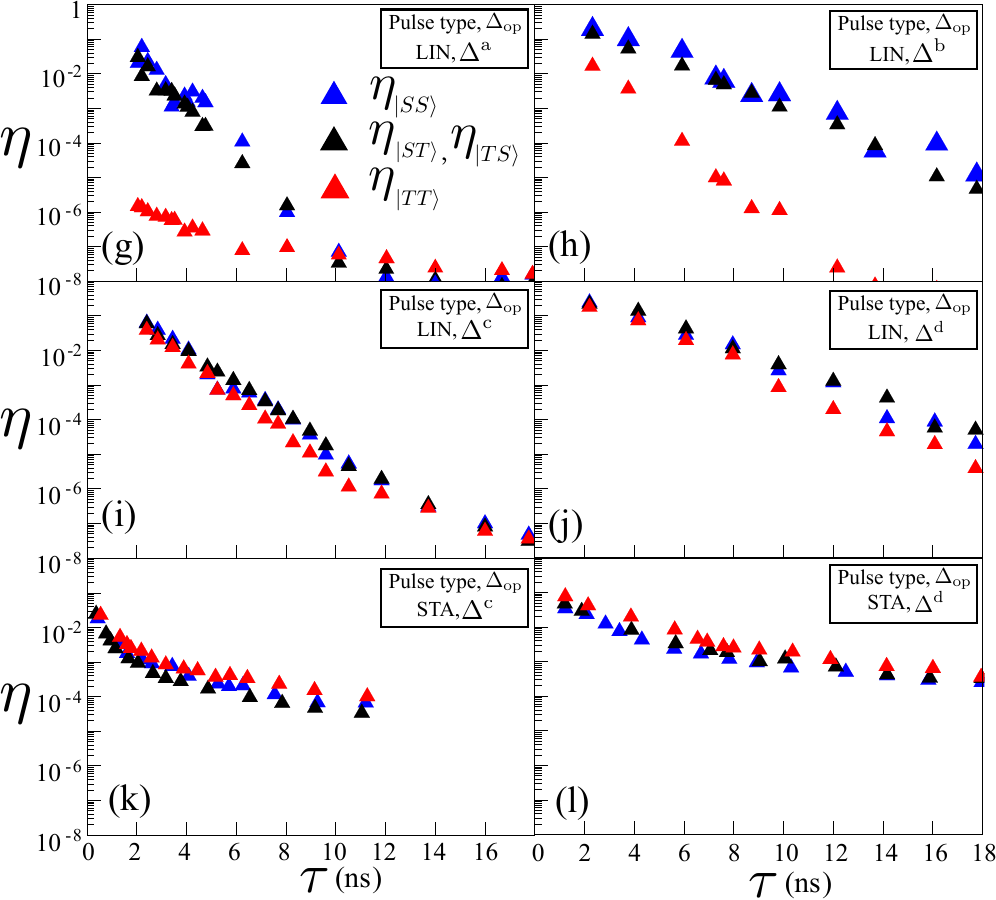}
	}
	\caption{Contribution of decoherence effect, $1-F_\gamma$, and leakage, $\eta$, to the gate infidelities, $1-F$, as functions of $\tau$ with $\Delta_\text{op}$ set at (a) $\Delta^\text{a}$ using LIN, (b) $\Delta^\text{b}$ using LIN, (c) $\Delta^\text{c}$ using LIN, (d) $\Delta^\text{d}$ using LIN, (e) $\Delta^\text{c}$ using STA and (f) $\Delta^\text{d}$ using STA. Leakage as functions of $\tau$ for different Hamiltonian blocks with $\Delta_\text{op}$ as (g) $\Delta^\text{a}$ using LIN, (h) $\Delta^\text{b}$ using LIN, (i) $\Delta^\text{c}$ using LIN, (j) $\Delta^\text{d}$ using LIN, (k) $\Delta^\text{c}$ using STA and (l) $\Delta^\text{d}$ using STA. Each leakage component is denoted as $\eta_\mathcal{H}$, where the subscript $\mathcal{H}$ indicates the Hamiltonian block. The results are obtained for GaAs system for $B=0$ T and $B=0.104$T.}
	\label{fig:infidelitiesCompt}
\end{figure}
Supplementary Fig.~\ref{fig:infidelitiesCompt} (a)-(f) show the contribution of decoherence effect (dephasing and relaxation), $1-F_\gamma$, and leakage,$\eta$ to the gate infidelities for a GaAs device. It can be observed that for short $\tau$, the gate fidelity is mainly limited by leakage while dephasing and relaxation effect does not contribute much to the gate infidelity relatively. On the other hand, for long $\tau$, the gate fidelity is mainly bounded by dephasing and relaxation while leakage is largely suppressed. In addition, Supplementary Fig.~\ref{fig:infidelitiesCompt} (c) and (e) show that the application of STA suppress the leakage effect for small $\tau$ before the accumulation of decoherence effect become prominent, resulting in the minimum gate infidelity among all cases. Fig.~\ref{fig:infidelitiesCompt} (g) shows that the leakage from $|TT\rangle$ is the lowest compared to other Hamiltonian blocks as performing a detuning ramp to $\Delta^\text{a}$ does not traverse through any charge transition point for $|TT\rangle$ subspace. On the other hand, a detuning ramping to  $\Delta^\text{b}$ undergoes a charge transition for $|TT\rangle$ from $|(1,1,1,1)\rangle$ to $|(2,0,0,2)\rangle$, thus the leakage is comparable for all Hamiltonian blocks at small $\tau$. Supplementary Fig.~\ref{fig:infidelitiesCompt} (i) and (k), where  $\Delta_\text{op} =\Delta^\text{c}$, show that the application of STA results in suppression of leakage for all Hamiltonian blocks at small $\tau$ as compared to linear ramping. The same goes for $\Delta^\text{d}$ when comparing Supplementary Fig.~\ref{fig:infidelitiesCompt} (j) and (l). Overall, for GaAs system, significant proportion of leakage occurs for $|SS\rangle$, $|ST\rangle$ and $|TS\rangle$ for linear ramping, while $|TT\rangle$ contributes the marginally higher leakage for STA.
\subsection{Contribution of decoherence effect and leakage to the gate infidelity for silicon system}
\begin{figure}[t]
	\centering{
		\includegraphics[width=0.45\columnwidth]{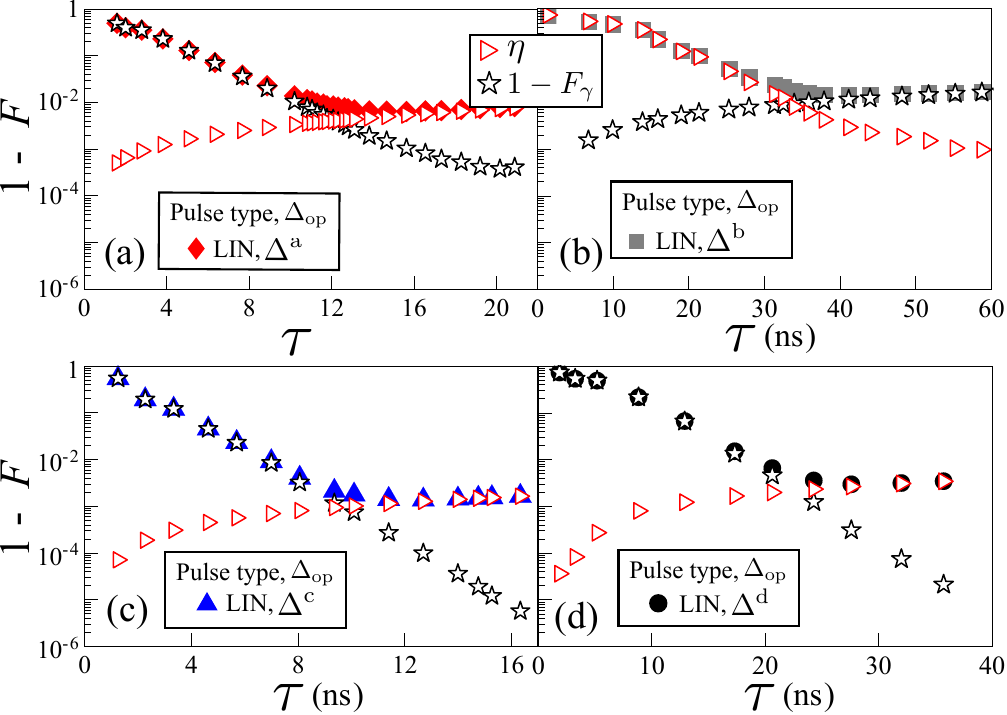}
		\includegraphics[width=0.45\columnwidth]{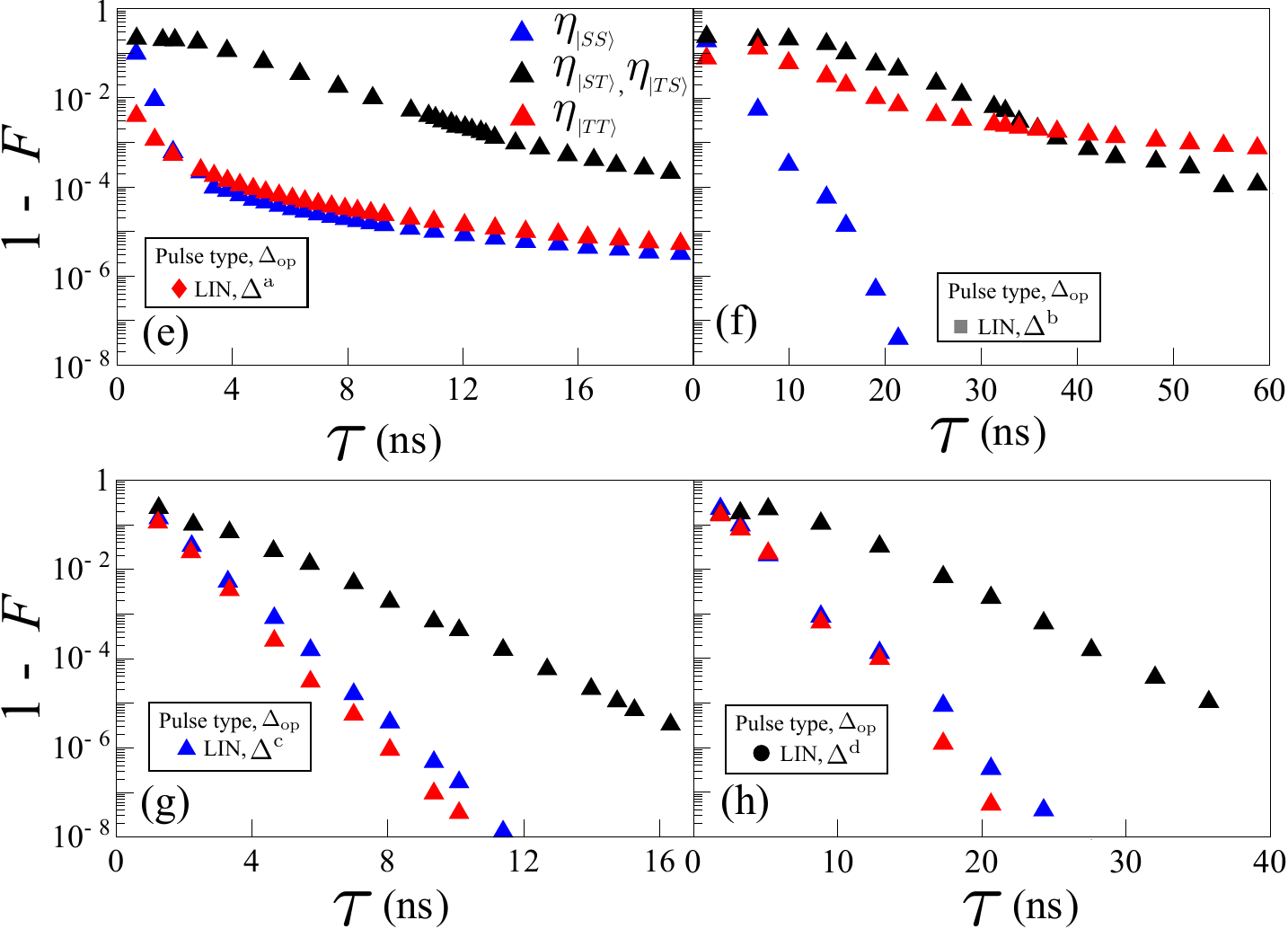}
	}
	\caption{Contribution of decoherence effect, $1-F_\gamma$, and leakage, $\eta$, to the gate infidelities, $1-F$, as functions of $\tau$ with $\Delta_\text{op}$ set at (a) ((b),(c),(d)) $\Delta^\text{a}$ ($\Delta^\text{b}$,$\Delta^\text{c}$,$\Delta^\text{d}$). Leakage as functions of $\tau$ for different Hamiltonian blocks with $\Delta_\text{op}$ as (g) $\Delta^\text{a}$ using LIN, (h) $\Delta^\text{b}$ using LIN, (i) $\Delta^\text{c}$ using LIN, (j) $\Delta^\text{d}$ using LIN, (k) $\Delta^\text{c}$ using STA and (l) $\Delta^\text{d}$ using STA. Each leakage component is denoted as $\eta_\mathcal{H}$, where the subscript $\mathcal{H}$ indicates the Hamiltonian block. The results are obtained for silicon system for $B=0$ T and $B=0.296$T.}
	\label{fig:InFComptsilicon}
\end{figure}
Supplementary Fig.~\ref{fig:InFComptsilicon} (a)-(d) show the contribution of decoherence effect and leakage to the gate infidelities for a silicon device. The results are similar to those shown for GaAs device, cf. Supplementary Fig.~\ref{fig:infidelitiesCompt} (a)-(f). Supplementary Fig.~\ref{fig:InFComptsilicon} (e)-(h) discern the leakage probabilities in different Hamiltonian blocks. In contrast to GaAs device, the total leakage probabilities are mostly committed by $|ST\rangle$ and $|TS\rangle$, while $|SS\rangle$ quickly decays as functions of the total gate time, $\tau$. This is attributed to the dissimilarity of the energy levels between GaAs and silicon system. Another notable difference is the leakage probability by $|TT\rangle$ is prominent at small $\tau$ for $\Delta^\text{a}$ since $\Delta^\text{a}$ has to be placed very close to the anticrossing between $|T^{11}\rangle|T^{11}\rangle$ and $|\widehat{TT}\rangle$, cf. Supplementary Fig.~\ref{fig:STEvalSi} (d).
\subsection{Hyperfine noise}\label{subsec:hyperfineNoise}
\begin{table}[t]
	\centering
	\begin{tabular}{|c|c|c|c|c|c|}
		\hline
		$\Delta_\text{op}$ & Pulse type & $\tau$ (ns) & $1- F_{\sigma_B}$ & $1- F_{\varphi}$& $\eta$\\
		\hline
		\hline
		& & & & &\\[-1em]
		$\Delta^{\text{a}}$ & \multirow{4}{*}{LIN} & $4.04$ &  $2.31\times10^{-3}$ & $2.12\times10^{-2}$& $2.11\times10^{-3}$ \\
		\cline{1-1} \cline{3-6}
		& & &  & &\\[-1em]
		$\Delta^\text{b}$ & & $8.72$ &  $1.45\times10^{-3}$  & $3.57\times10^{-2}$& $7.95\times10^{-3}$\\
		\cline{1-1} \cline{3-6}
		& & & & &\\[-1em]
		$\Delta^\text{c}$ & & $7.16$ &  $1.04\times10^{-3}$  & $1.15\times10^{-2}$& $1.14\times10^{-3}$\\
		\cline{1-1} \cline{3-6}
		& & & & &\\[-1em]
		$\Delta^\text{d}$ & & $14.16$ &  $4.54\times10^{-3}$  &$2.07\times10^{-2}$& $1.01\times10^{-3}$\\
		\hline
		& & & & &\\[-1em]
		$\Delta^\text{c}$ & \multirow{2}{*}{STA} &2.59&$6.20\times10^{-4}$  & $3.48\times10^{-3}$&$2.18\times10^{-3}$ \\
		\cline{1-1} \cline{3-6}
		& & & & &\\[-1em]
		$\Delta^\text{d}$& & 8.36 & $2.67\times10^{-3}$& $7.50\times10^{-3}$& $3.15\times10^{-3}$ \\
		\hline
	\end{tabular}
	\caption{Gate fidelities induced by the hyperfine noise, $1- F_{\sigma_B}$, for different $\Delta_{\text{op}}$, calculated using 500 noise realizations. Charge-noise induced gate infidelities, $1-F_\varphi$, and leakage, $\eta$, are shown in correspondence with $1- F_{\sigma_B}$. The gate time $\tau$ is chosen such that it produces the minimal gate infidelity as indicated in Fig.~\ref{fig:gateSummary} (b) of the main text.}
	\label{tab:tableHyperfineNoiseGateInF}
\end{table}
In GaAs system, dephasing due to fluctuation in the nuclear bath is a hurdle for gate operation \cite{Dial.13,Shulman.12,Nichol.17,Bluhm.10}. To obtain its effect on the gate fidelities, we numerically simulate the master equation Eq.~\eqref{eq:masterEq} over a discrete Gaussian distribution of quasistatic hyperfine noise with mean $\Delta B_z$ (zero in our case) and standard deviation $\sigma_{B}$. We define the hyperfine noise induced gate infidelities as
\begin{equation}
	1- F_{\sigma_B} = 1-\langle \Psi_\eta | (\mathcal{N}_{\sigma_B}\otimes I)[\rho_{\Psi_0}] |\Psi_\eta\rangle,
\end{equation}
where $|\Psi_\eta\rangle$ is the resulting state including the leakage for non-adiabatic ramping and $\mathcal{N}_{\sigma_B}$ is the evolution encapsulating only hyperfine noise, allowing us to focus on the effect of hyperfine noise. We use $\sigma_B = 0.5 $ mT, based on the decoherence time limited by hyperfine noise $\sim$$100$ ns \cite{Cerfontaine.20,Nichol.17,Dial.13}. Table~\ref{tab:tableHyperfineNoiseGateInF} shows that hyperfine induced gate infidelities are well below the gate infidelities contributed by charge noise and leakage, cf. Fig.~\ref{fig:gateSummary} in the main text and Supplemental Fig.~\ref{fig:infidelitiesCompt}. The insignificance of dephasing by hyperfine noise arises from the fact that 
the exchange energies and capacitive coupling are large in the range where the entangling gate is performed \cite{Shulman.12,Dial.13}.

\section{Elliptical Confinement Potential}
\begin{figure}[t]
	\centering{
		\includegraphics[width=0.75\columnwidth]{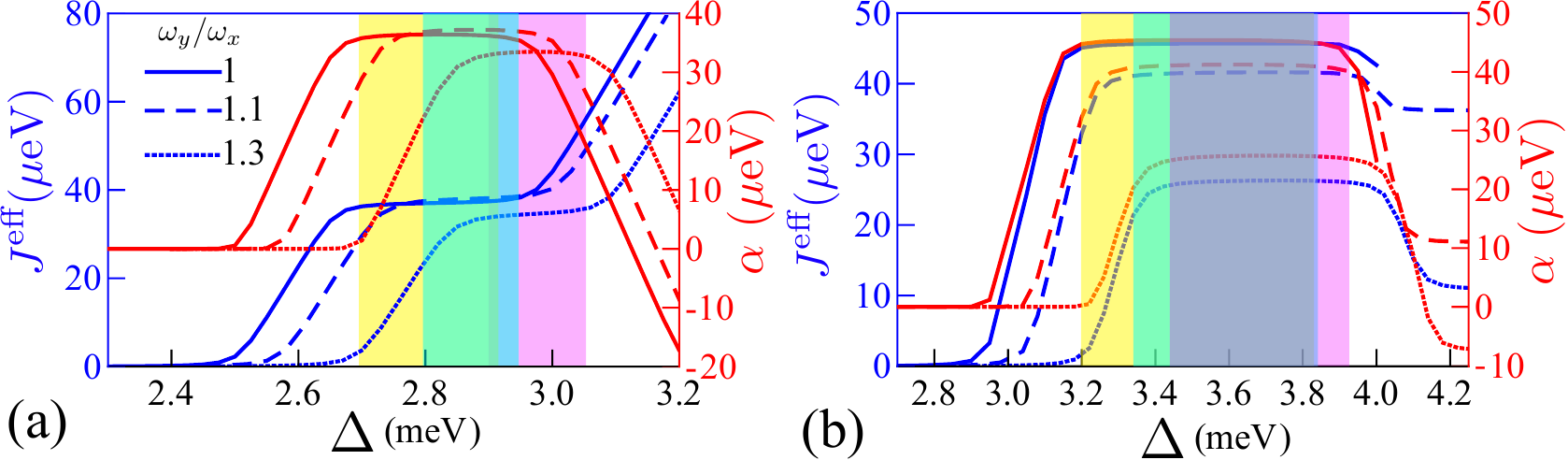}
	}
	\caption{Exchange energies, $J^\text{eff}$, and capacitive coupling, $\alpha$, as functions of the detuning $\Delta$ for (a) GaAs system at $B=0.186$ T and (b) silicon system at $B=0.395$ T for  elliptical confinement potential. The parameters are $\omega_x = \omega_0$, $\hbar\omega_0 = 1$ meV. Solid (dashed, dotted) line shows results for $\omega_y/\omega_x= \text{1 (1.1, 1.3)}$}
	\label{fig:ellip}
\end{figure}
Previous results are obtained for parabolic confinement potential, cf. Eq.~\eqref{eq:Vpotential} in the main text. Here, we performed calculation based on elliptical confinement potential \cite{Ezaki.97}, i.e.
\begin{equation}
	V(x,y) = \frac{1}{2} m^* \left\{\omega_x^2 \text{ Min}\Big[\left(x-R_{x \textit{1}}\right)^2+\Delta_{\it 1},\left(x-R_{\it x2}\right)^2+\Delta_{\it 2},
	\left(x-R_{\it x3}\right)^2+\Delta_{\it 3},\left(x-R_{x\textit{4}}\right)^2+\Delta_{\it 4}\Big]+\omega_y^2 y^2\right\},
\end{equation}
where $\omega_y/\omega_x \neq 1$.
Supplementary Fig.~\ref{fig:ellip} shows the effective exchange energies, $J^{\text{eff}}$ and capacitive coupling, $\alpha$, for ``Outer" detuning scheme. It can be observed that when the nearly-sweet-spot regime persists for elliptical confinement potential, cf. cyan and pink area in Supplementary Fig.~\ref{fig:ellip}, where $\partial J^{\text{eff}}/\partial \Delta$ is suppressed at which $\alpha$ at its maximal point.

\section{Magnetic Gradient}
\begin{figure}[t]
	\includegraphics[width=0.75\columnwidth]{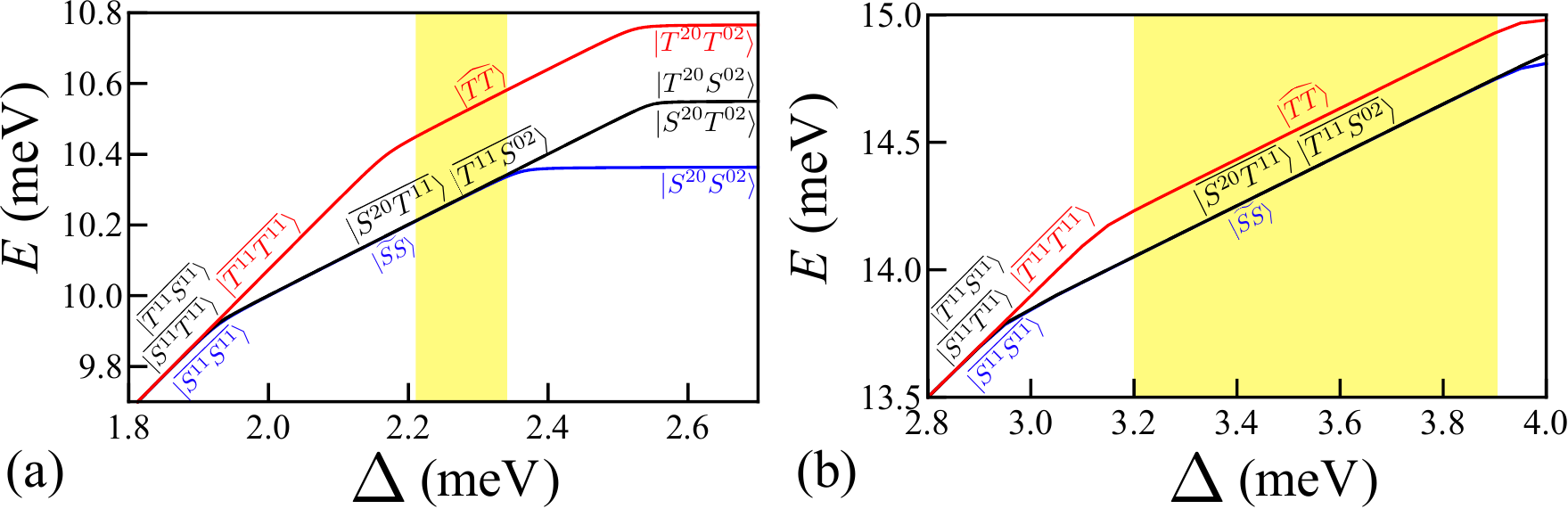}
	\caption{Energy levels as functions of detuning, $\Delta$, in presence of the magnetic gradient. (a) GaAs DDQD system at $B=0.104$ T and $\Delta E_\text{z} = 30.39$ MHz \cite{Shulman.12}. (b) Silicon DDQD system at $B=0.394$ T and $\Delta E_\text{z} = 48.6$ MHz (32 neV) \cite{Wu.14}.}
	\label{fig:MagGrad}
\end{figure}
In experiments, a magnetic gradient is applied to perform two-axis control on a singlet-triplet qubit. Here, we perform calculation in the presence of the magnetic gradient and the results are shown in Supplementary Fig.~\ref{fig:MagGrad}. We set a symmetric magnetic gradient on both DQDs, i.e. $E_{\text{z}_1}-E_{\text{z}_2} = E_{\text{z}_4}-E_{\text{z}_3} = \Delta E_\text{z}$, where $E_{\text{z}_j}$ is the Zeeman energy on the $j$-th dot from the left. The application of a magnetic gradient induces mixing between singlet and triplet states, hence, we denote the mixed states with an overline, e.g. $|\overline{S^{11}S^{11}}\rangle$ ($|\widetilde{SS}\rangle$ and $|\widehat{TT}\rangle$ are mixed with their triplet and singlet counterpart respectively, but we keep the same notations to signify the nearly-sweet-spot region). Using $|S^{11}S^{11}\rangle$ as an example, in presence of the magnetic gradient,
\begin{equation}
	|\overline{S^{11}S^{11}}\rangle = c_1 |S^{11}S^{11}\rangle + c_2 |S^{11}T^{11}\rangle  + c_3 |T^{11}S^{11}\rangle + c_4 |T^{11}T^{11}\rangle.
\end{equation}
Near the charge transition points, the energy splitting between $|S^{11}S^{11}\rangle$ and the other three logical states is large compared to $\Delta E_\text{z}$, resulting in $|c_1| \gg |c_2| \approx |c_3| \gg |c_4|$. Supplementary Fig.~\ref{fig:MagGrad} shows that the nearly-sweet-spot region persist in presence of the magnetic gradient, cf. yellow shaded region.

\section{Asymmetric dots}
\begin{figure}[t]
	\includegraphics[width=0.75\columnwidth]{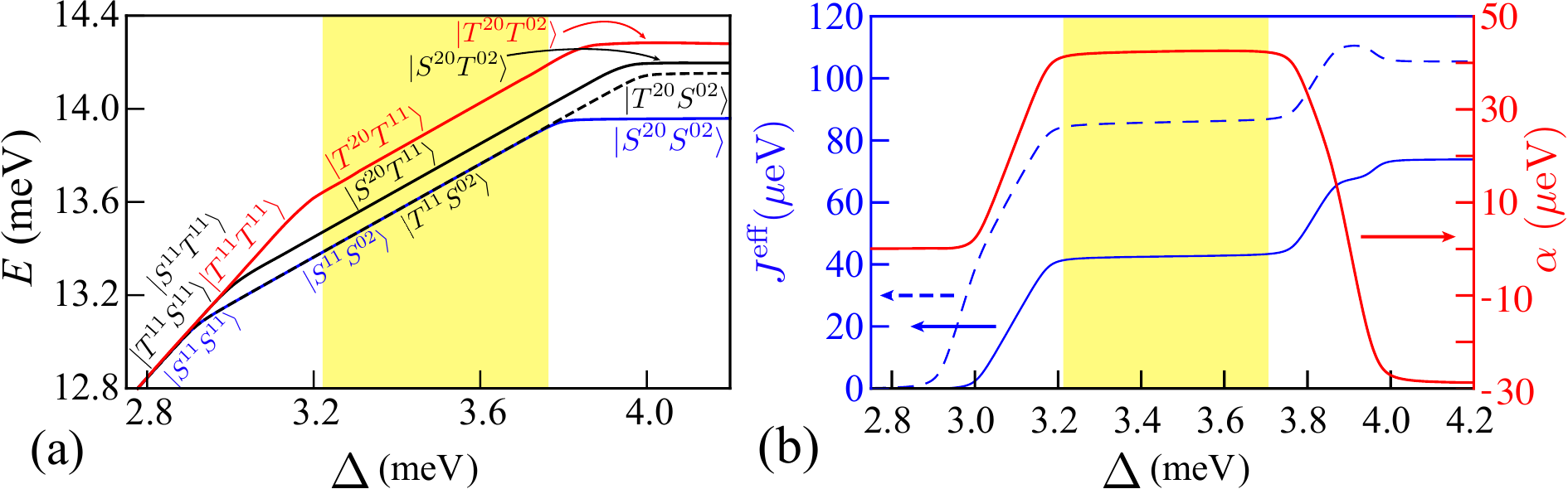}
	\caption{Nearly sweet spots for asymmetric dots of which the confinement energies of the quantum dots are $\{\hbar \omega_1,\hbar \omega_2,\hbar \omega_3,\hbar \omega_4\}$ = $\{0.93 \text{ meV}, 0.97\text{ meV}, 1.03\text{ meV}, 1.07\text{ meV}\}$ for a silicon DDQD device. The detunings are asymmetric as well, where $\Delta_1=\Delta_4=0$ while $\Delta_2 = \Delta_3 - 0.6$ meV, while $B=0.296$ T. (a) Energy levels v.s. detuning $\Delta$. (b) Exchange energies, $J^\text{eff}_\mathbb{L}$ and $J^\text{eff}_\mathbb{R}$, and capacitive coupling, $\alpha$, v.s. detuning.}
	\label{fig:asymDots}
\end{figure}
Most of the discussion in this work focus on symmetric case, i.e. constant value of $\omega_0$ across four quantum dots, which can be experimentally enforced by manually choosing the suitable gate tunings \cite{Li.14,Oxton.13,Angus.07,Henry.13test}. For completeness, we present here the result of calculation based on asymmetric dots. The confinement potential is defined as
\begin{equation}\label{eq:Vpotential2}
	V(\mathbf{r}) = \frac{1}{2} m^* \text{Min}\Big[\omega_{1}^2\left(\mathbf{r}-\mathbf{R}_{\it 1}\right)^2+\Delta_{\it 1},\omega_{2}^2\left(\mathbf{r}-\mathbf{R}_{\it 2}\right)^2+\Delta_{\it 2}+\omega_{3}^2\left(\mathbf{r}-\mathbf{R}_{\it 3}\right)^2+\Delta_{\it 3},\omega_{4}^2\left(\mathbf{r}-\mathbf{R}_{\it 4}\right)^2+\Delta_{\it 4}\Big].
\end{equation}
Supplementary Fig.~\ref{fig:asymDots} (a) shows the energy levels as functions of the detuning for asymmetric dots, i.e. the confinement strengths are all different for all quantum dot. It can be observed that the nearly-sweet-spot regime exists for asymmetric case, cf. the yellow shaded area. Supplementary Fig.~\ref{fig:asymDots} (b) shows that the region yields a relatively strong capacitive coupling, $\alpha$, while the susceptibility to the charge-noise is suppressed, $\partial J^{\text{eff}}/\partial \Delta \rightarrow 0$.

\begin{thebibliography}{88}
\bibitem [{s20()}]{s20t02Note}%
	\BibitemOpen
	\href@noop {} {}\bibinfo {note} {There are complications involving the
		$|T^{02}\rangle$ state and all such notations in the main text should be
		understood as two electrons occupying the lowest two levels respectively,
		i.e. $|T^{02}_\mathrm{L}\rangle$ in \cite{sm}.}\BibitemShut {Stop}%
\bibitem [{Fid()}]{FidelityNote}%
	\BibitemOpen
	\href@noop {} {}\bibinfo {note} {The definition of gate fidelity $F$ and the
		initial input state are explained in Sec. XI of \cite{sm}.}\BibitemShut
	{Stop}%
\end{thebibliography}
\end{document}